\documentclass[prb,pre,aps,twocolumn,amsmath,amssymb,floatfix,
superscriptaddress]{revtex4-1}
\usepackage[dvips]{graphics}
\usepackage{graphicx}
\usepackage{color}
\usepackage{blindtext}
\usepackage{amsmath}
\definecolor{dred}{rgb}{0.75,0,0}
\usepackage{graphicx}
\usepackage{dcolumn}
\usepackage{bm}
\usepackage{xcolor}
\usepackage{braket}
\usepackage{physics}
\usepackage{appendix}
\usepackage{hyperref}
\hypersetup{
 colorlinks=true,
citecolor=magenta,filecolor=blue,
 linkcolor=magenta,
 }

\usepackage{graphicx}
\usepackage{dcolumn}
\usepackage{bm}
\usepackage{xcolor}

\definecolor{codegreen}{rgb}{0,0.6,0}
\definecolor{codegray}{rgb}{0.5,0.5,0.5}
\definecolor{codepurple}{rgb}{0.58,0,0.82}
\definecolor{backcolour}{rgb}{0.95,0.95,0.92}
 
\begin{document}

\preprint{APS/123-QED}

\title{\textcolor{blue}{Topological phase transition and its stability against an applied magnetic field in a class of low dimensional decorated lattices}} 

\author{Sougata Biswas}
\affiliation{Department of Physics, Presidency University, 86/1 College Street, Kolkata, West Bengal - 700 073, India}
\affiliation{sougata.rs@presiuniv.ac.in}

\date{\today}

\begin{abstract}
The possibility of topological phase transition with or without a magnetic flux trapped in the cells of a class of decorated lattices is explored in details. Using a tight binding Hamiltonian and a real space decimation scheme we analytically obtain the non-dispersive and dispersive energy bands, and exactly locate the eigenvalues at which  energy gaps close. We find that despite the local breaking of the time reversal symmetry, as a magnetic field is turned `on',  the topological invariant exhibits quantization, and a clear appearance of edge localized modes is observed exactly at the gap-closing energy eigenvalues in the topologically non-trivial insulating phase. The bulk boundary correspondence is obeyed. The protection of the edge states by a chiral symmetry is confirmed for both the presence and absence of magnetic flux. Our studies have been extended to other kinds of decorated lattices where topological phase transition is possible only above a threshold value of the hopping integral describing these lattices. Our results are analytically exact.
\end{abstract}

\maketitle

\section{Introduction}
\label{intro}
The Su-Schrieffer-Heeger (SSH) model~\cite{su,heeger,asboth} have recently drawn immense attention from the condensed matter physics community as a fundamental and illustrative example of a one-dimensional (1D) topological insulator~\cite{thouless}. Topological effects have been realized recently through exciting experiments involving photonic systems~\cite{hening,malzard,weimann,yang,alex}, or using lasers~\cite{bandres,harder}. Topology-induced interface states are observed in dielectric resonator chains~\cite{poli}, electronics systems~\cite{qi, hasan,ezawa}, optical and acoustical systems~\cite{ozawa, segev, huang, ma}, for example. \par


\begin{figure}[ht]
\centering
\includegraphics[width=.9\columnwidth]{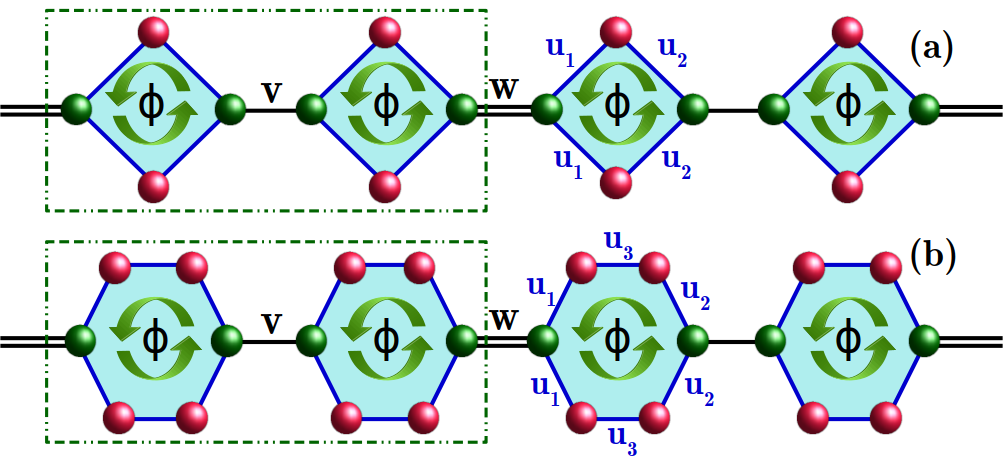}
\caption{(Color online) Schematic diagram of (a) Su-Schrieffer-Hegger Diamond (b) Su-Schrieffer-Hegger Hexagon lattices. Each diamond or hexagonal cavity traps a magnetic flux $\phi$. The unit cells are marked by the green-shared boxes. The distribution of overlap integrals is explained in the text.  }  
\label{fig1}
\end{figure}

Let us refer back to the SSH model that was primarily significant for shedding light on the properties of conjugated polymers~\cite{lu,baeriswyl}. The Hamiltonian of the SSH model is described conveniently within a tight binding formalism containing two periodically alternating overlap integrals $v$ and $w$, for example. Depending upon whether $v > w$ or $v < w$ the model exhibits topologically trivial and non-trivial phases. This change of phase is described by a topological invariant, viz, the Zak phase~\cite{zak} that flips its value from $0$ to  $\pi$. The appearance of robust edge localized states speaks about the bulk-boundary correspondence~\cite{asboth}. 

There are several examples where the topological phase transition (TPT) has been investigated using certain extensions (variants) of the basic SSH structure~\cite{xu,sougata}. Other examples include coupled SSH chains~\cite{li, sougata1}, SSH `trimer' lattice~\cite{anastasia,alvarez}, four-bond SSH model~\cite{bid}, chains with non-local coupling~\cite{mirosh}, topological insulators with non-centered inversion symmetry~\cite{ricardo1}, $2^n$ root topological insulator~\cite{ricardo3}, topological properties of two bosons~\cite{ricardo2}, and a study of the topological phases in a multistrand Creutz ladder~\cite{amrita} to name a few.\par
Almost all of these studies related to the TPT  have excluded the consideration of the influence of a perturbing field, a magnetic field for example. This is where we focus our attention to. In this communication, we will analyze issues related to the TPT in a class of Su-Schrieffer-Hegger (SSH) inspired decorated lattices both in the presence and in the absence of a uniform magnetic field. To simplify things we prefer to study in detail cases where the magnetic field pierces each individual plaquette in the lattices shown in  Fig.~\ref{fig1}. Two decorated SSH-Diamond (SSHD) and SSH-Hexagon (SSHH) geometries are depicted here. The magnetic field results in a flux trapped in the structural units, as shown, breaking the time-reversal symmetry (TRS) of the electron-hop around the plaquettes. Along the main linear backbone, in between the plaquettes there is no trapped flux and hence the TRS is preserved. Our first aim is to see the possibility of any TPT, existence of any edge mode, followed by a thorough investigation of the band structures in such lattices.  In general, this class of lattices exhibits non-dispersive flat bands that result out of destructive quantum interference, triggered by the local environment around a cluster of sites~\cite{leykam,derzhko1,bergholtz}. 

A series of natural questions floats up at this point. They can be framed as follows:

$(i)$ Is there any topological invariant associated with the flat band? \\
$(ii)$ What happens if a flat band and an edge state occur at the same eigenvalue? \\
$(iii)$ Is it possible to distinguish the edge state from the flat band energy at the topological non-trivial insulating phase?, and finally, \\
$(iv)$ Are there any special criteria using which one can avoid any possible merging of the flat band and edge state energies? \\
In this piece of work, using such decorated lattices, we intend to scrutinize and find an answer to these questions. This is our first point of interest. \par  
The second and an equally important concern for us is the observation that the trapping of a magnetic flux in a polygonal loop such as shown in Fig.~\ref{fig1} breaks the TRS locally, around the arms of the polygon. In general, a topological system obeys a group of symmetry operations along with the TRS. The effect of a non-zero magnetic flux on the energy bands and the possibility of a TPT under such a broken TRS  is not obviously understood, and thus it stands out as one of the main motivations of this present study. 

Finally, we extend our study to other similar kinds of decorated lattices. For SSH-Square-Hexagon (SSHSH)and SSH-Square-Octagon(SSHSO) lattices it is found that to see a TPT the external hopping parameter must be set above a {\it threshold value}. This is discussed in detail.  Towards the end of this paper, we discuss the impossibility of a TPT under the influence of a local magnetic field such as chosen in this work using an additional example of an SSH-Triangular (SSHT) lattice model. \par
The arrangement of our research work is as follows. In section~\ref{band} we introduce the model systems and write down the corresponding Hamiltonian using a tight binding formalism. After that with the help of a decimation technique, we evaluate gap-closing energies, flat bands, and the dispersion curves. A detailed analysis of band structures is done by direct diagonalizing of the $k$-space Hamiltonian. The calculation of the 
{ \it topological invariant} associated with each band and {\it symmetry operations} are explained in section~\ref{invariant}. In section~\ref{edge} the appearance of {\it edge localized states} in the topologically non-trivial insulating phase, the distribution of the amplitudes of the wavefunction, and the robustness of an edge state against disorder are discussed in detail. The topological behaviours of a similar class of lattices are investigated in section~\ref{model}. Finally, in section~\ref{conclusion}, we draw our conclusion.

\begin{figure}[ht]
\centering
(a)\includegraphics[width=.44\columnwidth]{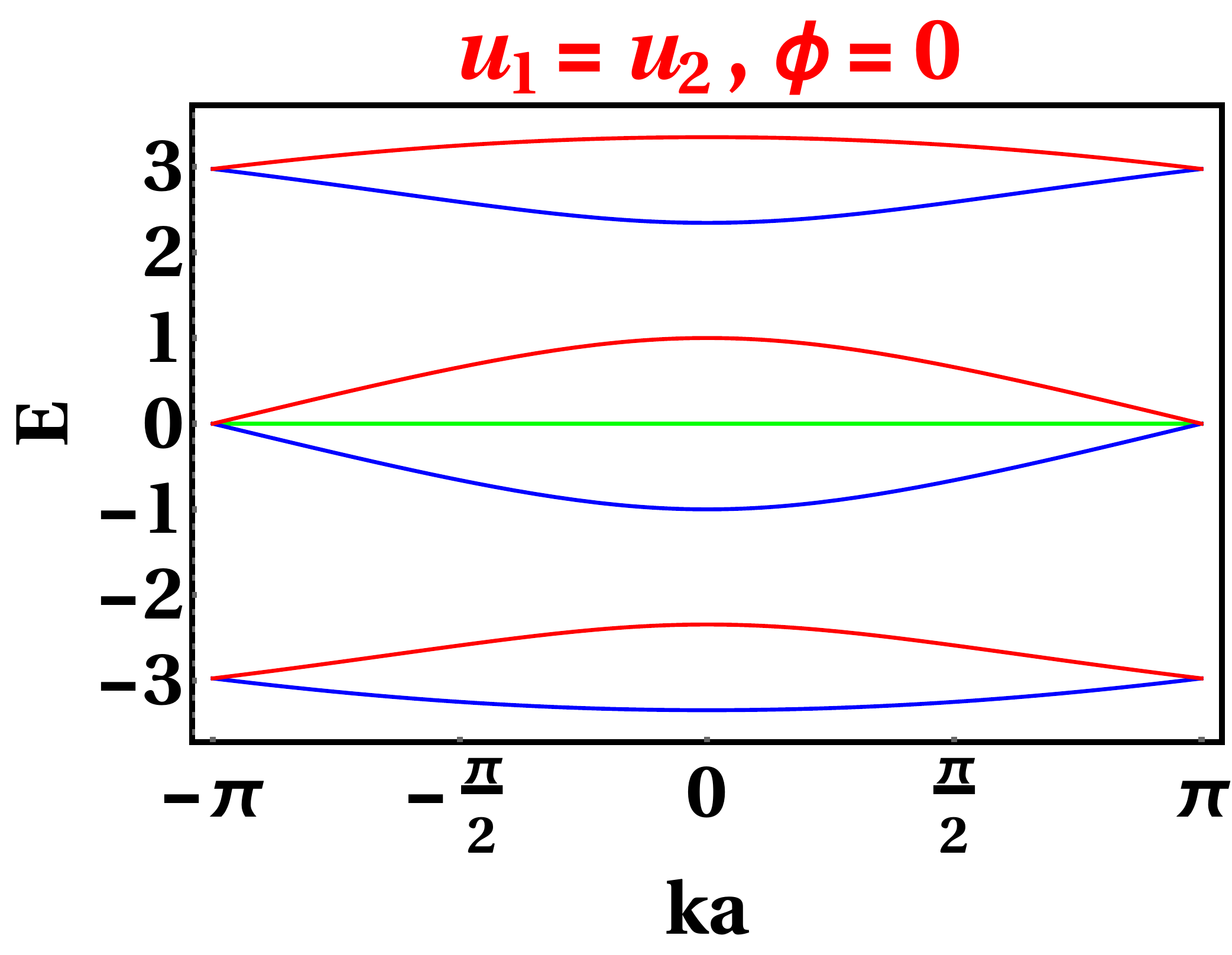}
(b)\includegraphics[width=.44\columnwidth]{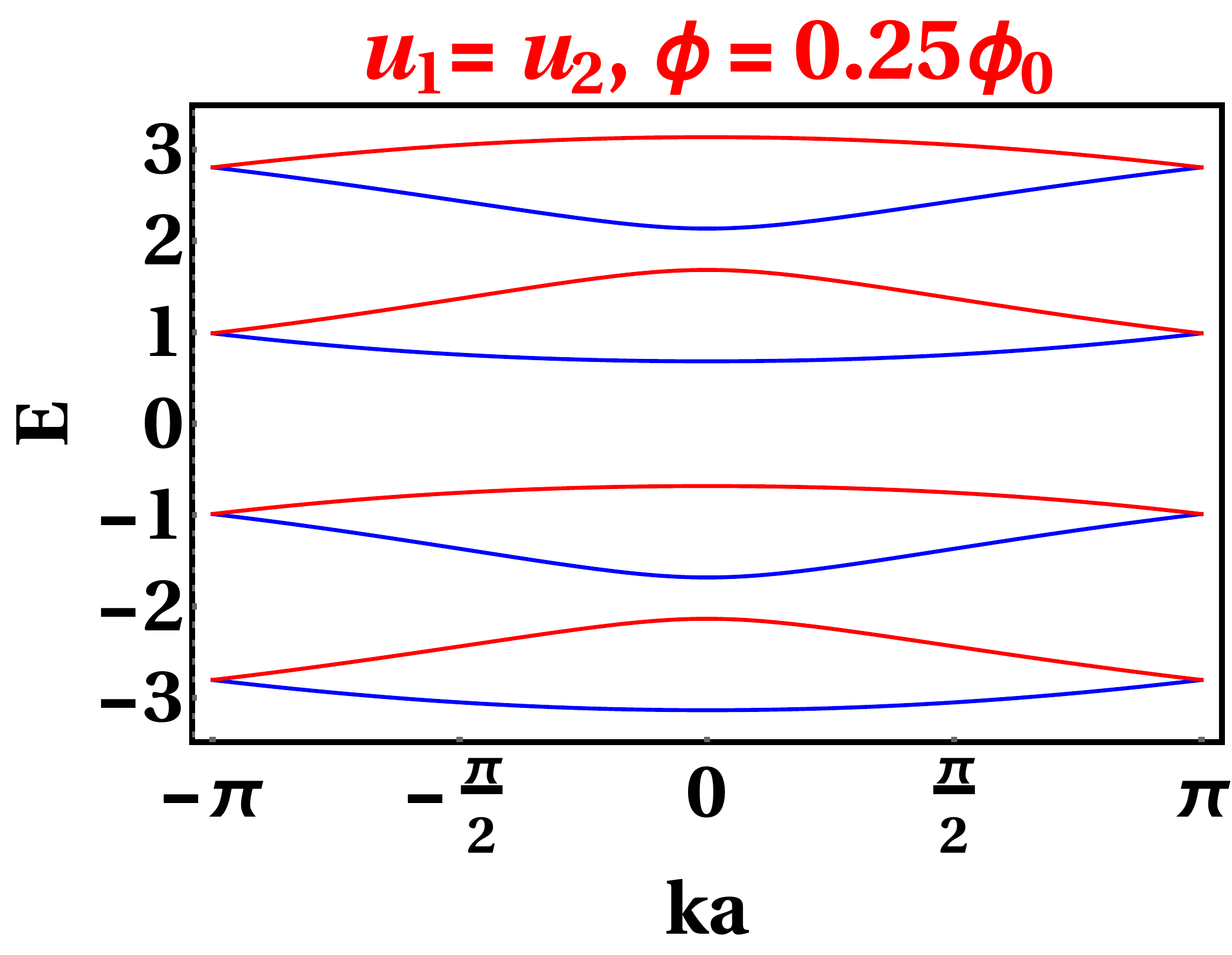}
(c)\includegraphics[width=.44\columnwidth]{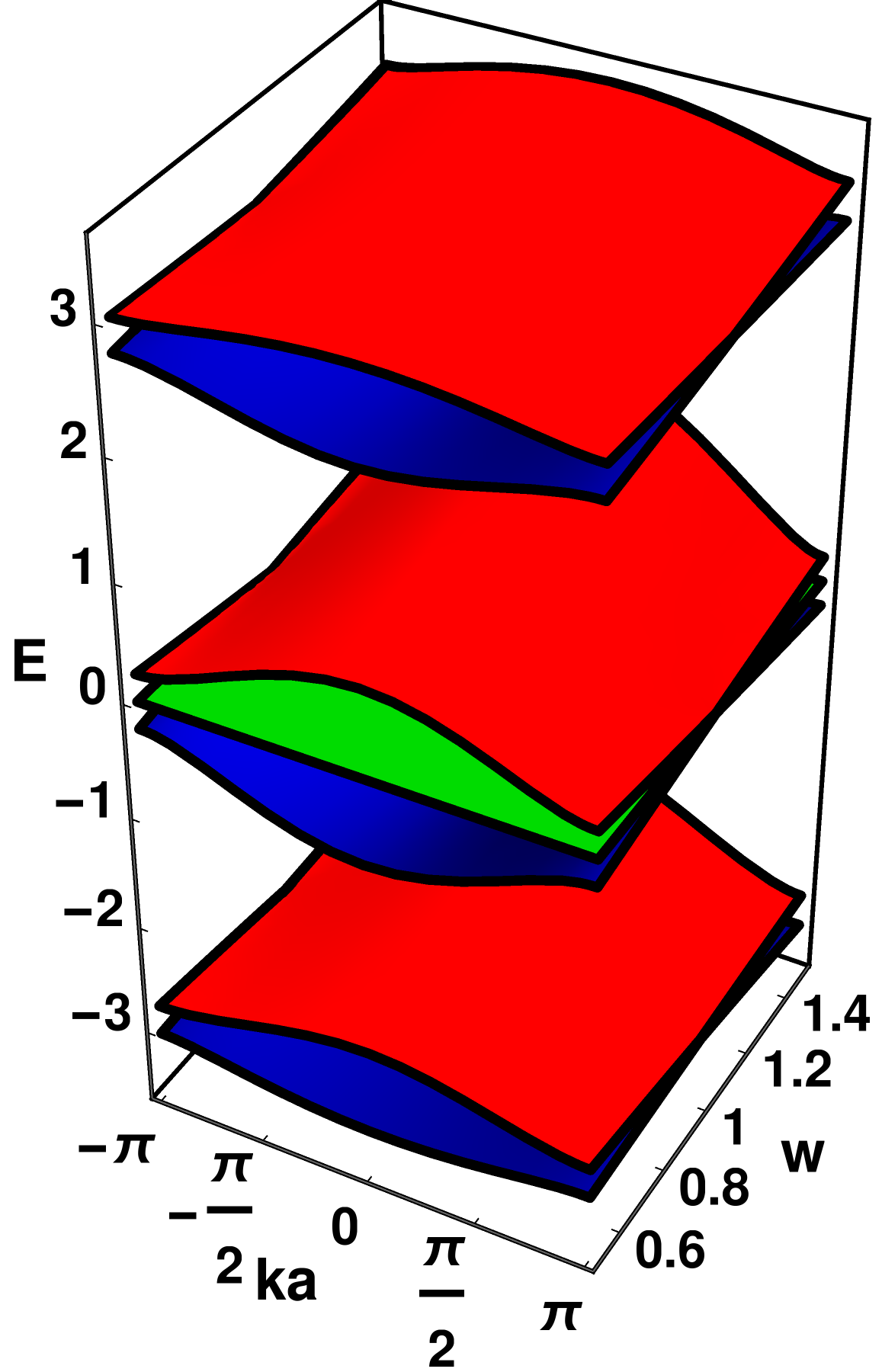}
(d)\includegraphics[width=.44\columnwidth]{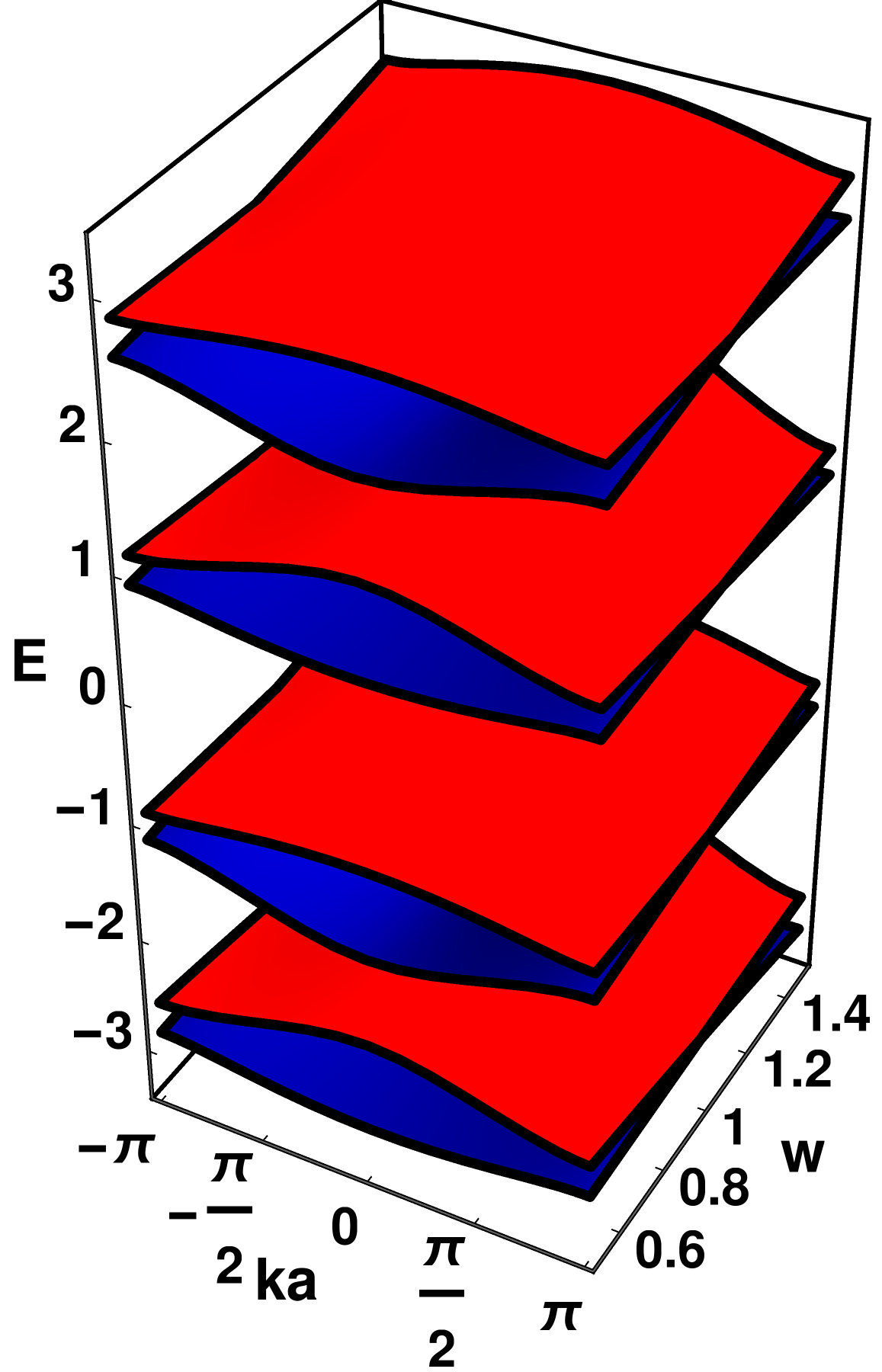}
\caption{(Color online) Energy vs wave vector (E vs ka) dispersion relations for the SSH-Diamond lattice. (a,c) are for zero flux and (b,d) show the results in presence of magnetic flux. The parameters are chosen as (a) $ \epsilon = 0, u_{1} = u_{2} = 1.4, v = 1, w = 1,  \phi = 0$, (b)  $ \epsilon = 0, u_{1} = u_{2} = 1.4, v = 1, w = 1,  \phi = \frac{1}{4} \phi_{0}$. The flat bands (doubly degenerate) are marked in green. (c) and (d) are the corresponding 3-d plots of (a) and (b) with different intercell hopping $w$. }  
\label{ek1}
\end{figure}


\section{Energy Bands}
\label{band}
\subsection{The model and Hamiltonian}
The tight binding Hamiltonian is written as, 
\begin{equation}
    H = \sum_j \epsilon_j c_j^\dag c_j + \sum_{<j,k>} t_{jk} c_j^\dag c_k + h.c.
    \label{ham}
\end{equation}
Here $c_j^\dag$ and $c_j$ present the creation and annihilation operator at the $j$-th atomic site. The `on-site' potential $\epsilon_j$ is chosen as $\epsilon$  for all sites, and we will set $\epsilon=0$ throughout the numerical calculation. The nearest neighbor overlap integrals are denoted by $t_{jk}$. In our system of interest, it is named $u_{l}$ ($l = 1, 2, 3$), $v$ or $w$ as shown in Fig.~\ref{fig1}. \par
When a magnetic flux is applied inside the diamond or hexagonal plaquettes (as depicted in Fig.~\ref{fig1}), the  time-reversal symmetry is {\it locally} broken. This implies that the electron-hopping in the `forward' or `backward' directions along the arms of the plaquettes differ in phase. The hopping integrals along the arms of these plaquettes, viz, ($u_{l}$'s) are associated with the {\it Peierls' phase factor} ($\pm\theta$)~\cite{hofs}. For SSH-diamond and SSH-hexagon lattices, we choose $\theta = \frac{2\pi\phi}{4 \phi_{0}}$ and  $\theta = \frac{2\pi\phi}{6 \phi_{0}}$ respectively. Here $\phi_{0} = \frac{hc}{e}$, the fundamental flux quantum. Without loss of generality, we select the positive sign for $u_{l} e^{ i \theta}$ when the electron ``hops" in an anti-clockwise direction along one arm of the cavity. We will refer to this as ``forward" hopping. Conversely, the term ``backward" hopping will be used for $u_{l} e^{- i \theta}$, corresponding to the electron hopping in a clockwise direction.\par

\subsection{Flat, Dispersive energy bands and Gap closing energy: Real space decimation scheme approach }

To determine the energy bands and to identify the eigenvalues where the energy gaps close at the boundaries of the Brillouin zone, we will diagonalize the Hamiltonian in $k$-space. Simultaneously, we will apply a real-space decimation method, which is built on a set of difference equations that correspond to a discretized form of the Schrödinger equation on a lattice. These difference equations are given by,

\begin{equation} 
(E - \epsilon_j) \psi_j = \sum_k t_{jk} \psi_k
\label{diff}
\end{equation}
where, the amplitude of the wavefunction at the $j$-th atomic site is $\psi_{j}$, and $k$ runs over the nearest neighbors of the $j$-th site. 

First, we decimate out the red-colored atomic sites in both decorated lattices as shown in Fig~\ref{fig1}. Now the `renormalized' onsite potentials and the hopping amplitudes  turn out to be, 
\begin{eqnarray}
\epsilon_{A} = \epsilon_{1} + \frac{t_{12}^2+v^2}{(E-\epsilon_{2})} \nonumber\\
\epsilon_{B} = \epsilon_{1} + \frac{t_{12}^2+w^2}{(E-\epsilon_{2})} \nonumber\\
t_{A} = \frac{t_{12} v}{(E-\epsilon_{2})} \nonumber\\
t_{B} = \frac{t_{12} w}{(E-\epsilon_{2})}
\label{eq1}
\end{eqnarray}
For SSH-Diamond Lattice (SSHDL) $\epsilon_{1}=\epsilon+(2 u_{1}^2)/(E-\epsilon)$, $ \epsilon_{2}=\epsilon+(2 u_{2}^2)/(E-\epsilon)$ and $ t_{12}=(2 u_{1} u_{2} \cos{2 \theta})/(E-\epsilon)$. For the SSH-Hexagon Lattice (SSHHL) it can be written as $\epsilon_{1}=\epsilon+ 2 u_{1}^2/(E-\Tilde{\epsilon})$, $\epsilon_{2}=\epsilon+2 u_{2}^2/(E-\Tilde{\epsilon})$ and $t_{12}=(2 u_{1}u_{2}u_{3} \cos{3 \theta})/(E-\epsilon)(E-\Tilde{\epsilon}) $, where $\Tilde{\epsilon}=\epsilon+u_{3}^2/(E-\epsilon)$.\par
From the above equations, it is clear at $v=w$ that the energy bands are closed at the Brillouin Zone (BZ) boundaries. The particular energies at which the Gaps are closed are the solutions of the equation $E - \epsilon_{A(B)} = 0$.\par

Using Eq.~\ref{eq1} the gap-closing energies are calculated for SSHDL. Under a non-zero magnetic flux, it becomes,
\begin{equation}
    E =  \frac{1}{2}\left(2 \epsilon \pm \sqrt{2} \sqrt{2u_{1}^2+2u_{2}^2+v^2 \pm \xi_{1}} \right)
    \label{gapSSHD}
\end{equation}
where $\xi_{1} = (4u_{1}^4+4u_{2}^4 + 4u_{1}^2v^2 + 4u_{2}^2v^2+v^4 + 8u_{1}^2u_{2}^2 \cos{2 \pi \phi})^{\frac{1}{2}}$. When no magnetic flux is trapped ($\phi = 0$) the energies for the SSH-diamond lattice turn out as, $E = \epsilon, \epsilon \pm \sqrt{2 u_{1}^2 + 2 u_{2}^2 +v^2}$. This result completely matches the dispersion curves as shown in Fig.~\ref{ek1}\par
Similarly, the gap-closing energies of the SSH-hexagon lattice can also be obtained. At $\phi = 0$, these are written as,
\begin{equation}
    E = \frac{1}{2} \left(2\epsilon \pm \sqrt{2} \sqrt{2u_{1}^2+2u_{2}^2+u_{3}^2+v^2 \pm \xi_{2}} \right)
    \label{gapSSHH}
\end{equation}
Here $\xi_{2} = (4 u_{1}^4-8u_{1}^2u_{2}^2+4u_{2}^4+4u_{1}^2u_{3}^2+4u_{2}^2u_{3}^2+u_{3}^4+4u_{1}^2v^2+4u_{2}^2v^2-2u_{3}^2v^2+v^4)^{\frac{1}{2}}$. At $u_{1} = u_{2} = u_{3}$ the energies become $E = \epsilon \pm u_{3}$ and $E = \epsilon \pm \sqrt{4u_{3}^2+v^2}$. In the presence of magnetic flux, these gap-closing energies are also calculated from the solution of $E -\epsilon_{A(B)} = 0$ with the help of Eq.~\ref{eq1}. Corresponding graphical results are shown in Fig.~\ref{ek2}. \par


\begin{figure}[ht]
\centering
(a)\includegraphics[width=.44\columnwidth]{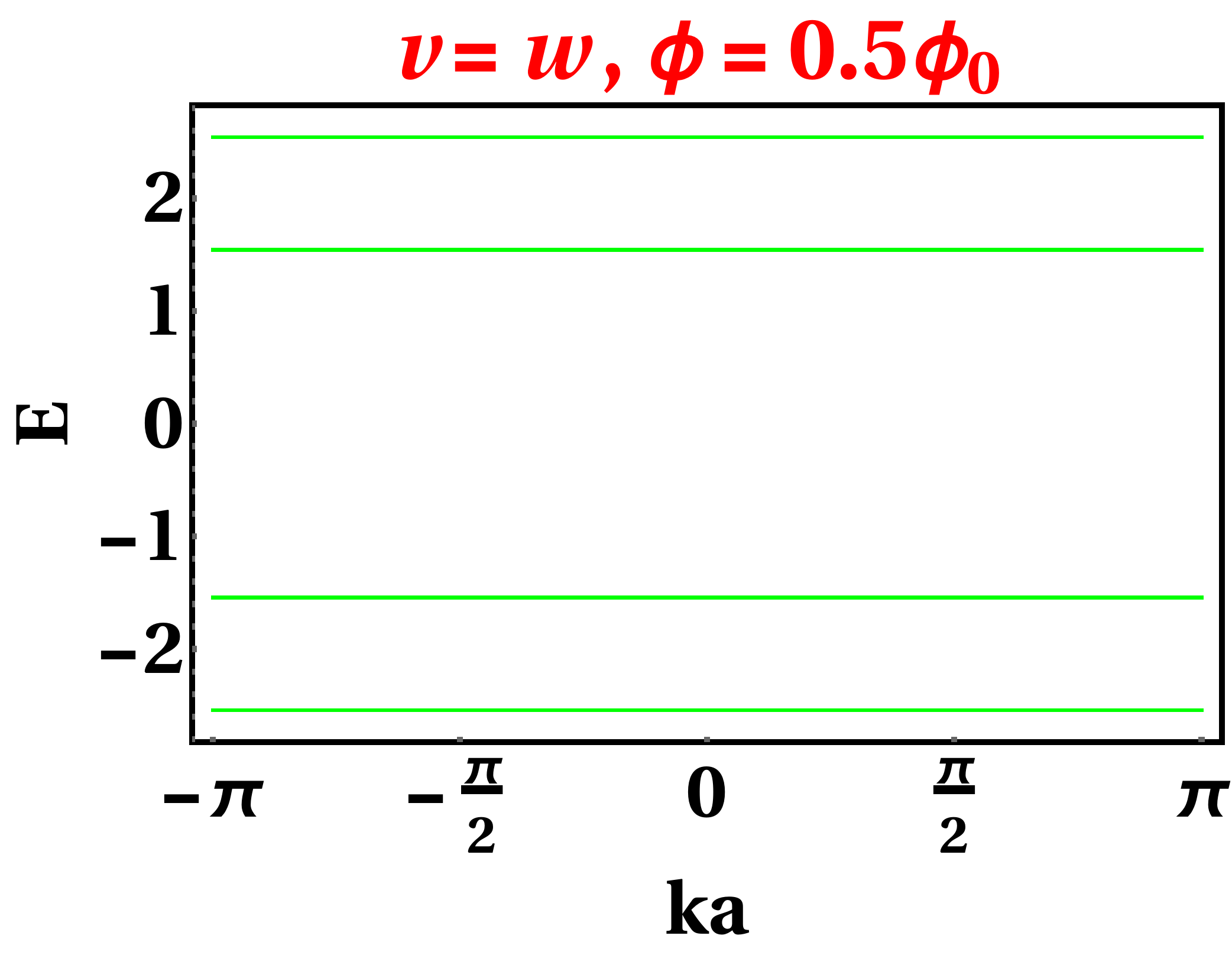}
(b)\includegraphics[width=.44\columnwidth]{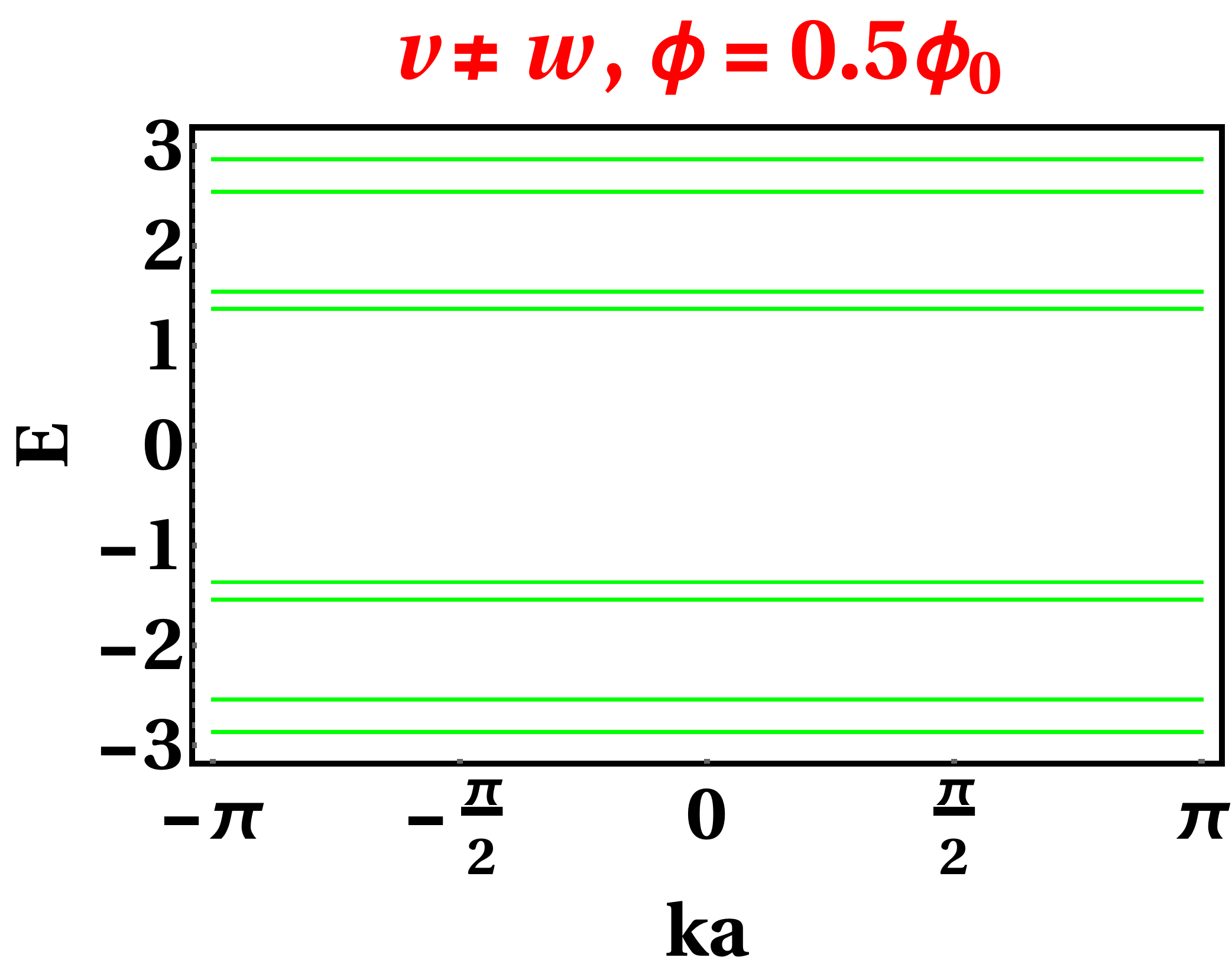}
(c)\includegraphics[width=.44\columnwidth]{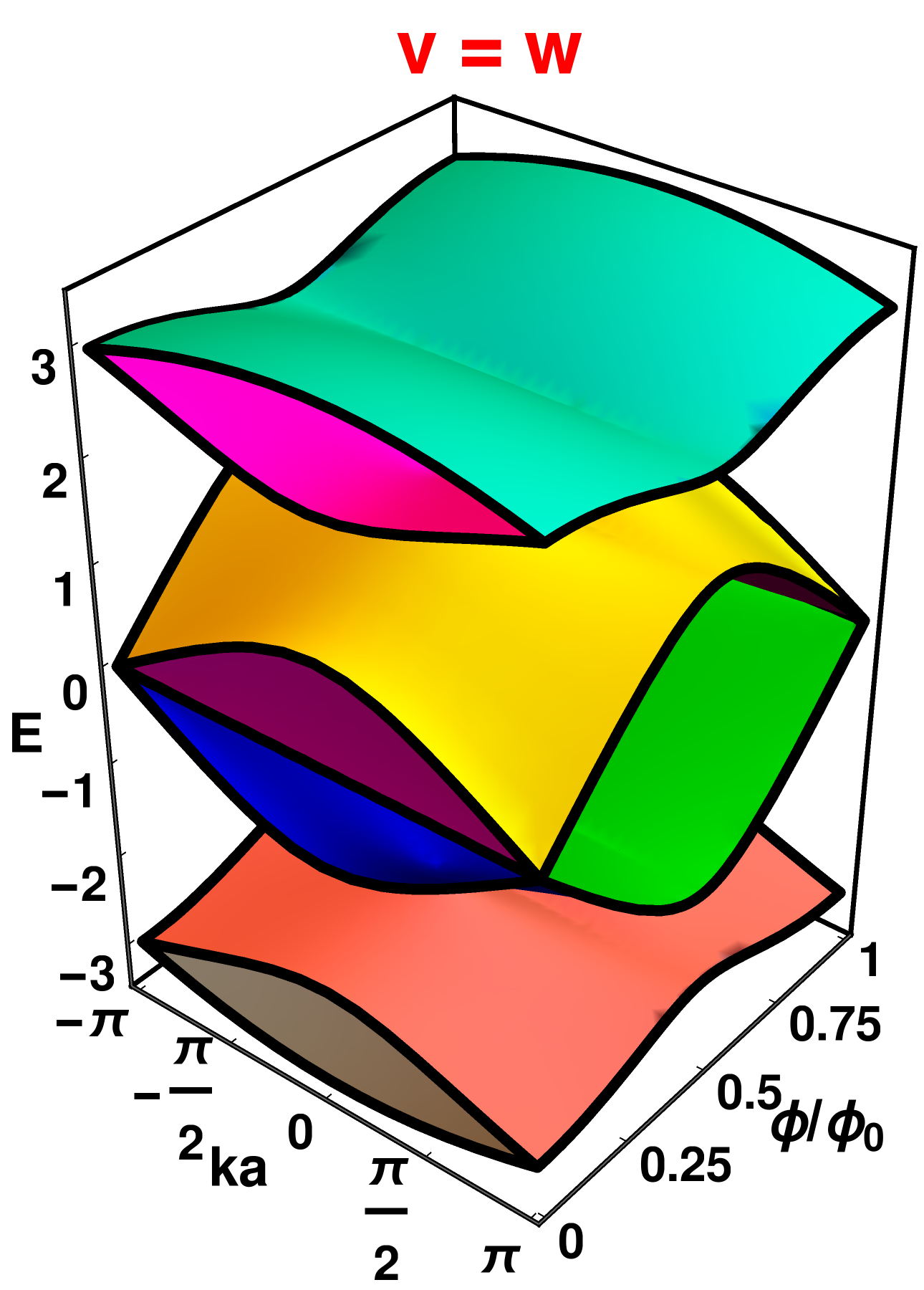}
(d)\includegraphics[width=.44\columnwidth]{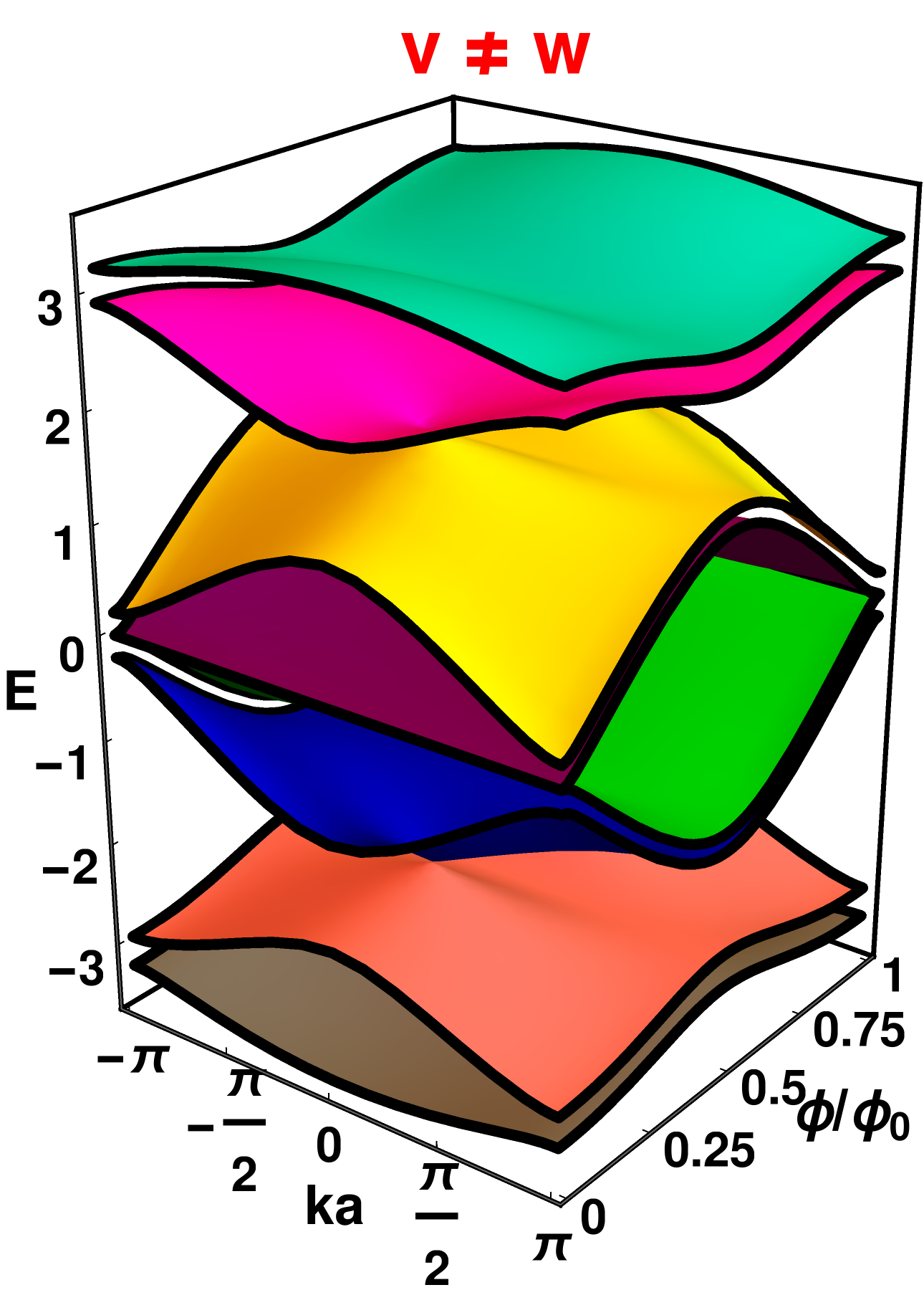}
\caption{(Color online) Energy vs wave vector (E vs ka) dispersion relations for the SSH-Diamond lattice (a,b) in the presence of magnetic flux $\phi = 0.5 \phi_{0}$. The parameters are chosen as (a) $ \epsilon = 0, u_{1} = u_{2} = 1.4, v = 1, w = 1$, (b)  $ \epsilon = 0, u_{1} = u_{2} = 1.4, v = 1, w = 1.5 $. (c) and (d) are the corresponding 3-d plots of the dispersion relations with magnetic flux $\phi$. }  
\label{ek-flat}
\end{figure}


One more decimation can map these staggered lattices into a simple one-dimensional periodic chain with a new onsite potential $\epsilon_{eff}$ and a uniform hopping integral $t_{eff}$, given by,
\begin{eqnarray}
\epsilon_{eff} & = & \epsilon_{A} + \frac{{t_A}^2 + {t_B}^2}{E - \epsilon_B} \nonumber \\
t_{eff}  & = & \frac{t_A t_B}{E - \epsilon_B}
\label{a}
\end{eqnarray}
It is now simple to work out the dispersion relations for the effective 1-d  chains. The dispersion relation is written as, 
\begin{eqnarray}
 E & = & \epsilon_{eff} + 2 t_{eff} \cos~ka'
\label{}
\end{eqnarray}

Here, `$a'$' represents the effective lattice spacing of the renormalized 1-d periodic chain and $k$ is the wave vector.\par
In the absence of magnetic flux the dispersion relation for the SSHDL and SSHHL can be written as, 
\begin{widetext}
\begin{eqnarray}
    (E - \epsilon)^2 \left[\beta_{1} (\alpha_{1}\beta_{1}-\gamma_{1})(\alpha_{1}\beta_{1} - \delta_{1})-(E-\epsilon)^2\beta_{1}((4u_{1}^2u_{2}^2(v^2+w^2)+8u_{1}^2u_{2}^2vw\cos{ka}))\right] = 0\nonumber\\
    \left[(E - \epsilon)^2-u_{3}^2 \right]^2 \left[\beta_{2} (\alpha_{2}\beta_{2}-\gamma_{2})(\alpha_{2}\beta_{2} - \delta_{2})-((E-\epsilon)^2-u_{3}^2)^2\beta_{2}((4u_{1}^2u_{2}^2u_{3}^2(v^2+w^2)+8u_{1}^2u_{2}^2u_{3}^2vw\cos{ka}))\right] = 0
    \label{eq2}
\end{eqnarray}
\end{widetext}
where $\alpha_{1} = (E-\epsilon)^2-2u_{1}^2$, $\beta_{1} = (E-\epsilon)^2-2u_{2}^2$, $\gamma_{1} = 4u_{1}^2u_{2}^2+v^2(E-\epsilon)^2$, $\delta_{1} = 4u_{1}^2u_{2}^2+w^2(E-\epsilon)^2$ (for SSHDL) and $\alpha_{2} = (E- \epsilon)((E- \epsilon)^2-u_{3}^2)-2 (E-\epsilon)u_{1}^2$, $\beta_{2} = (E- \epsilon)((E- \epsilon)^2-u_{3}^2)-2 (E-\epsilon)u_{2}^2$, $\gamma_{2} = 4 u_{1}^2u_{2}^2u_{3}^2 + v^2 ((E- \epsilon)^2-u_{3}^2)^2$, $\delta_{2} = 4 u_{1}^2u_{2}^2u_{3}^2 + w^2 ((E- \epsilon)^2-u_{3}^2)^2$ (for SSHHL).  The solutions of the first part of Eq.~\ref{eq2} give the locations of non-dispersive flat bands at $E = \epsilon$ (for SSHDL, Fig.~\ref{ek1}) and at $E = \epsilon \pm u_{3}$ (for SSHHL, Fig~\ref{ek2}). The appearance of power factor $2$ indicates that these energies are doubly degenerate. The solutions of the second part give the locations of all the dispersion curves. In a similar technique, one can find the dispersion relation of such decorated lattices under a non-zero magnetic field.\par
For the SSH-diamond lattice, it is seen that at $E = \epsilon$ there is a flat band, also it is one energy eigenvalue where the energy gap closes (as shown in Eq.~\ref{gapSSHD}, Eq.~\ref{eq2}) at the BZ boundary. Similarly, in the SSH-hexagon lattice, the gap closes around the flat band energies $E = \epsilon \pm u_{3}$ when $u_{1} = u_{2} = u_{3}$ (see Eq.~\ref{gapSSHH}, Eq.~\ref{eq2}). So in this particular case, the flat band energy and edge state energy coexist in the same energy eigenvalue. In the edge state section, we will discuss the mechanism of distinguishing edge state energy from flat band energy in the topological non-trivial insulating phase. This coexistence can be destroyed in SSH-hexagon lattice by setting the parameter $u_{3}$ with appropriate numerical value different from $u_{1(2)}$. In that case, the gap-closing energy is shifted (see Eq.~\ref{gapSSHH}), but the flat band still remains at $E = \epsilon \pm u_{3}$.\par
Before ending this section, we will discuss the effect of magnetic flux on the dispersion relation when the applied magnetic flux is set to a value $\phi = 0.5 \phi_{0}$. In this situation in both SSHD and SSHH lattices, the renormalized hopping term $t_{12}$ (as shown in Eq.~\ref{eq1}) becomes zero. Due to this the dispersion relation is now completely independent of the wave vector $k$ and corresponding energy eigenvalues for SSHD lattice are calculated by the equations,

\begin{equation}
   E = \epsilon \pm \frac{1}{\sqrt{2}}\sqrt{2u_{1}^2+2u_{2}^2+v^2 \pm \Gamma}
   \label{flateq}
\end{equation} 
where $\Gamma = (4u_{1}^4 - 8u_{1}^2u_{2}^2+4u_{2}^4+4u_{1}^2v^2+4u_{2}^2v^2+v^4)^{\frac{1}{2}}$. At $v=w$ the flat bands appear in these four doubly degenerate eigenvalues (corresponding energy band diagram in shown in Fig.~\ref{ek-flat} (a,c)). The degeneracy of the flat bands is broken when $v \neq w$. In that case, the non-degenerate flat bands are observed with eight different eigenvalues which can be calculated by Eq.~\ref{flateq} just by putting different values $v$ (gives four eigenvalues) and $w$ (gives four eigenvalues) as depicted in fig.~\ref{ek-flat}(b,d). Using the same technique all flat band energies can be obtained for the SSHH lattices.\par

\subsection{Results from the direct diagonalization of the Hamiltonian}


After diagonalizing the respective $k$-space Hamiltonians the dispersion relations are obtained for both the decorated lattices (Eq.~\ref{ham1} for SSHDL and Eq.~\ref{ham2} for SSHHL). In Fig.~\ref{ek1}(a) the energy band diagram of the SSH-diamond lattice is depicted when no magnetic flux is trapped. Along with the dispersive bands there is a doubly degenerate flat band (green colored) at $E = \epsilon$ (here we choose $\epsilon$ as zero). If the inter-cell overlap integral is varied, energy gaps open up (for $v>w$) at first and then close(for $v=w$) and then open up again (for $v<w$) at the BZ boundary as shown in Fig.~\ref{ek1}(c). When a magnetic flux (here $\phi = \dfrac{1}{4}\phi_{0}$) is trapped in each diamond cavity of such SSHD lattices, the non-dispersive flat bands disappear (Fig.~\ref{ek1}(b)). However, the gap opening-closing-opening situation remains completely unaffected (see Fig~\ref{ek1}(d)) in the presence of magnetic flux.\par


\begin{figure}[ht]
\centering
(a)\includegraphics[width=.44\columnwidth]{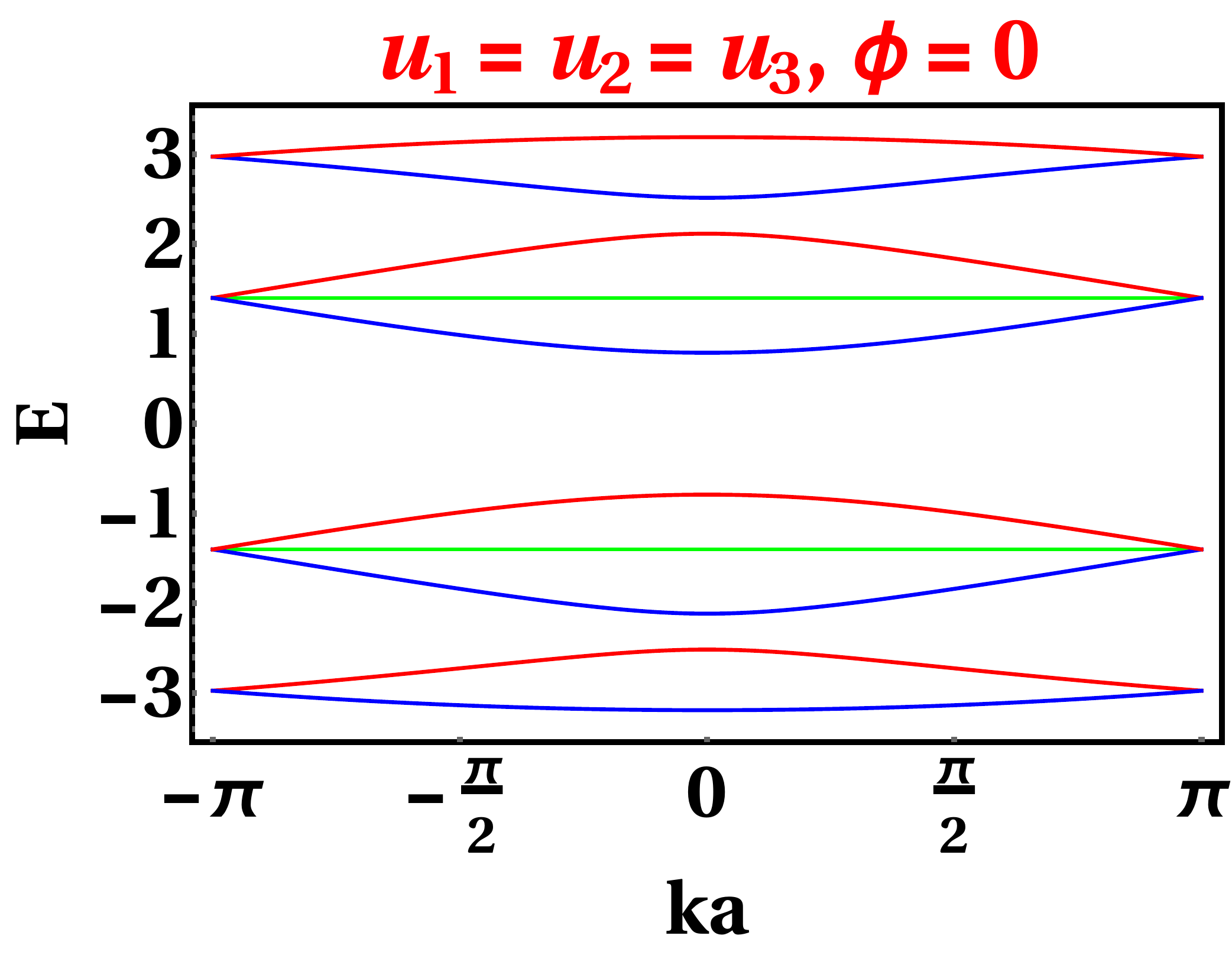}
(b)\includegraphics[width=.44\columnwidth]{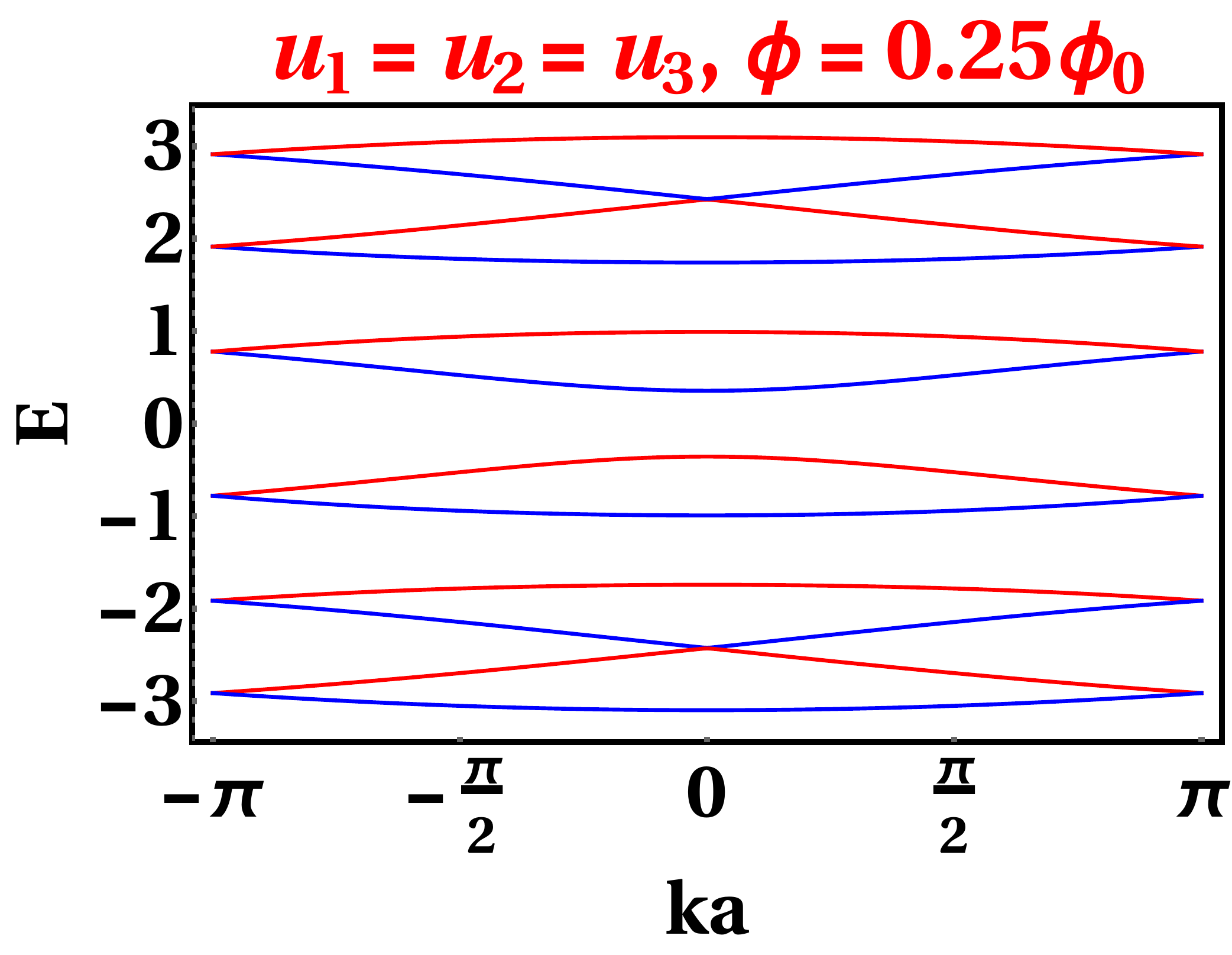}
(c)\includegraphics[width=.44\columnwidth]{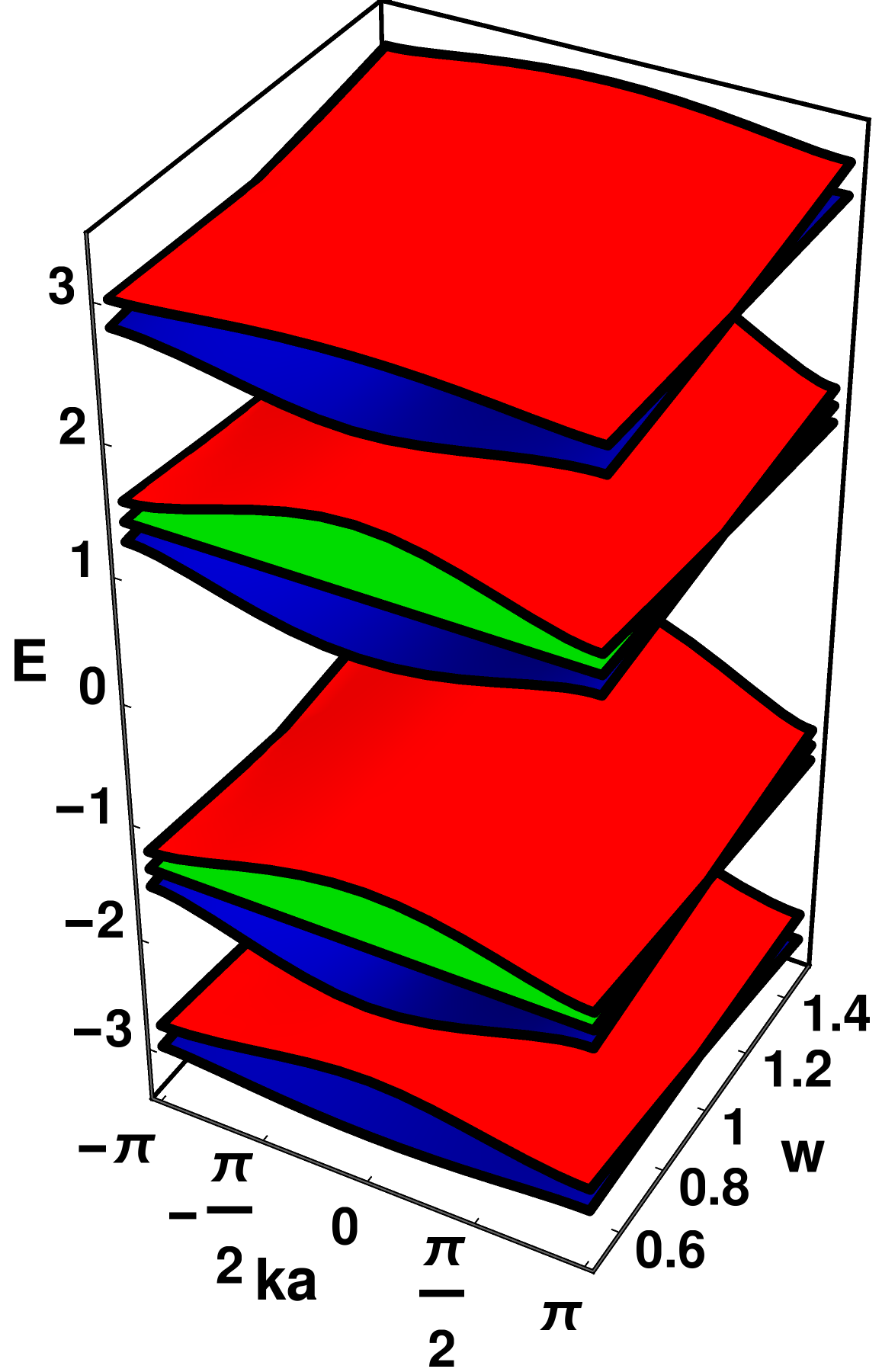}
(d)\includegraphics[width=.44\columnwidth]{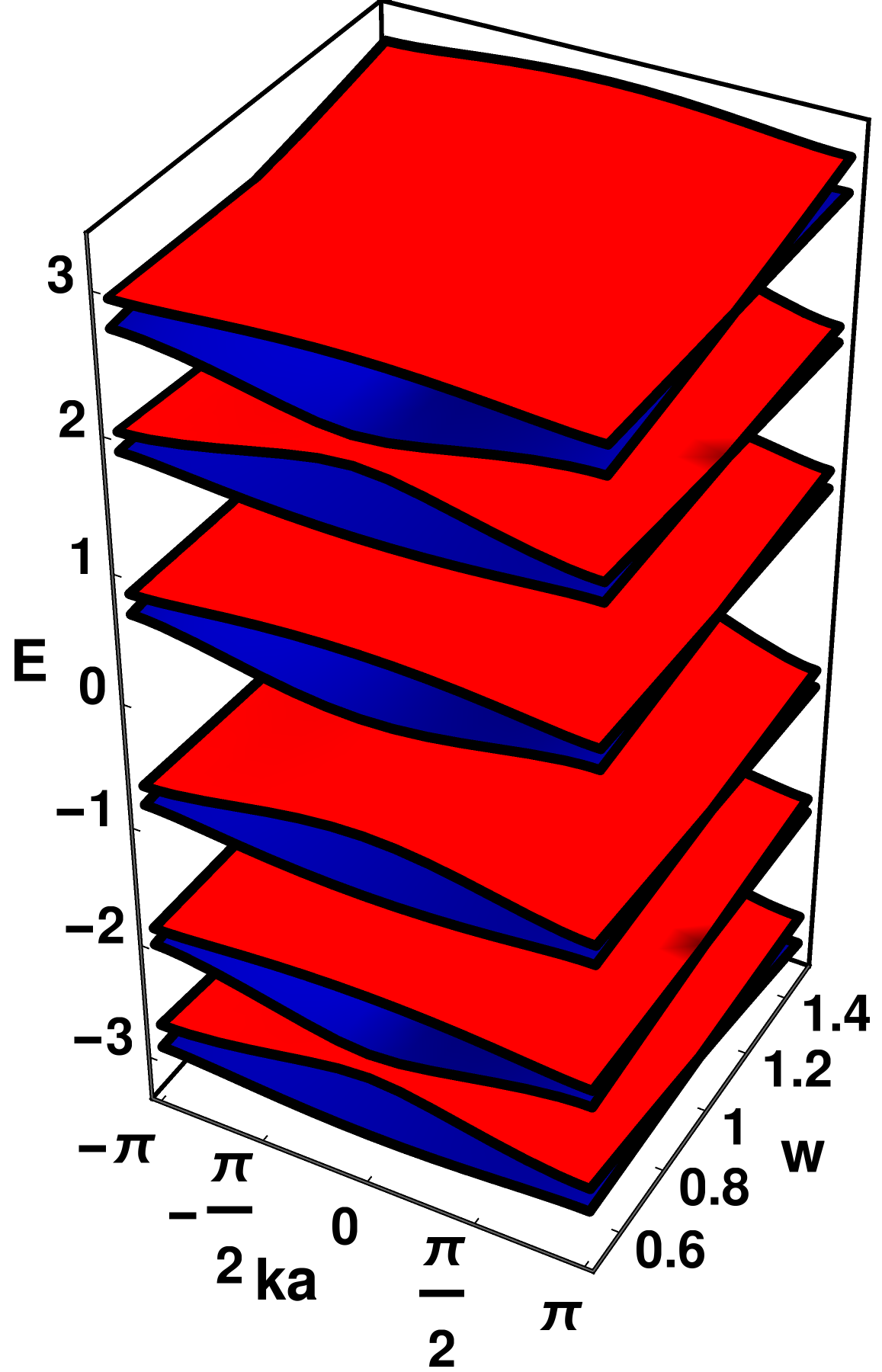}
\caption{(Color online)  Energy vs wave vector (E vs ka) dispersion relations for the SSH-Hexagonal lattice (a,c) absence of magnetic flux and (b,d) presence of magnetic flux. The parameters are chosen as (a) $ \epsilon = 0, u_{1} = u_{2} = u_{3} = 1.4, v = 1, w = 1,  \phi = 0$, (b)  $ \epsilon = 0, u_{1} = u_{2} = u_{3} = 1.4, v = 1, w = 1,  \phi = \frac{1}{4} \phi_{0}$. The flat bands (doubly degenerate) are marked in green. (c) and (d) are the corresponding 3-d plots of (a) and (b) with different intercell hopping $w$. }  
\label{ek2}
\end{figure}

The dispersion curves for the SSH-hexagon lattice are plotted in Fig.~\ref{ek2}(a) at $\phi = 0$. The flat bands appear at $E = \epsilon \pm u_{3}$ as shown in Fig.~\ref{ek2}(a) (both are doubly degenerate). Under a magnetic flux (here $\phi = \dfrac{1}{4}\phi_{0}$) all energy bands become dispersive (see Fig.~\ref{ek2}(b)). In both the situations, in the presence or absence of magnetic flux, all dispersive bands participate in gap opening-closing-opening phenomena at the BZ boundary as shown in Fig.~\ref{ek2}(c),(d). \par

So far, in both the cases, we have set the external hopping parameter as $u_{1} = u_{2}$ (for SSHDL) and $u_{1} = u_{2} =u_{3}$(for SSHHL). Due to this typical distribution of hopping integrals in the absence of magnetic flux, in both the cases the energy-gaps open up or close around the flat band energy ($E = \epsilon$ for SSHDL and $E = \epsilon \pm u_{3}$ for SSHHL). So, if TPT occurs, corresponding edge states in the non-trivial insulating phase appear at that flat band energy. Now, we want to see, if it is possible to distinguish the edge state-energy from the flat band-energy. We will discuss it elaborately in the `edge state' section. There is an alternative way to separate the two energy regimes for SSH-hexagon lattice, just by setting $u_{1} = u_{2} \neq u_{3}$. With an appropriate choice of $u_{3}$, the gap closing energy then separates out from the flat band energy, such that the overlapping situation of the edge state with flat band energy is completely absent. But for the SSH-diamond lattice it is not possible. \par

\begin{figure}[ht]
\centering
(a)\includegraphics[width=.44\columnwidth]{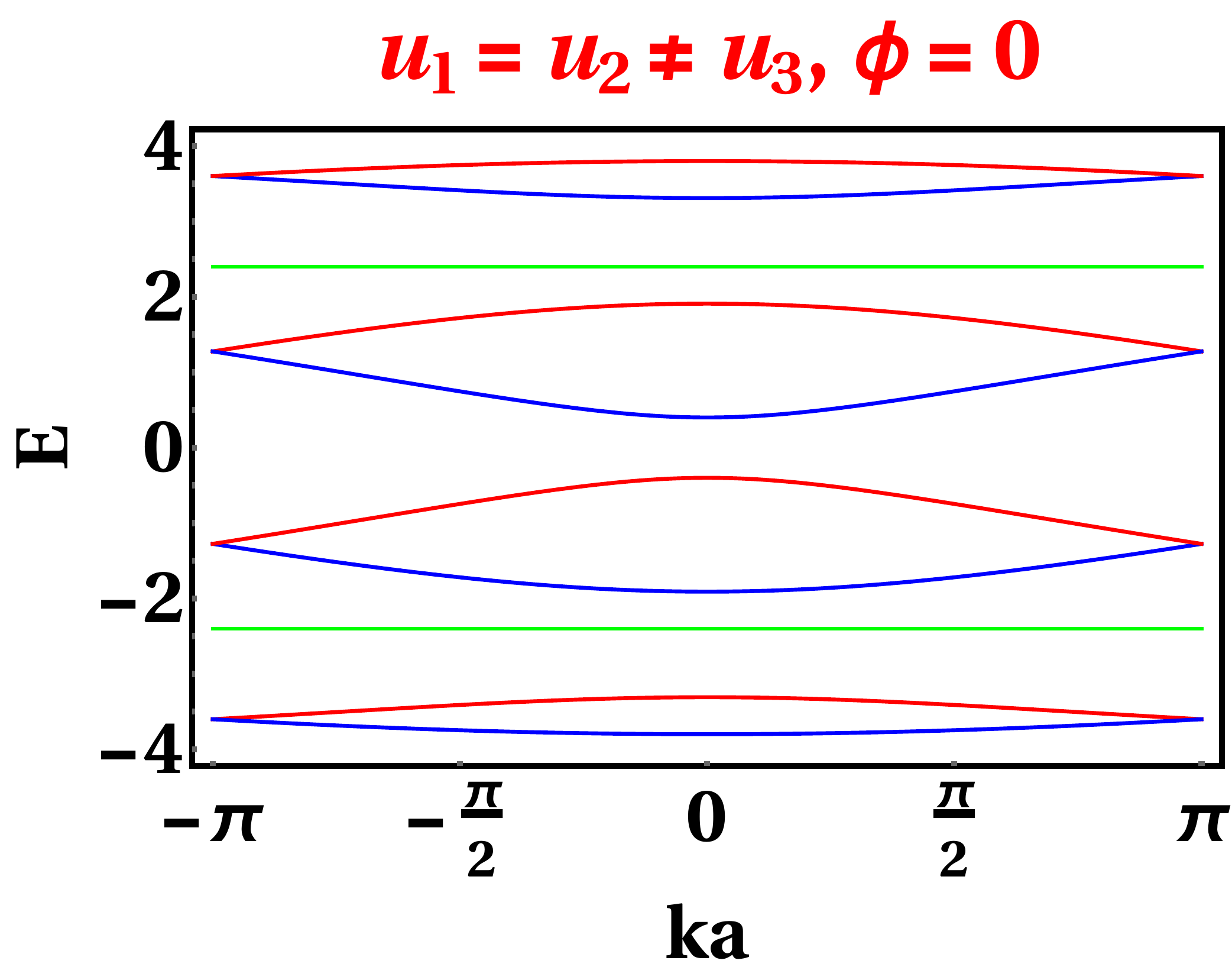}
(b)\includegraphics[width=.44\columnwidth]{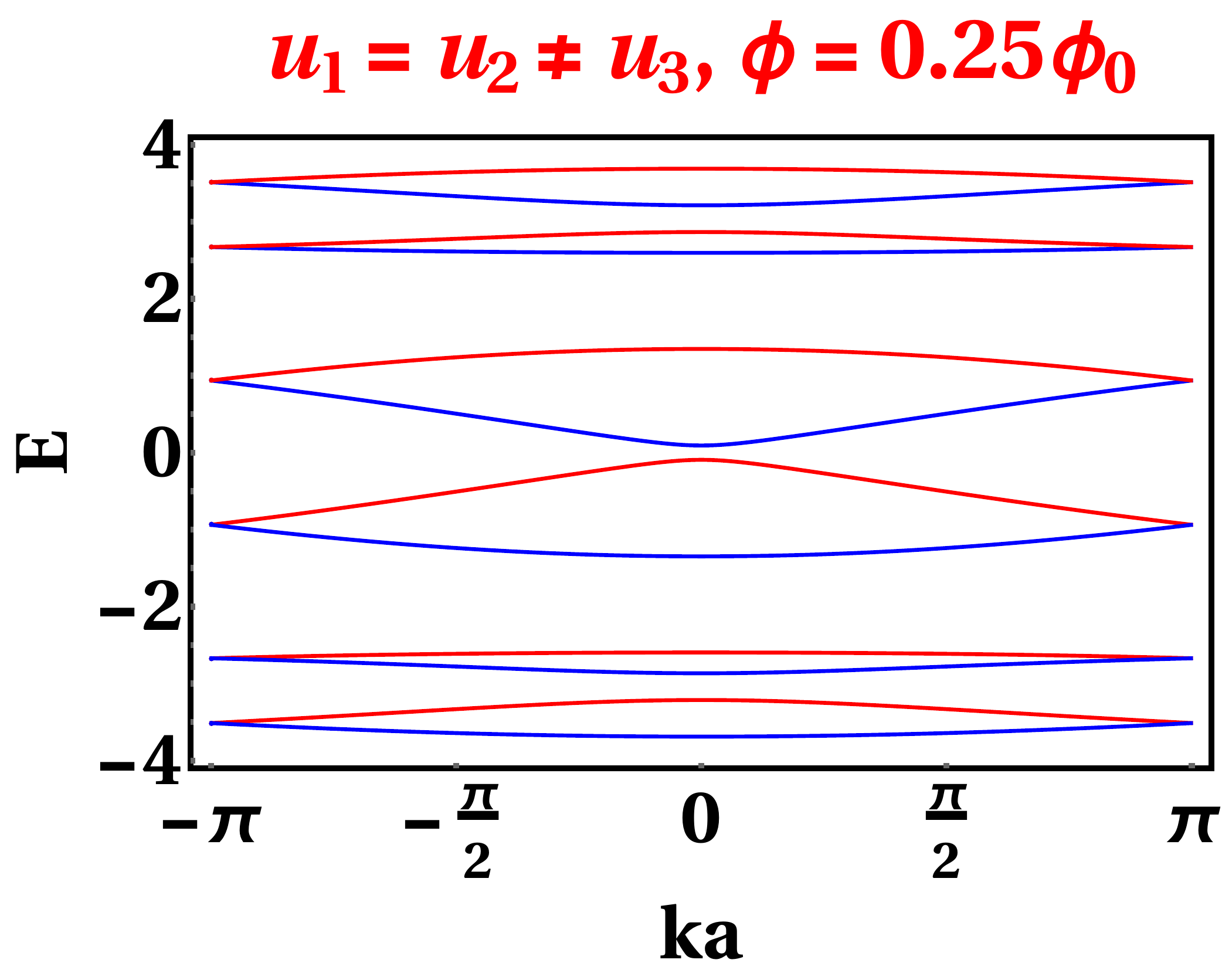}
(c)\includegraphics[width=.44\columnwidth]{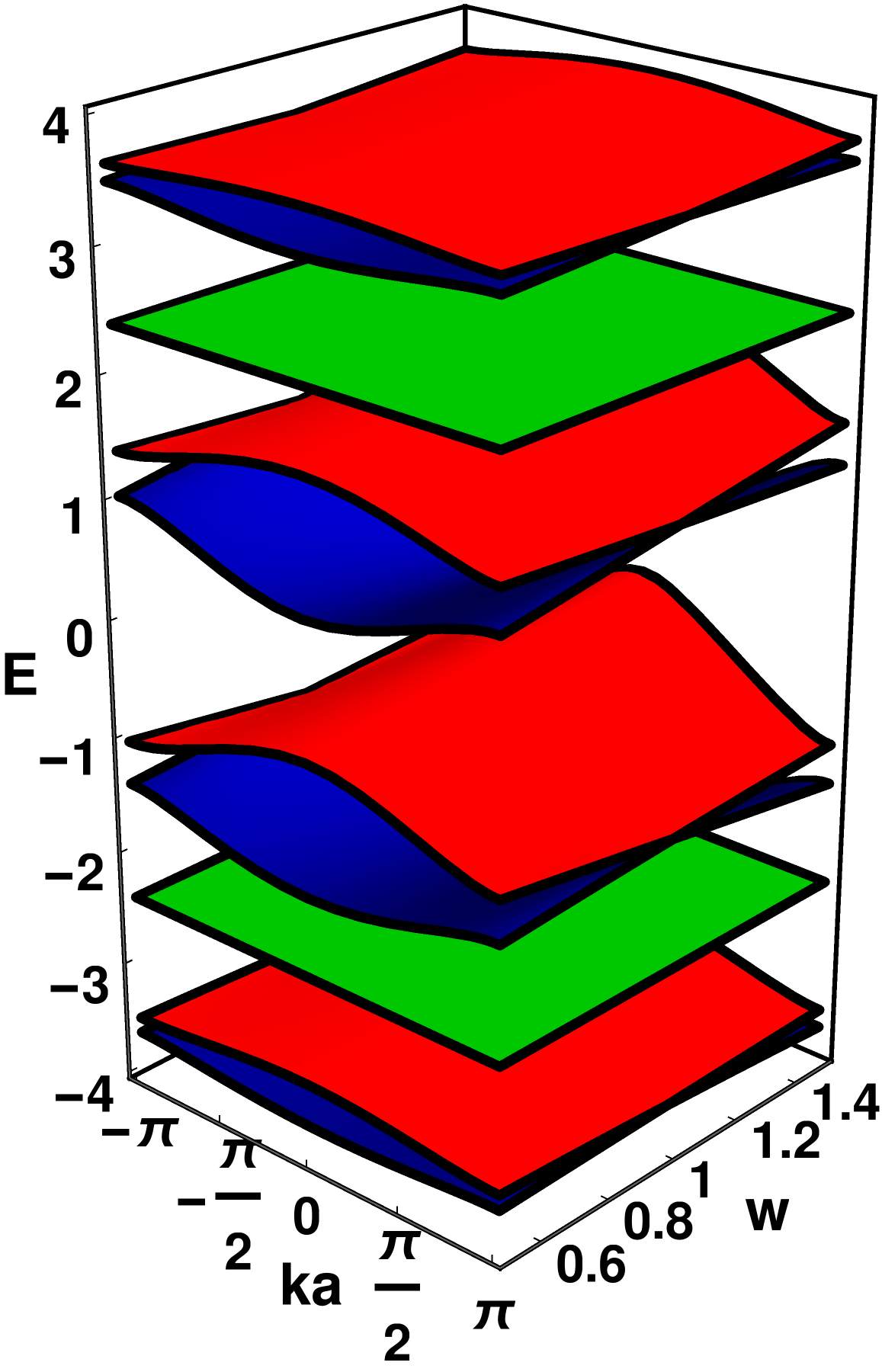}
(d)\includegraphics[width=.44\columnwidth]{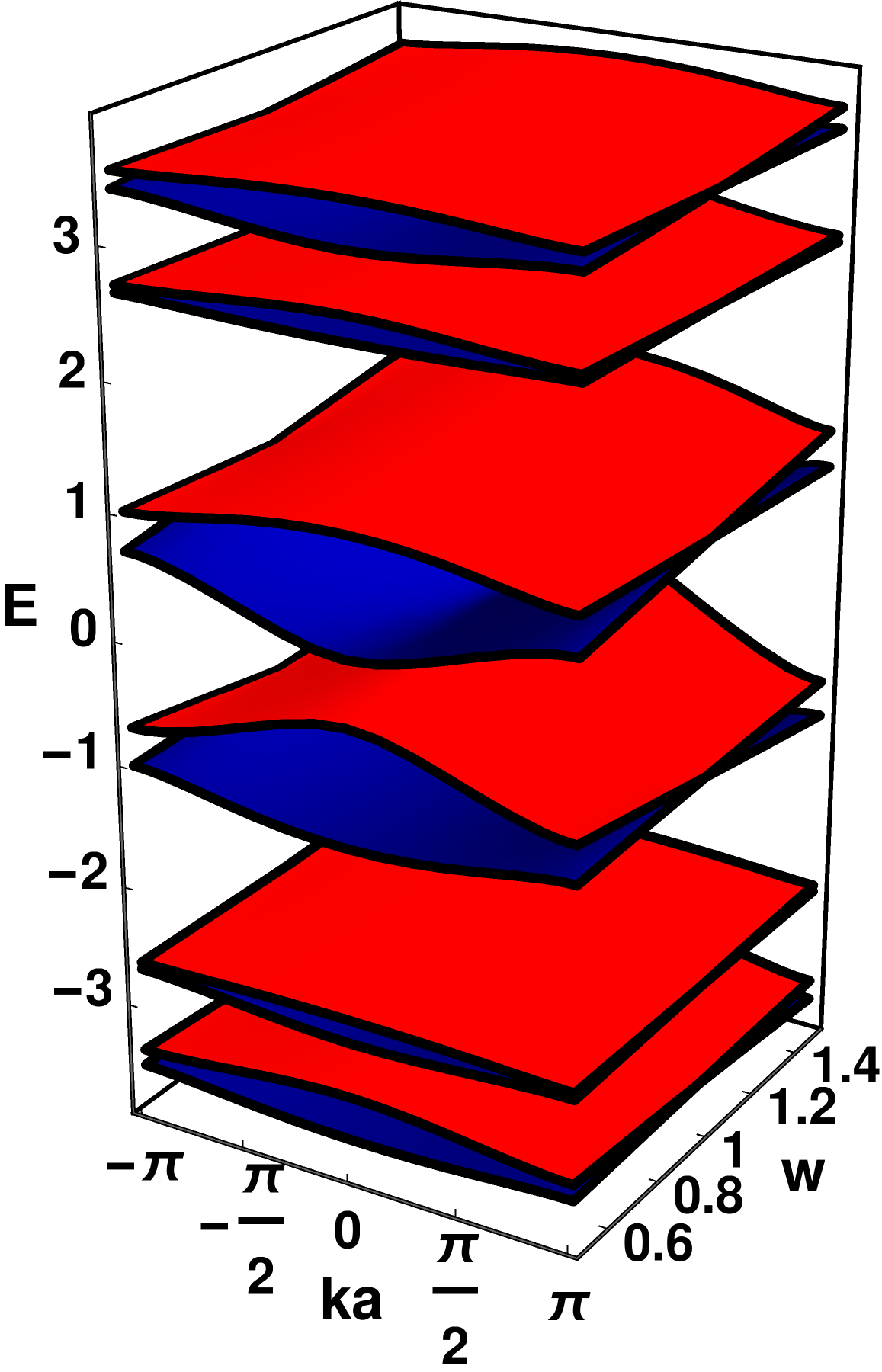}
\caption{(Color online)  Energy vs wave vector (E vs ka) dispersion relations for the SSH-Hexagonal lattice - (a,c) in absence of magnetic flux and (b,d) in  presence of magnetic flux. The parameters are chosen as (a) $ \epsilon = 0, u_{1} = u_{2} = 1.4, u_{3} = 2.4, v = 1, w = 1,  \phi = 0$, (b)  $ \epsilon = 0, u_{1} = u_{2} = 1.4, u_{3} = 2.4, v = 1, w = 1,  \phi = \frac{1}{4} \phi_{0}$. The flat bands (doubly degenerate) are marked in green. (c) and (d) are the corresponding 3-d plots of (a) and (b) with different intercell hopping $w$. }  
\label{ek3}
\end{figure}


In Fig.~\ref{ek3}(a) the dispersion curves of a SSHHL are plotted with a parameter choice $u_{1} = u_{2} \neq u_{3}$ (here $u_{1} = u_{2} = 1.4, u_{3} = 2.4$) under the absence of magnetic flux. As discussed earlier the flat band always appears at the energy $E = \epsilon \pm u_{3}$, it is unaffected by any choice of $u_{1}$ or $u_{2}$. Now the gap-closing energy is shifted from the flat band energy. As a result, the co-existence of the flat band energy with the gap-closing energy (as seen in $u_{1} = u_{2} = u_{3}$ case) completely disappears. When a non-zero magnetic flux is applied in each cavity of the hexagon, these doubly degenerate flat bands are destroyed and appear with a pair of dispersive bands as shown in Fig.~\ref{ek3}(b). Here also in both in presence or absence of magnetic flux the gap opening-closing-opening situation is observed as the inter-cell hopping $w$ is varied (see Fig.~\ref{ek3}(c),(d)).\par
As discussed earlier when a particular magnetic flux $\phi = 0.5 \phi_{0}$ is trapped in each cavity of SSHD or SSHH lattices all energy bands become non-dispersive. At $v=w$ the doubly degenerate flat bands of SSHDL are shown in Fig.~\ref{ek-flat}(a). The breaking of the degeneracy of the flat band energies at $v \neq w$ is demonstrated in Fig.~\ref{ek-flat}(b). All the energies of the SSHD lattice are also calculated analytically using Eq.~\ref{flateq}, already discussed in the previous section. The effect of magnetic flux in an entire range $0$ to $\phi_{0}$ on the band structure of SSHDL is shown in Fig.~\ref{ek-flat}(c) for $v=w$ and (d) for $v \neq w$. Here also the appearance of the flat bands at $\phi = 0.5 \phi_{0}$ is apparent. \par

So, these types of decorated lattices show a re-entrant insulator-metal-insulator behavior depending upon the choice of the intra(or, inter)-cell overlap integral. This indicates a primary signal for observing a topological phase transition (TPT). In the absence of magnetic flux non-dispersive flat bands are always present in the band diagram even in the metallic situation when all gaps are closed. As a result, although the gap closed in the metallic situation ($v = w$), the system shows localization at the flat band energies. Such non-dispersive character of the energy bands is completely destroyed by the magnetic flux and, in the metallic phase, all eigenstates become de-localized.\par

\section{Topological Invariant}
\label{invariant}
\subsection{The Zak Phase}
From the dispersion curves, it seemes that both the insulating phases ($v>w$ or $v<w$) are the same. But this  is not right. More information is needed to study the character of these insulating phases in detail. This information is hidden in the corresponding eigenvector of the $k$-space Hamiltonian matrix (containing the topological aspect of such decorated lattice model).\par
To distinguish these two insulating cases, a topological invariant is needed. Here the {\it Zak phase}~\cite{zak} must be calculated. If a system exhibits a topological phase transition, the corresponding Zak phase of each dispersive band will always have a quantized value (either $0$ or $\pi$). During the transition from one insulating phase to another it flips its quantized value from zero to $\pi$, thereby distinguishing  the trivial phase from the non-trivial one.\par

The Zak phase for the $n$-th bulk band can be written in the form,
\begin{equation}
    Z = -i \oint_{BZ}  \mathcal{A}_{nk}(k) dk
    \label{zak} 
\end{equation}
where $\mathcal{A}_{nk}$ is  the Berry curvature~\cite{asboth} of the $n$-th Bloch eigenstate,
\begin{equation}
    \mathcal{A}_{n{k}}(k)= \bra{\psi_{n{k}}}\ket{\frac{d\psi_{nk}}{dk}}
\end{equation}
This $\psi_{n,k}$ is actually the eigenvector corresponding to the $n^{th}$ eigenvalue of the $k$-space Hamiltonian. 

The Zak phase is computed along a closed loop within the Brillouin zone, with $(\ket{\psi_{nk}})$ representing the $n^{th}$ Bloch state. It is an intrinsic property of the bulk system. To determine the Zak phase of the bands, the { \it Wilson loop method}~\cite{fukui, wang} is employed, which transforms the integral into a summation over the Brillouin zone. This gauge-invariant approach ensures that the calculated Zak phase remains unaffected by any arbitrary phase shifts in the Bloch wavefunction. For a non-degenerate $s^{th}$ band, this summation is expressed as follows~\cite{fukui, wang}:
\begin{equation} 
Z_s = - Im ~ \left [ \log \prod_{k_n} \bra{\psi_{k_n,s}} \ket{\psi_{k_{n+1},s} } \right ]
\label{wilson}
\end{equation}
Using this Willson loop technique we can calculate the Zak phase associated with each dispersive band for such decorated lattice models.  
\subsubsection{SSH-Diamond Lattice (SSHDL)}
When no magnetic flux is trapped the SSHDL band diagram consists of both dispersive and non-dispersive bands. It is observed that the eigenvector corresponding to the doubly degenerate flat band energy $E= \epsilon$ is absolutely {\it independent of the wave vector $k$}. So, in this case, the calculation of the Zak phase of the flat band has no significance. All the dispersive bands associated with a quantized Zak phase and change their value from $0$ to $\pi$ from the topological trivial insulating phase ($v>w$) to the topological non-trivial insulating phase($v<w$). Under non-zero magnetic flux (here $\phi = \dfrac{1}{4}\phi_{0}$) these doubly degenerate flat bands transform into pairs of dispersive bands. Interestingly, the magnetic flux never affects the quantization of the Zak phase. All dispersive bands have quantized Zak phase-values, which turn out to be $0$ for $v>w$ and $\pi$ for $v<w$.\par


\begin{figure}[ht]
\centering
(a)\includegraphics[width=.44\columnwidth]{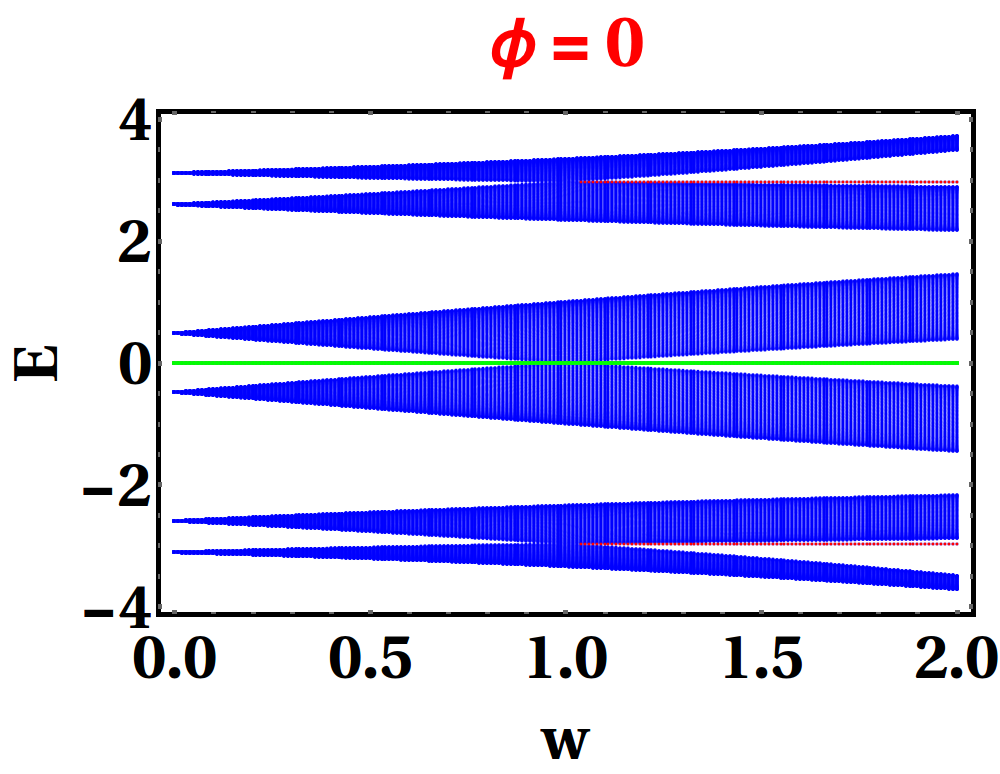}
(b)\includegraphics[width=.44\columnwidth]{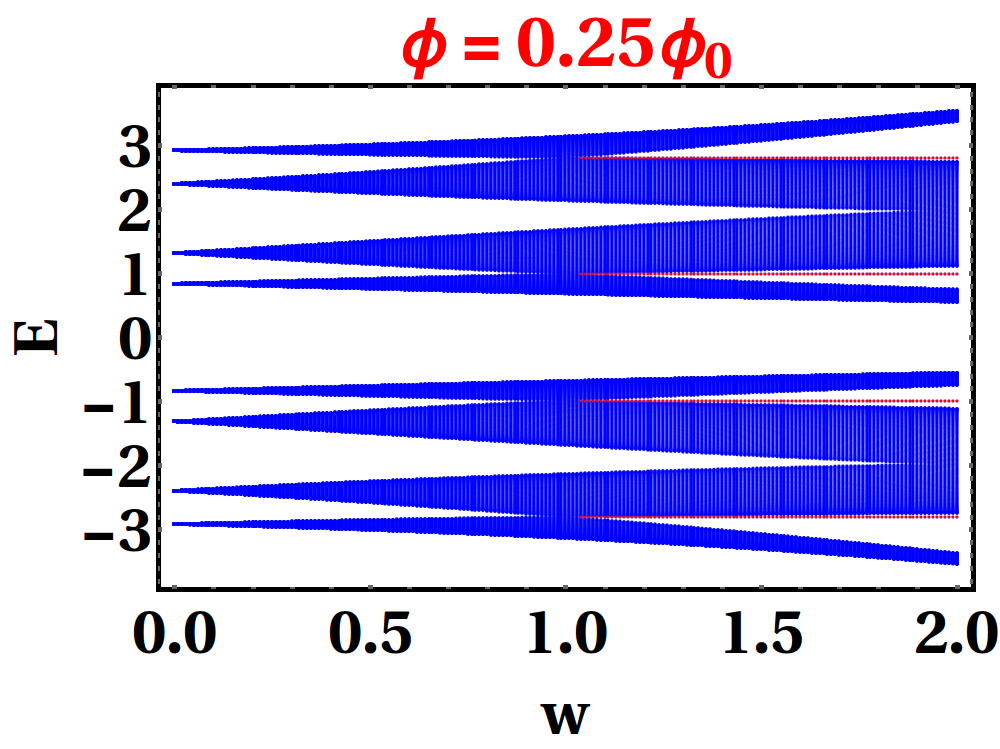}
(c)\includegraphics[width=.44\columnwidth]{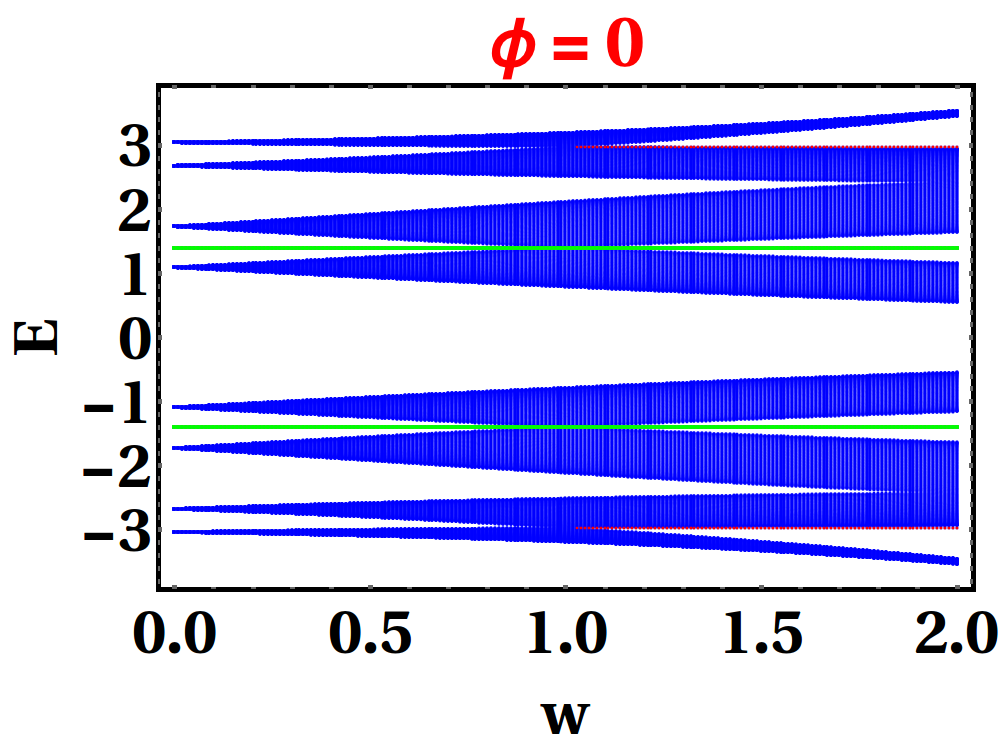}
(d)\includegraphics[width=.44\columnwidth]{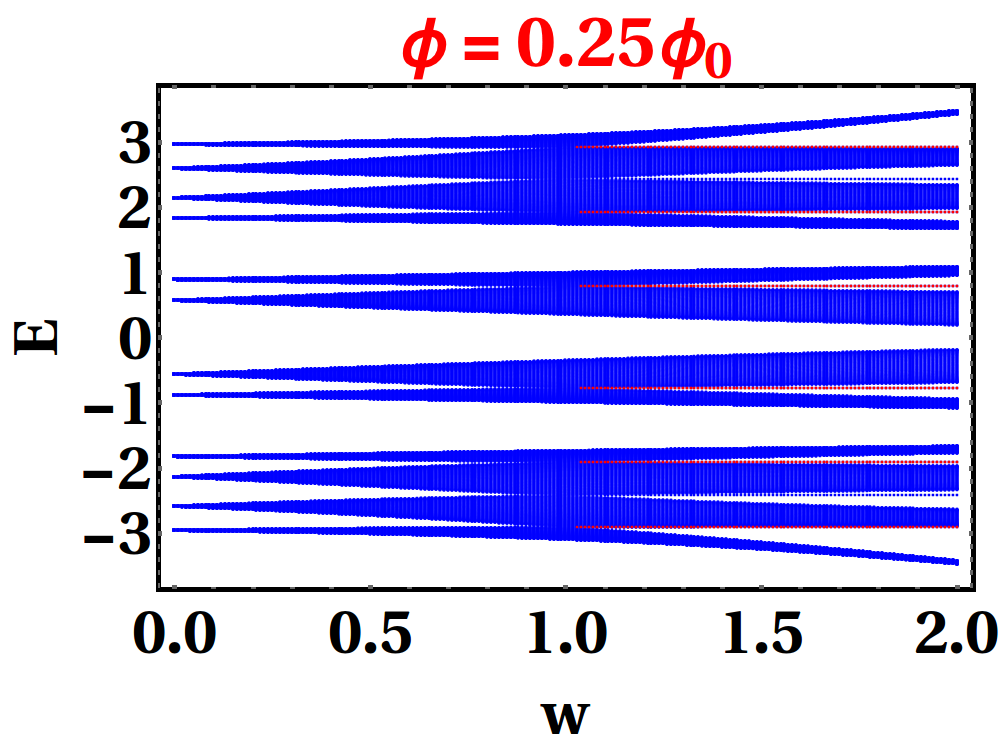}
(e)\includegraphics[width=.44\columnwidth]{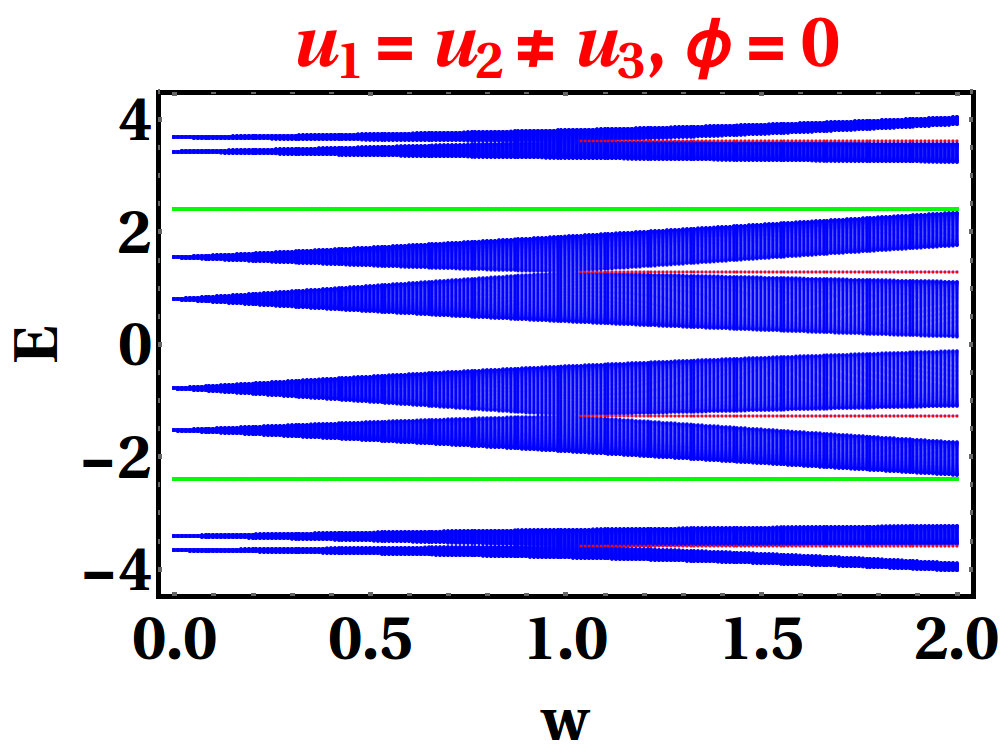}
(f)\includegraphics[width=.44\columnwidth]{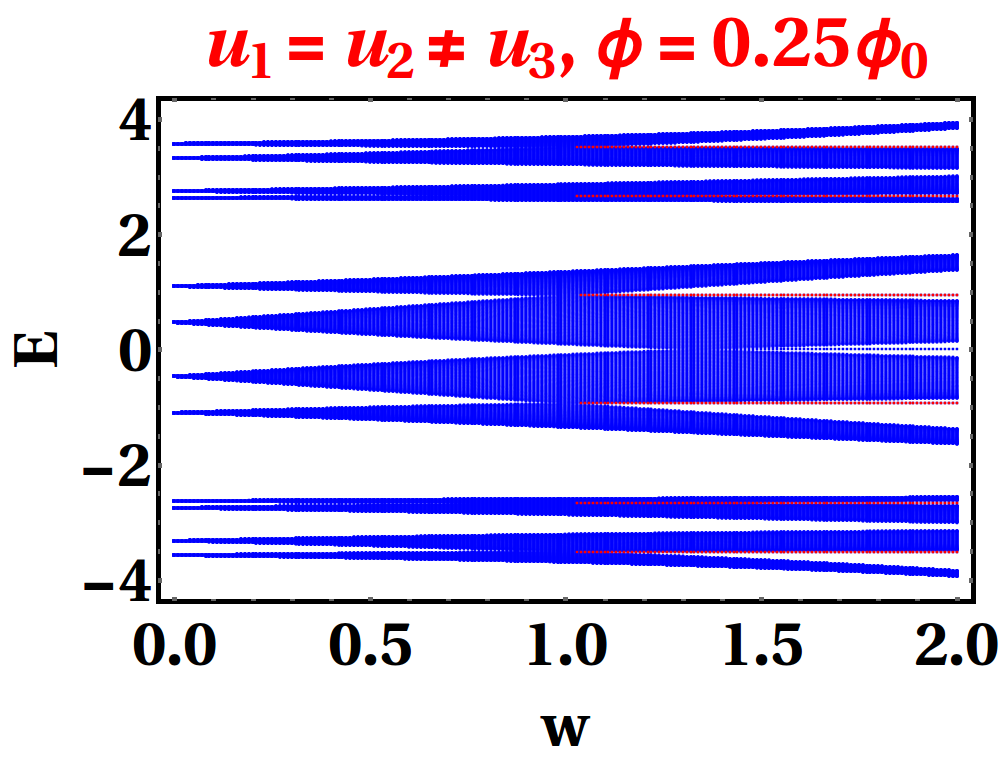}
\caption{(Color online) Energy spectra for (a,b)  SSH-Diamond and  (c-f) SSH-Hexagonal lattices with the coupling $w$ varies. We have used open boundary conditions for $N_x=200$, where $N_x$ denotes the number of unit cells taken along the $x$-direction. For both networks, (a,c,e) shows the energy spectrum when no magnetic flux is applied and (b,d,f) under a applied magnetic flux $\phi = \frac{1}{4}\phi_{0}$. The parameters are chosen as (a) $ \epsilon = 0, u_{1} = u_{2} = 1.4, v = 1,  \phi = 0$, (b) $ \epsilon = 0, u_{1} = u_{2} = 1.4, v = 1, \phi = \frac{1}{4}\phi_{0}$, (c) $ \epsilon = 0, u_{1} = u_{2} = u_{3} = 1.4, v = 1,  \phi = 0$, (d) $ \epsilon = 0, u_{1} = u_{2} = u_{3} = 1.4, v = 1,  \phi = \frac{1}{4}\phi_{0}$, (e) $ \epsilon = 0, u_{1} = u_{2} = 1.4, u_{3} = 2.4, v = 1,  \phi = 0$, and (f) $ \epsilon = 0, u_{1} = u_{2} = 1.4, u_{3} = 2.4, v = 1,  \phi = \frac{1}{4}\phi_{0}$. In all cases, the edge states are clearly apparent. Flat band energies are marked by green and the edge state energies are marked by red. The situation where edge state energy overlaps with the flat band energy in the topological non-trivial insulating phase, also is in green color due to highly degenerate flat band energy. }  
\label{e-vs-w}
\end{figure}


\begin{figure}[ht]
\centering
(a)\includegraphics[width=.44\columnwidth]{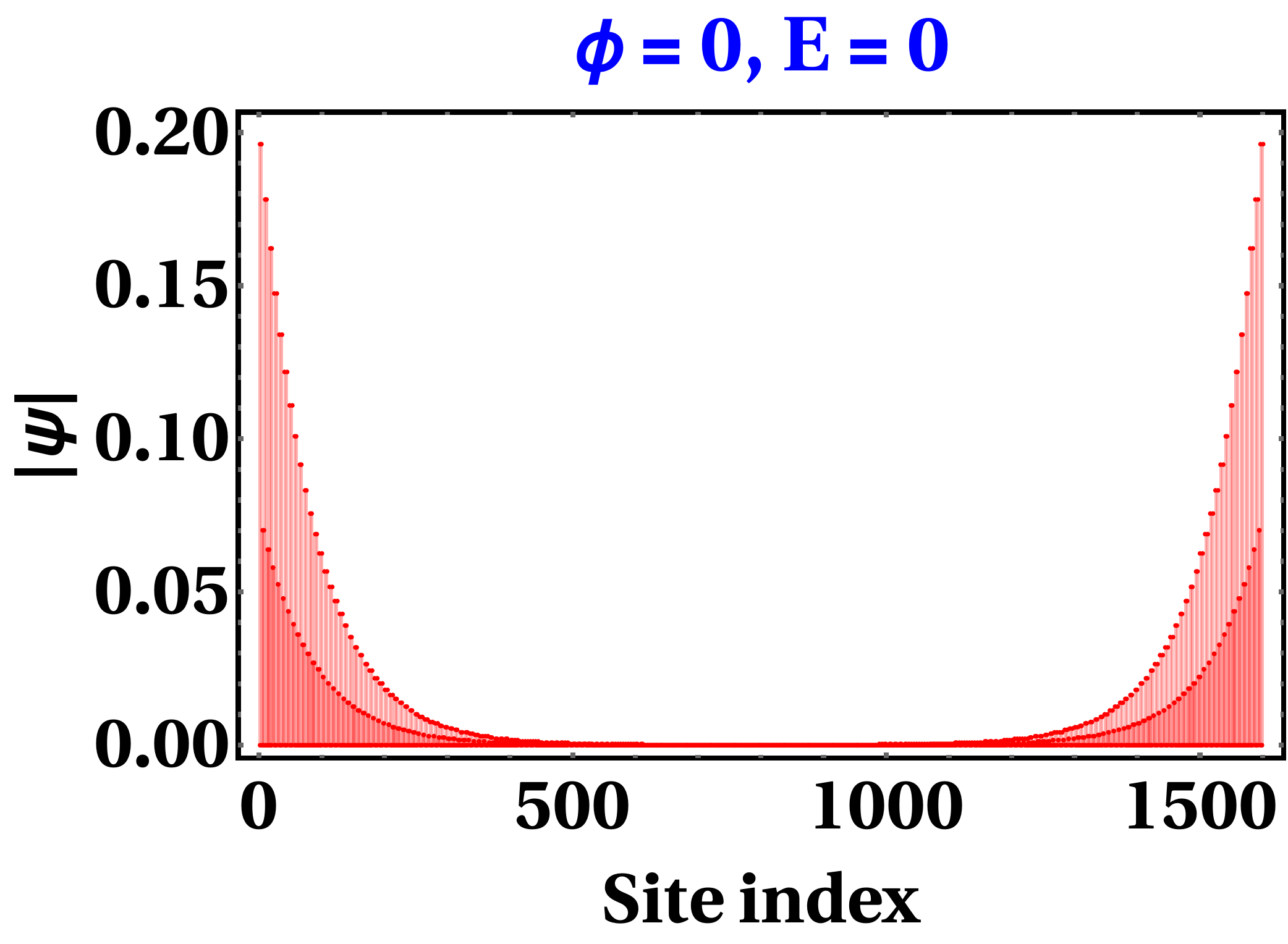}
(b)\includegraphics[width=.44\columnwidth]{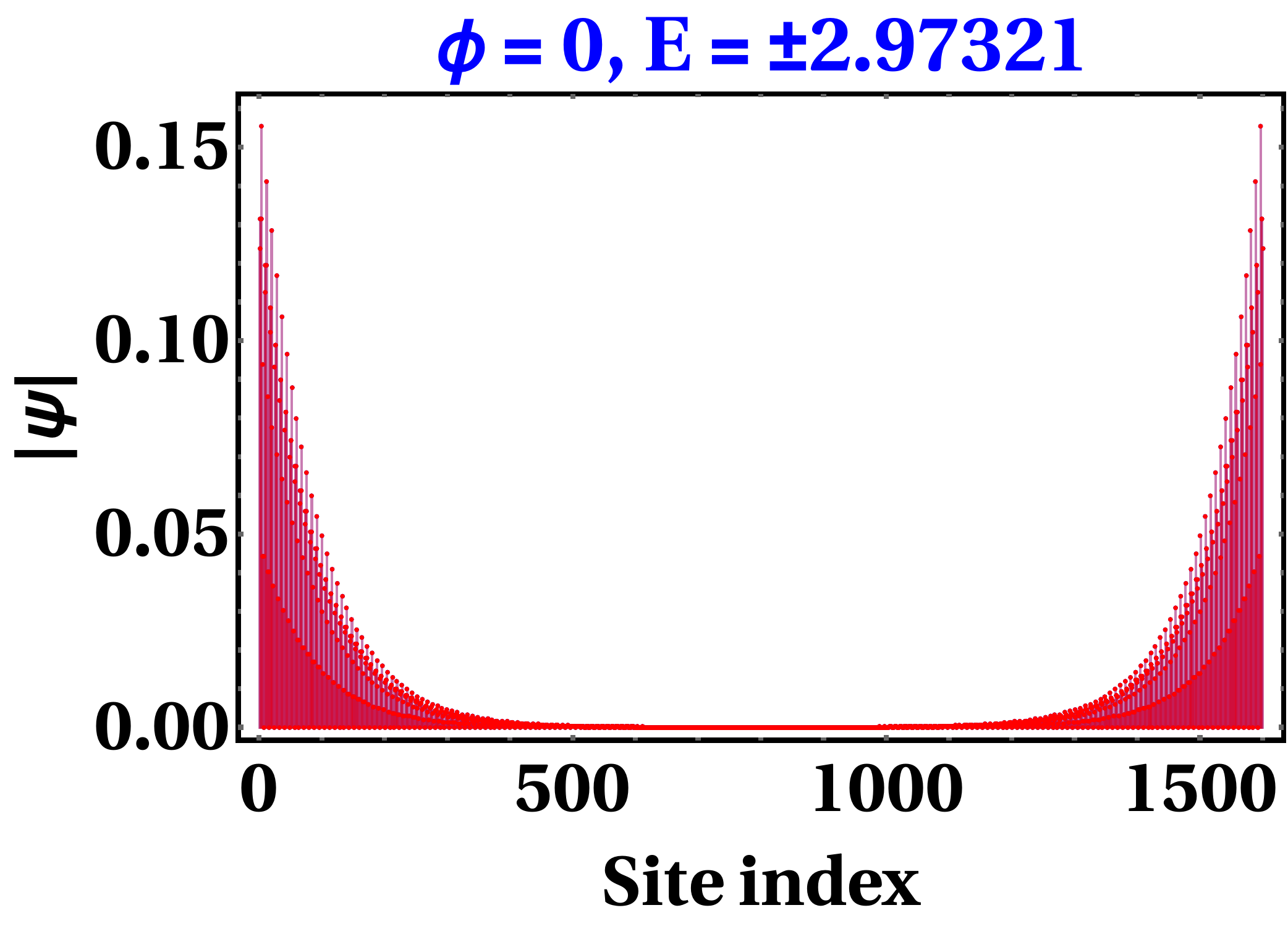}
(c)\includegraphics[width=.44\columnwidth]{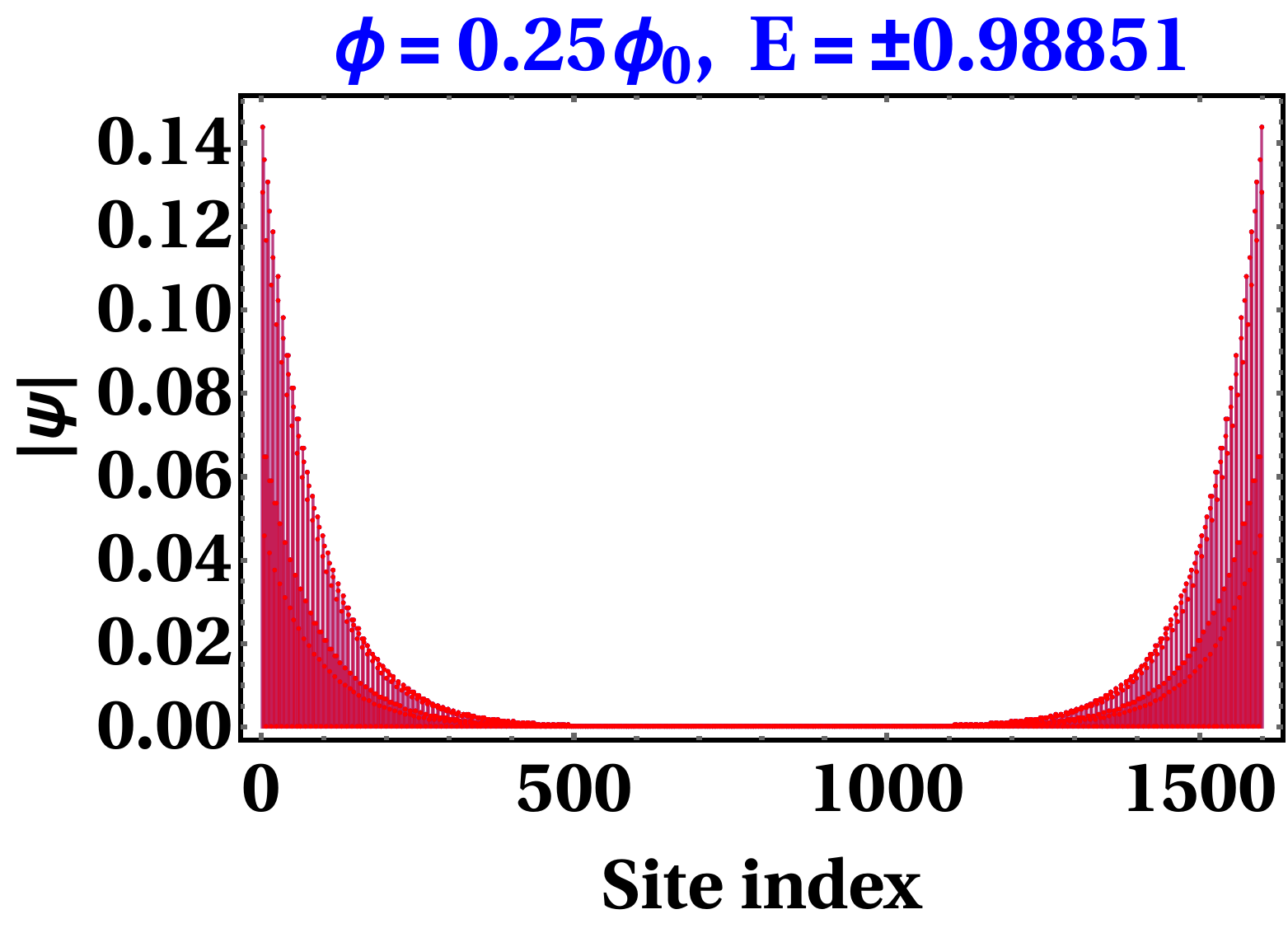}
(d)\includegraphics[width=.44\columnwidth]{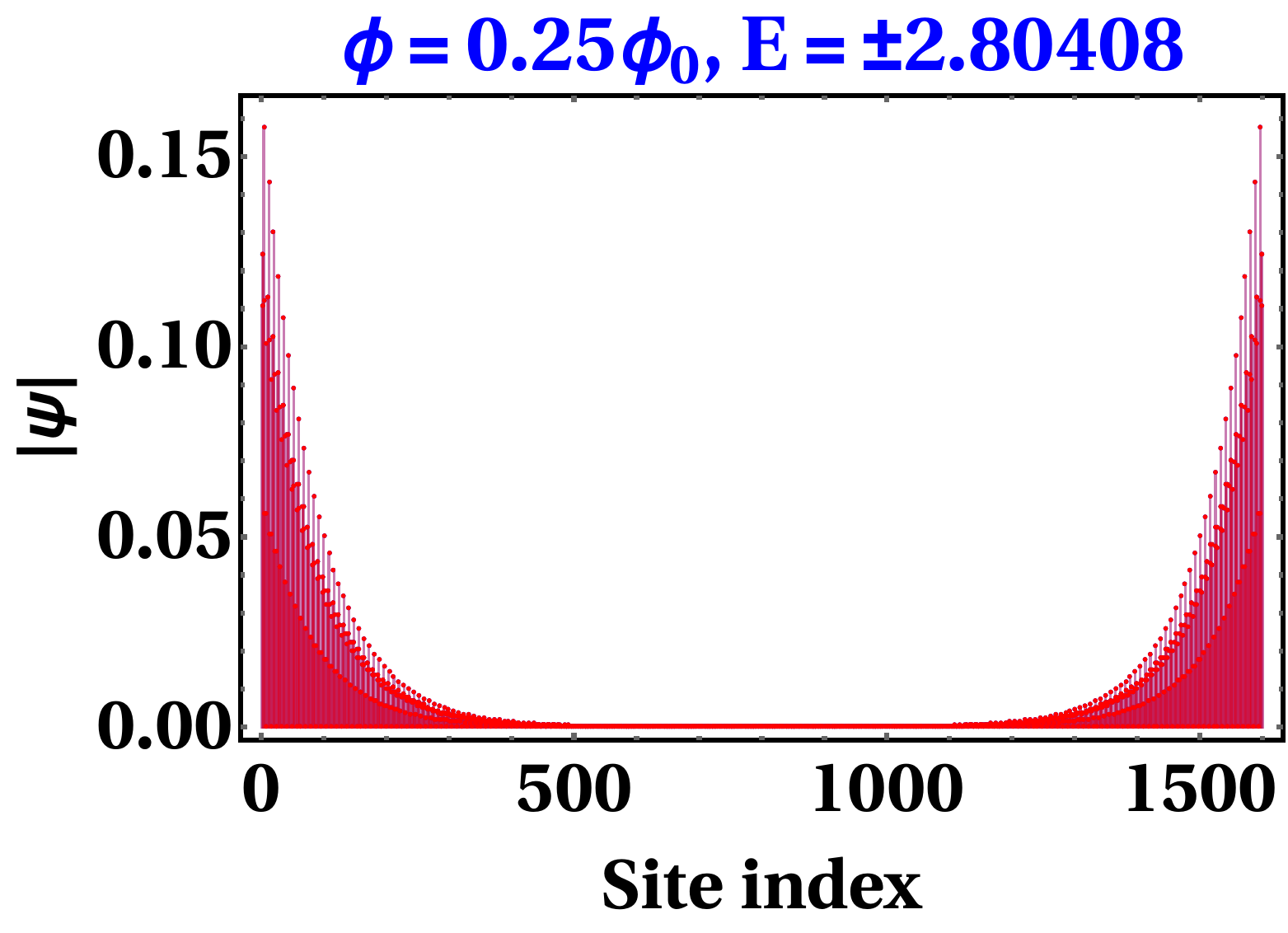}
\caption{(Color online) Distribution of the wavefunction of the edge state energies for SSH-Diamond lattices with energy (a) $E = 0$ (b) $E = \pm 2.97321$ (in the absence of magnetic flux, $\phi = 0$) (c) $E = \pm 0.98851$ and (d) $E = \pm 2.80408$ (under applied magnetic flux, $\phi = 0.25 \phi_{0}$). We have used open boundary conditions for $N_x=200$, where $N_x$ denotes the number of unit cells taken along the $x$-direction. The parameters are chosen as $ \epsilon = 0, u_{1} = u_{2} = 1.4, v = 1, w= 1.1$. Two different edge state energies ($\pm E$) are plotted in the same plot with red and blue color. All of these energies are doubly degenerate.}  
\label{wave1}
\end{figure}

\begin{figure}[ht]
\centering
(a)\includegraphics[width=.44\columnwidth]{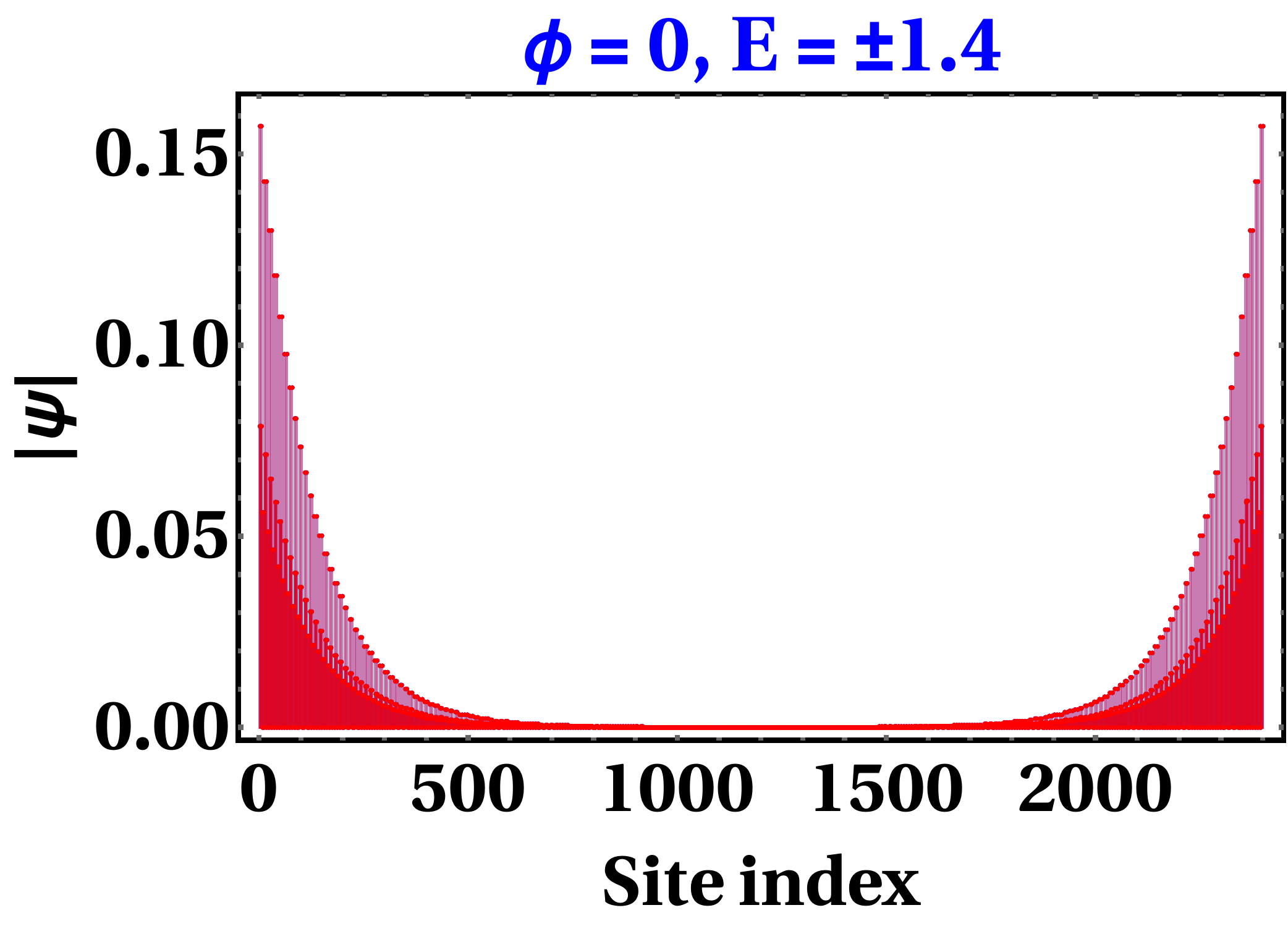}
(b)\includegraphics[width=.44\columnwidth]{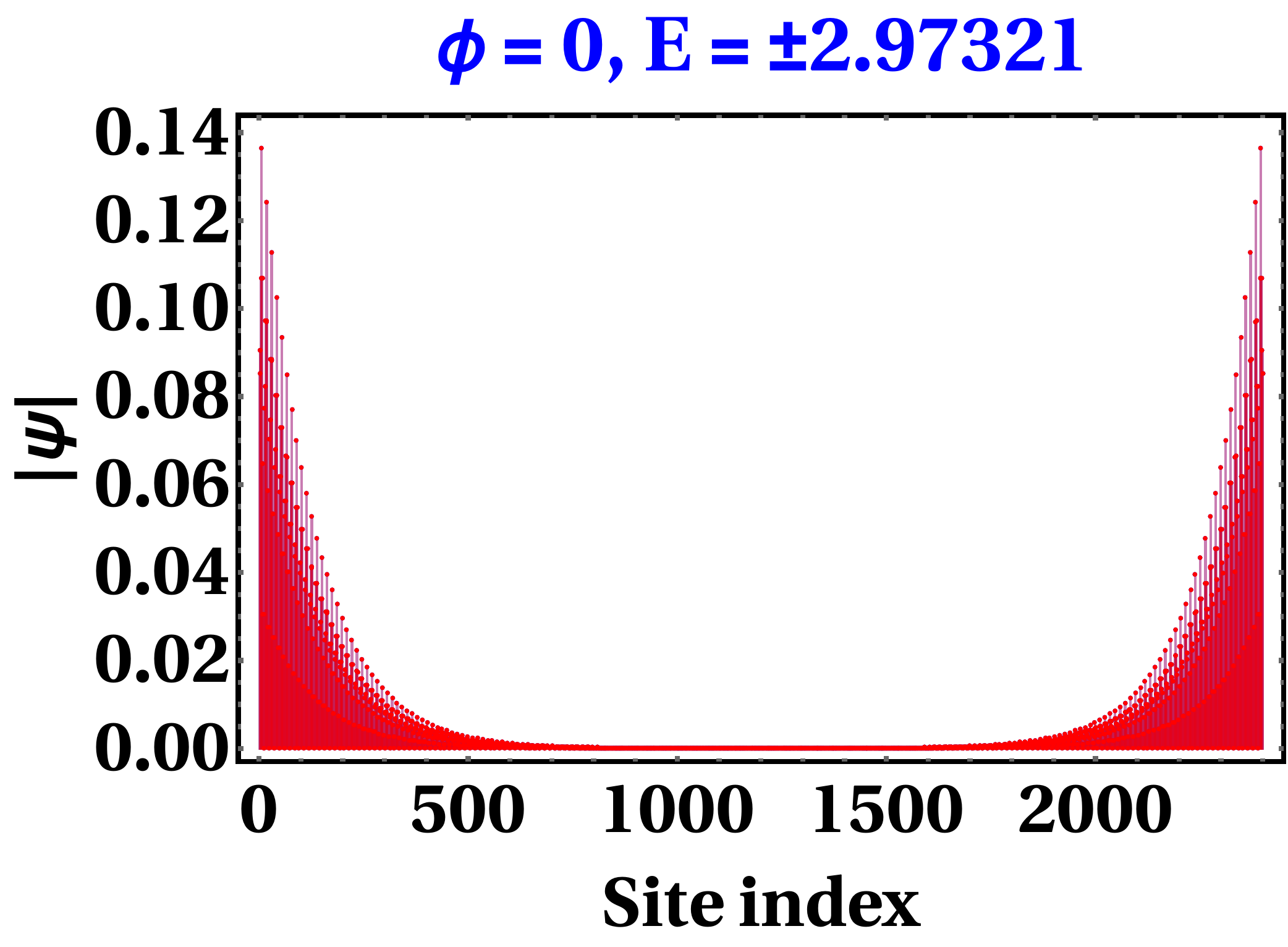}
(c)\includegraphics[width=.44\columnwidth]{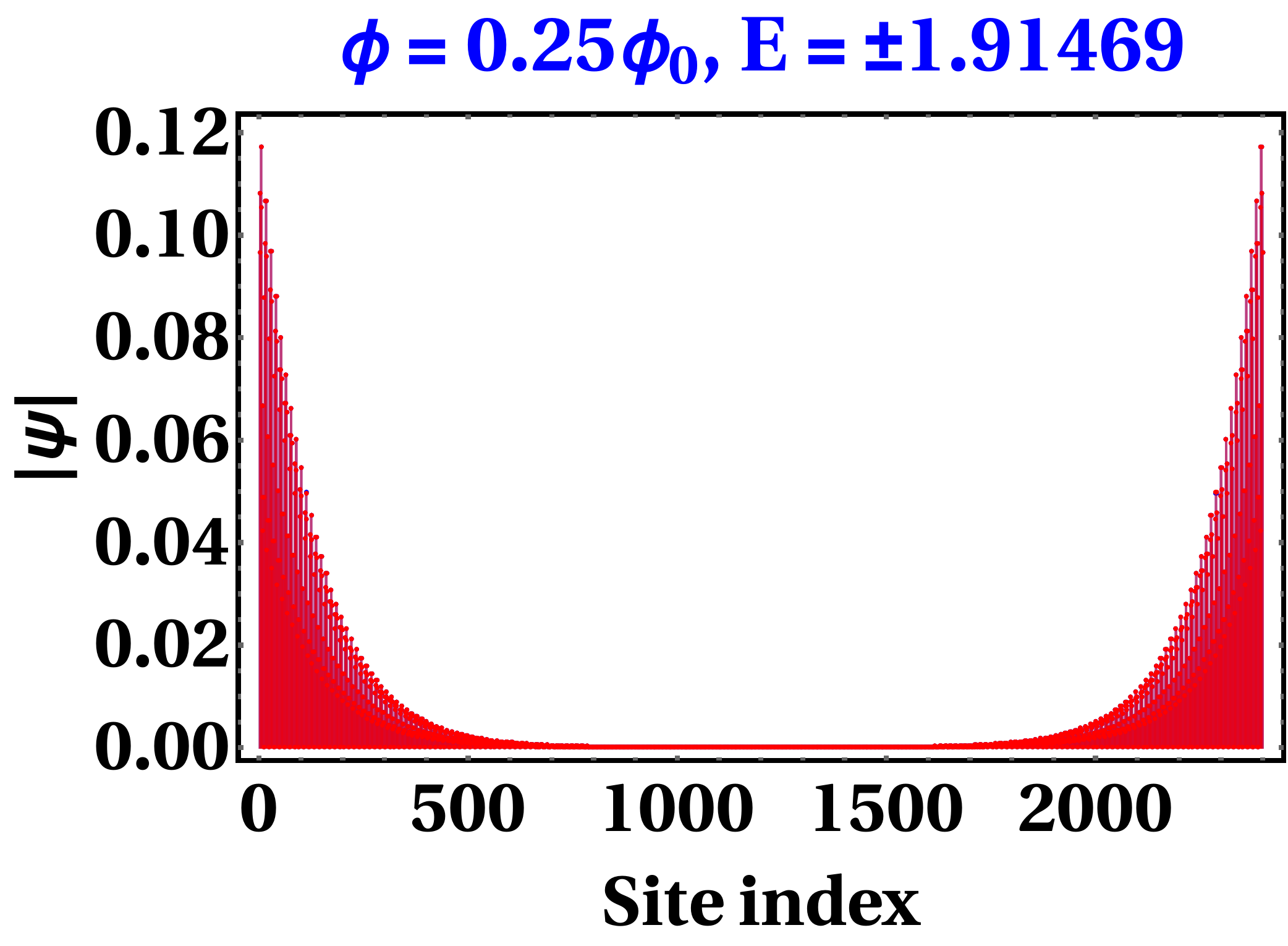}
(d)\includegraphics[width=.44\columnwidth]{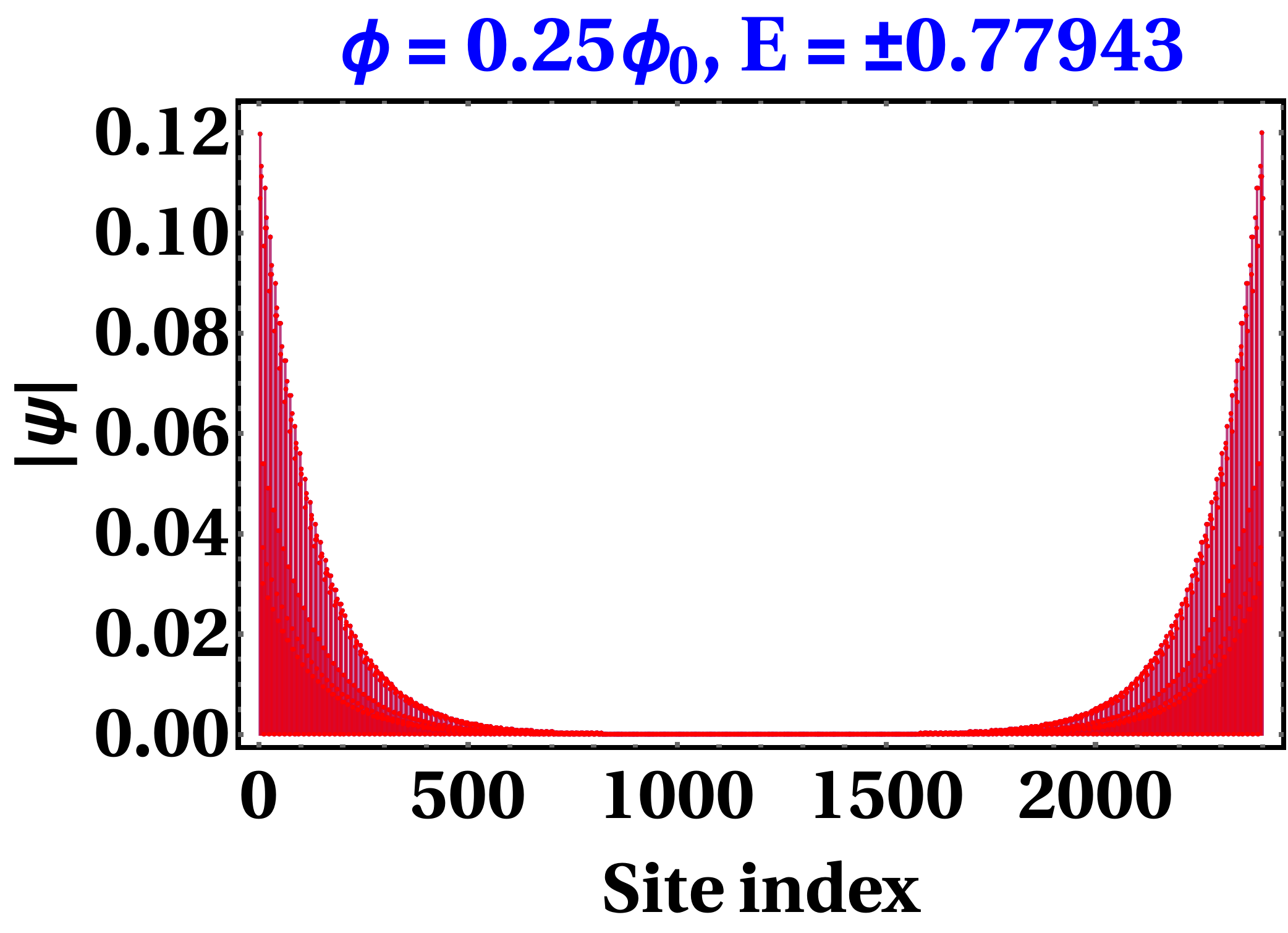}
(e)\includegraphics[width=.44\columnwidth]{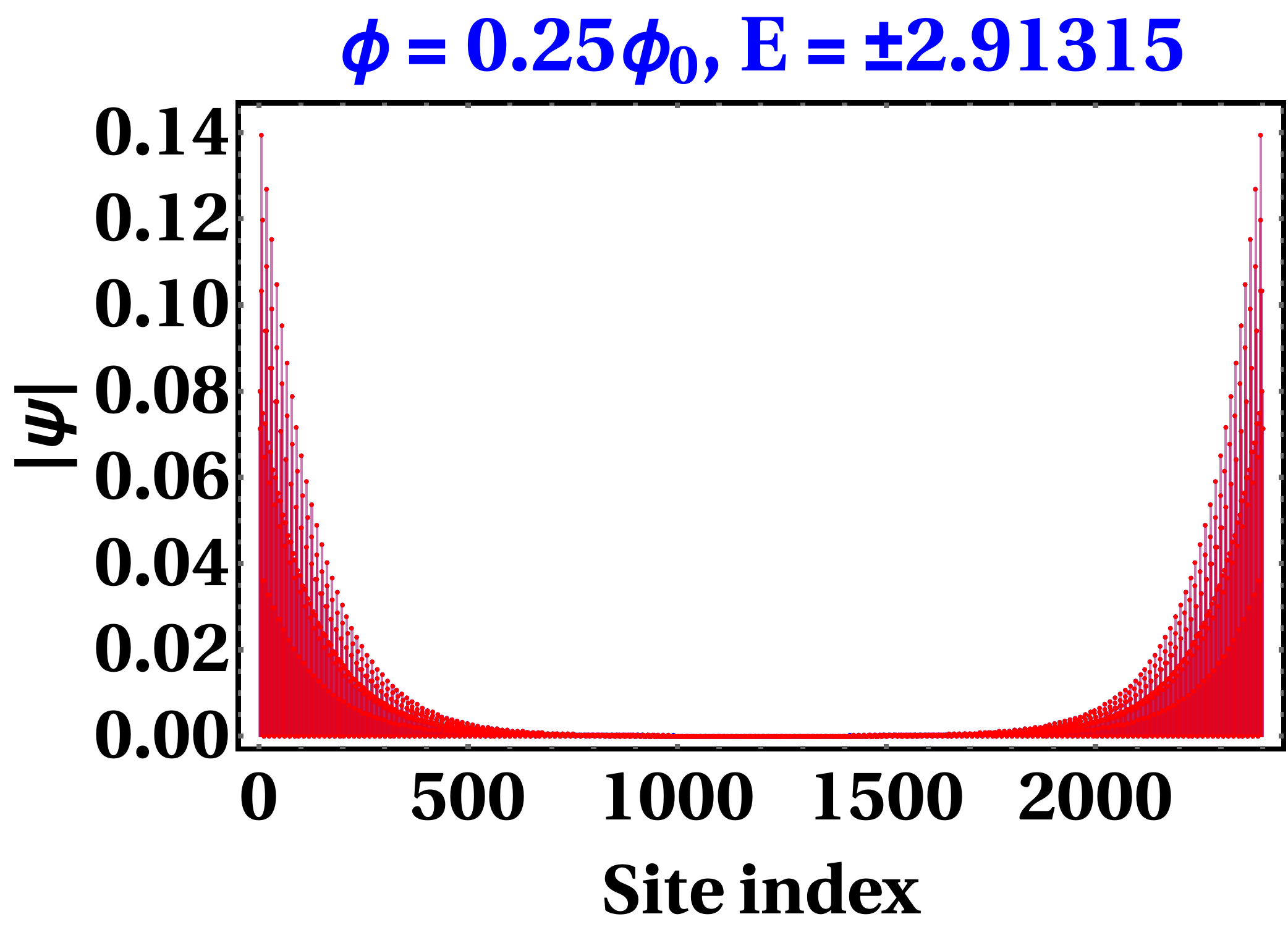}
\caption{(Color online) Distribution of the wavefunction of the edge state energies for SSH-Hexagonal lattices with energy (a) $E = \pm 1.4$ (b) $E = \pm 2.97321$ (in the absence of magnetic flux, $\phi = 0$) (c) $E = \pm 1.91469$, (d) $E = \pm 0.77943$ and (e) $E = \pm 2.91315$ (under applied magnetic flux, $\phi = 0.25 \phi_{0}$). We have applied open boundary conditions for $N_x=200$, where The number of unit cells taken along the $x$-direction is indicated by $N_x$. The parameters are chosen as $ \epsilon = 0, u_{1} = u_{2} = u_{3} = 1.4, v = 1, w= 1.1$. Two different edge state energies ($\pm E$) are plotted in the same plot with red and blue color. All of these energies are doubly degenerate.}  
\label{wave2}
\end{figure}

\subsubsection{SSH-Hexagon Lattice (SSHHL)}
For SSHHL the energy band diagram contains both dispersive and non-dispersive bands in the absence of magnetic flux in both $u_{1} = u_{2} = u_{3}$ (see Fig.~\ref{ek2}) and $u{1} = u_{2} \neq u_{3}$ (see Fig.~\ref{ek3}) situation. Here also the eigenvectors of the flat bands are $k$ independent, such that they never enter the Zak phase calculation. At the same time, in both cases, all dispersive bands show the flipping of the Zak phase from $0$ to $\pi$ under both the presence and absence of magnetic flux. \par
In all the above calculations for both SSHDL and SSHHL, we set the parameter as $u_{1} = u_{2}$. Now the question is, what happens at $u_{1} \neq u_{2}$ situation? If we deviate the numerical value of $u_{1}$ from that of $u_{2}$, the quantization of the Zak phase is affected and it no longer gives the quantized values. So, in that situation, the system has no topological invariant. Another situation must be skipped to see TPT when magnetic flux chooses $\phi = 0.5 \phi_{0}$ due to the flatness of all bands in the band structure.\par 

\subsection{Symmetry Operation}
In general, the lattice models which show topological phase transition exhibit time-reversal symmetry. Then the Hamiltonian must satisfy the condition,  $\hat{\mathcal{H}}(-k)^\ast = \hat{\mathcal{H}}(k)$. When no magnetic flux is applied both SSHDL and SSHHL shows time-reversal symmetry. Under magnetic flux, the Peierls' phase factor is introduced with each arm along the plaquette. Due to this, the time-reversal symmetry is broken. \par
Another important symmetry that is associated with the topological lattice models is {\it Chiral symmetry}. Corresponding chiral operator $\mathcal{C}$ is defined as, 
\begin{eqnarray}
{\mathcal{C}}^{-1}~\mathcal{H}(k)~{\mathcal{C}} & = &-\mathcal{H}(k)
     \label{ch}
 \end{eqnarray}
We have calculated the chiral operator for both SSHDL and SSHHL for both presence and absence of magnetic flux. Although the Hamiltonian is changed under magnetic flux, still we get a common chiral operator for both zero(non-zero) magnetic fluxes. The chiral operator for SSHDL ($\mathcal{C}_{1}$) and SSHHL ($\mathcal{C}_{2}$) can be written as,
\begin{eqnarray}
    \mathcal{C}_{1} & = & diag[1,-1,-1,1,-1,1,1,-1]\nonumber\\
     \mathcal{C}_{2} & = & diag[1,-1,-1,1,1,-1,1,-1,-1,1,1,-1]
\end{eqnarray}
So, in the absence of magnetic flux, SSHDL and SSHHL both obey the time-reversal as well as the chiral symmetries.  The presence of magnetic flux destroys time-reversal symmetry and only the chiral symmetry survives. As discussed earlier, in both these cases (presence and absence of magnetic flux) we get the topological invariant associated with each dispersive band, which indicates that even in the absence of time-reversal symmetry topological phase transition can take place in such decorated lattices. \par
\section{Edge States}
\label{edge}
The appearance of the edge states at the boundaries of the topologically non-trivial insulating phase gives a positive signal of topological phase transition.  To examine these edge states, we create a finite-sized SSHD (SSHH) chain by connecting two hundred unit cells along the $x$-axis. This construction terminates with a unit cell at the end. In Fig.~\ref{e-vs-w}(a) the variation of energy spectrum against the inter-cell hopping $w$ is demonstrated for SSH-Diamond lattice when no magnetic flux is trapped in the diamond cavity. The existence of the edge states at the topological non-trivial insulating phase is apparent. The flat band energy $E = \epsilon$ is present in both topological (non)-trivial insulating phase and highly degenerate. In this case, the edge state at $E = \epsilon$ is hidden in the topological non-trivial phase. \par 
A Similar plot for SSHHL is depicted in Fig.~\ref{e-vs-w}(c) when $u_{1} = u_{2} = u_{3}$ and no magnetic flux is applied. Here also edge states are apparent and the co-existence of edge state energy with flat band energy at $E = \epsilon \pm u_{3}$ is observed. When $u_{1} = u_{2} \neq u_{3}$ is satisfied, edge state energy is shifted from flat band energy, and the clear appearance of edge states are visible (as shown in Fig.~\ref{e-vs-w} (e)).\par
Under magnetic flux, the variation of the energy spectrum is plotted against the inter-cluster hopping $w$ for both SSHD and SSHH lattices Fig~\ref{e-vs-w}(b,d,f). The existence of edge states at each gap-closing energy is observed in topological non-trivial insulating phase $v<w$.

\begin{figure}[ht]
\centering
(a)\includegraphics[width=.44\columnwidth]{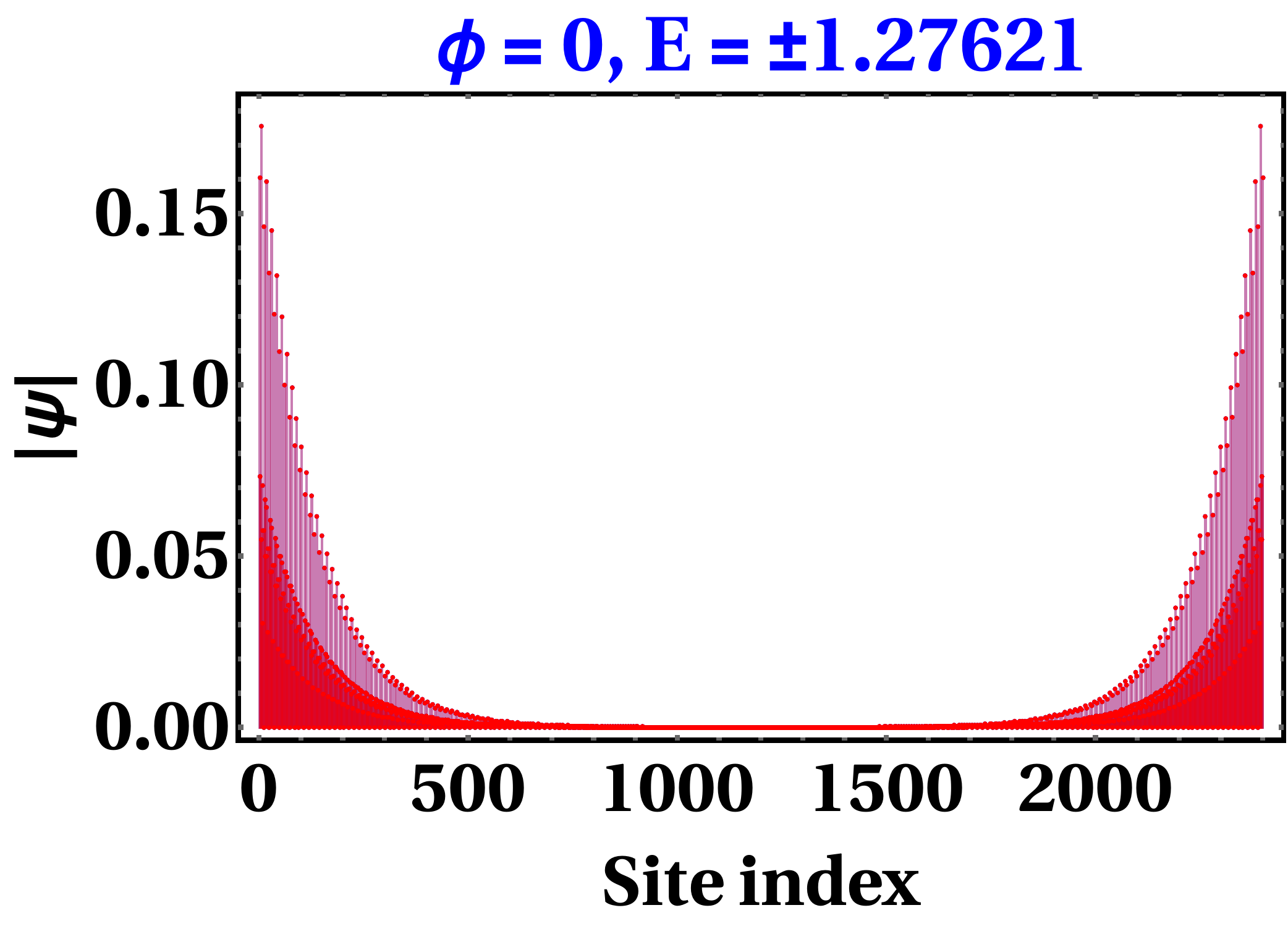}
(b)\includegraphics[width=.44\columnwidth]{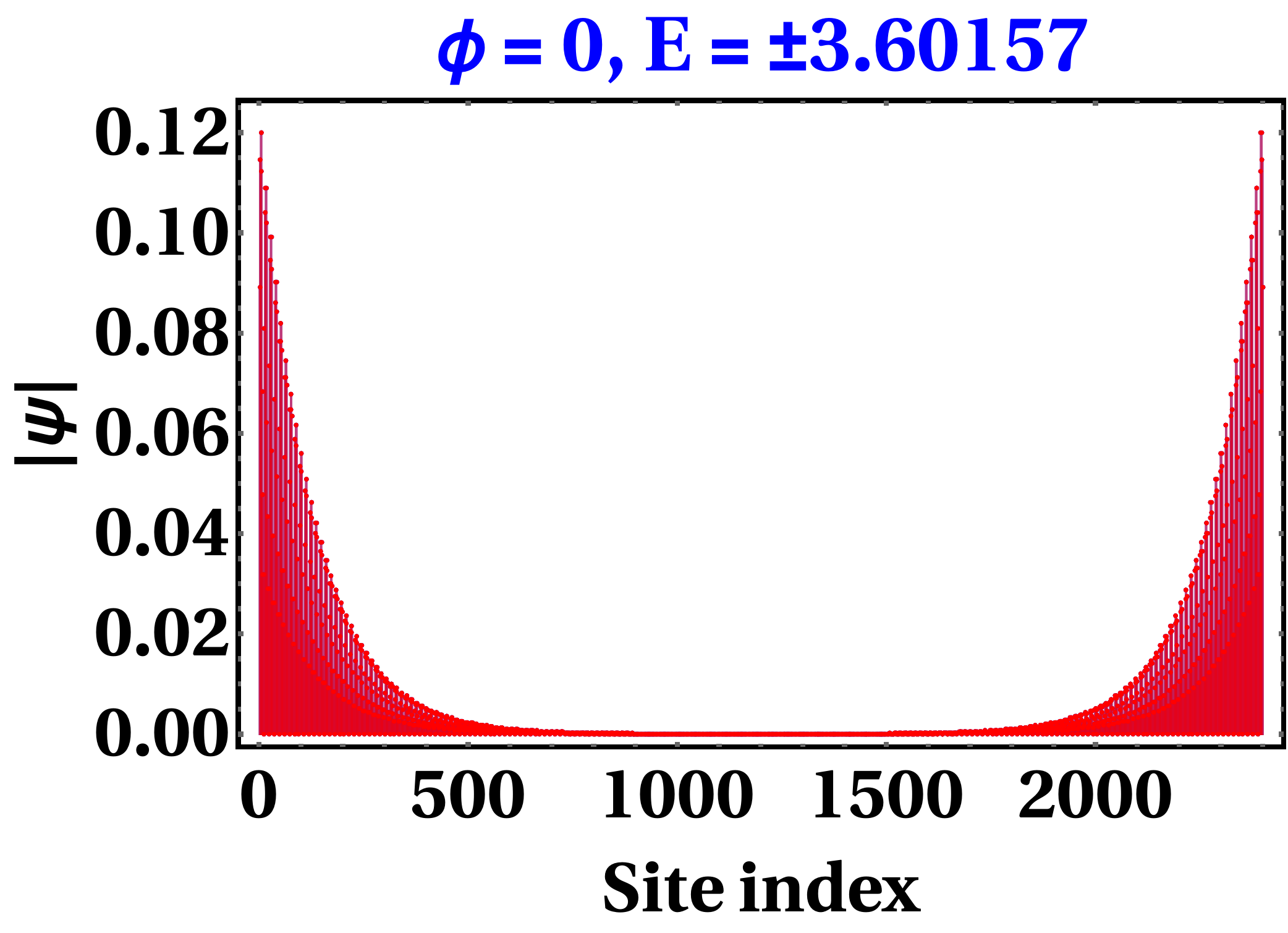}
(c)\includegraphics[width=.44\columnwidth]{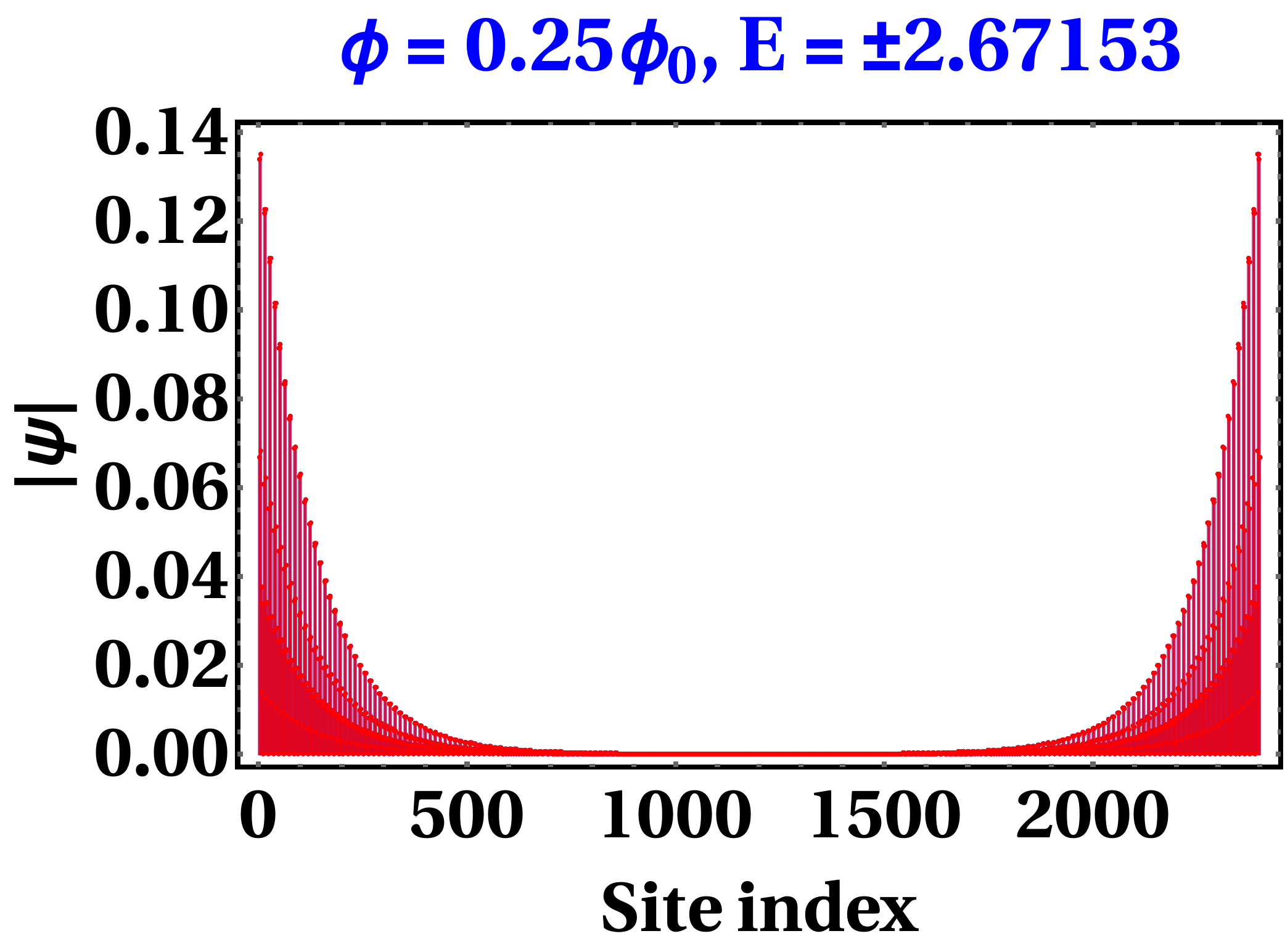}
(d)\includegraphics[width=.44\columnwidth]{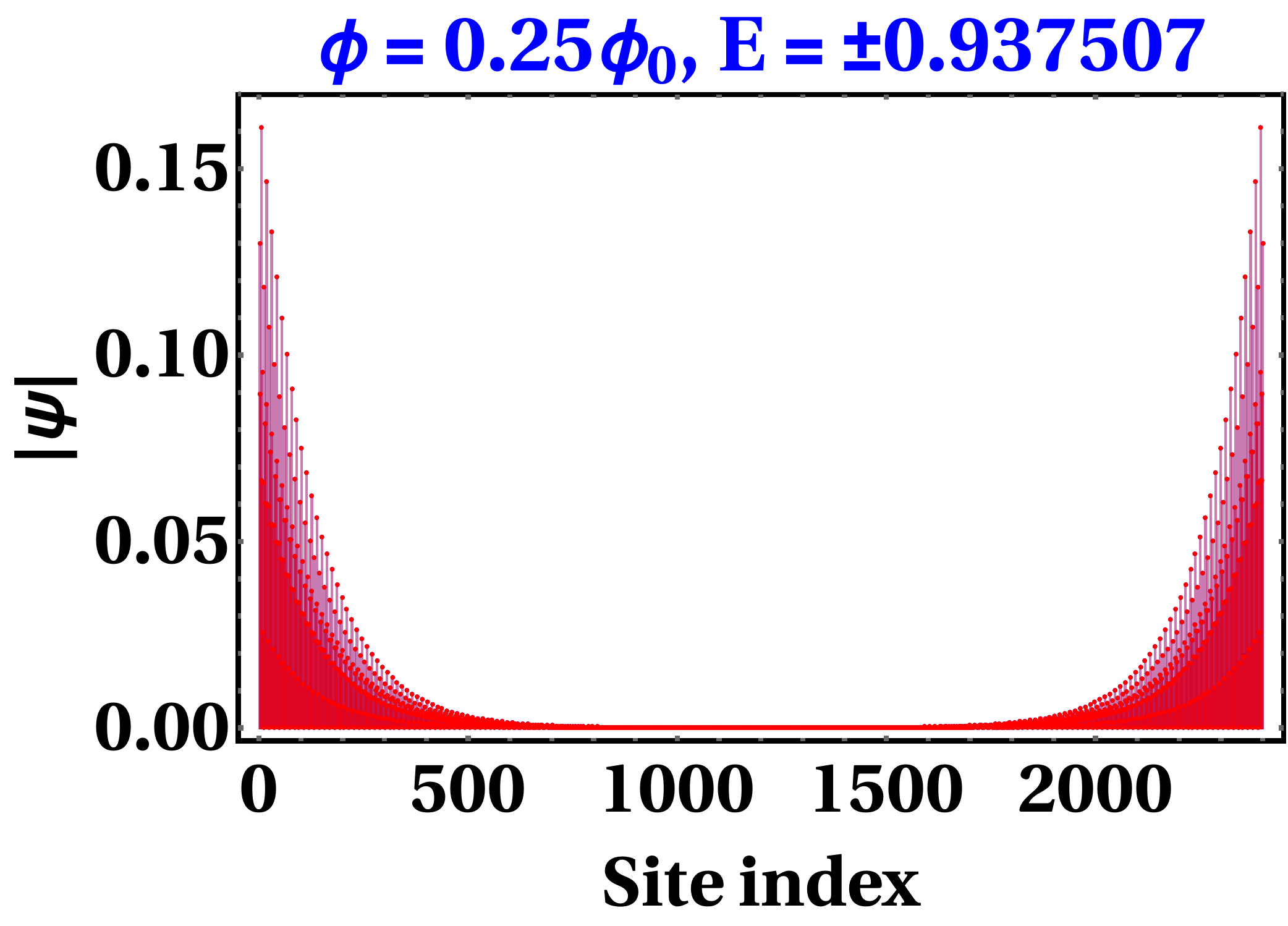}
(e)\includegraphics[width=.44\columnwidth]{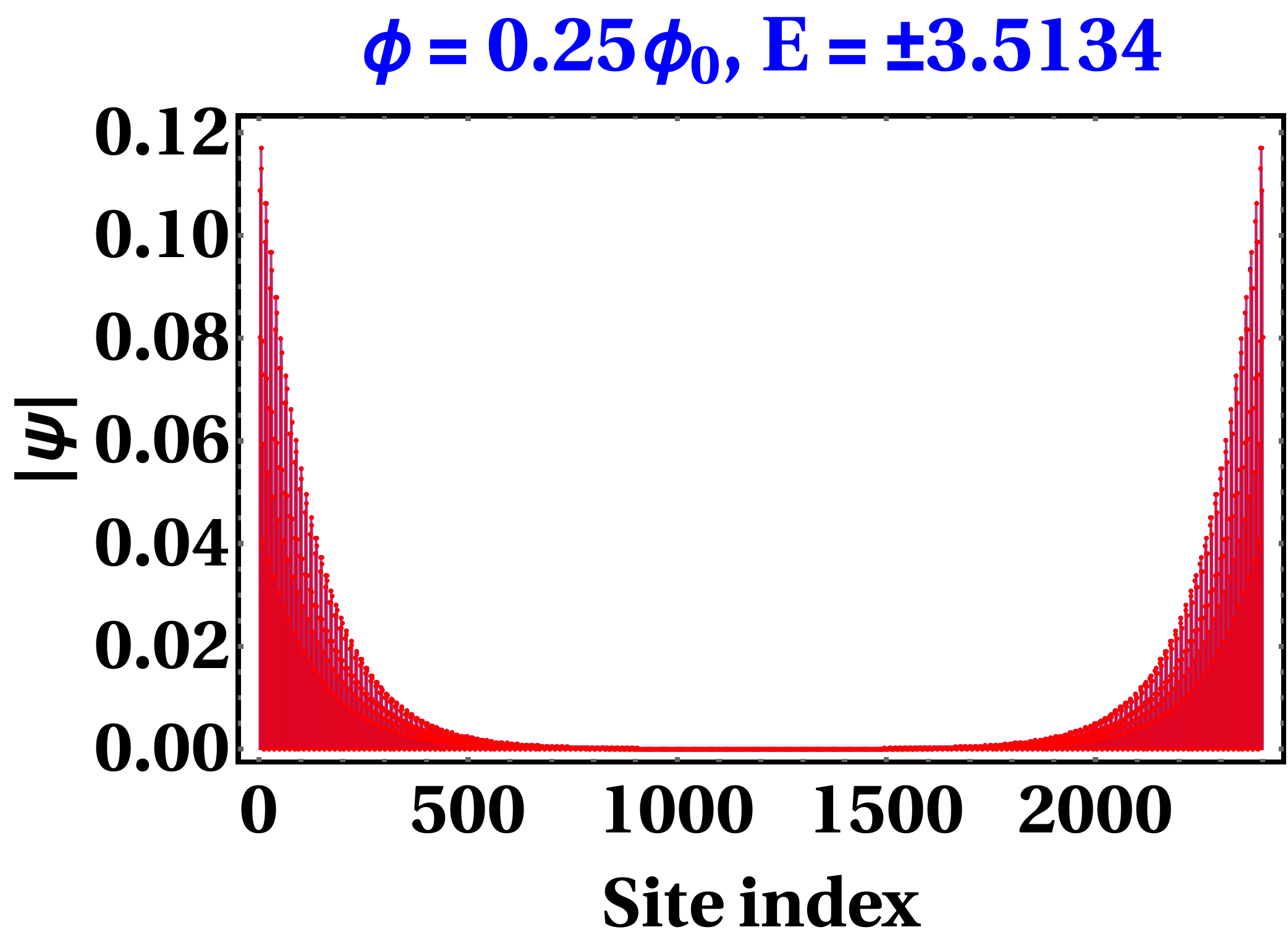}
\caption{(Color online) Distribution of the wavefunction of the edge state energies for SSH-Hexagonal lattices with energy (a) $E = \pm 1.27621$ (b) $E = \pm 3.60157$ (in the absence of magnetic flux, $\phi = 0$) (c) $E = \pm 2.67153$, (d) $E = \pm 0.937507$ and (e) $E = \pm 3.5134$ (under applied magnetic flux, $\phi = 0.25 \phi_{0}$). We have implemented open boundary conditions for $N_x=200$, where $N_x$ refers to the total number of unit cells arranged along the $x$-direction. The parameters are chosen as $ \epsilon = 0, u_{1} = u_{2} =1.4, u_{3} = 2.4, v = 1, w= 1.1$. Two different edge state energies ($\pm E$) are plotted in the same plot with red and blue color. All of these energies are doubly degenerate.}  
\label{wave3}
\end{figure}
\subsubsection{Distinguishing edge state energy from the flat band energy}
In flux-free case both SSHD and SSHH ( at $u_{1} = u_{2} = u_{3}$) lattice the energy gap is opened around the flat band energy $E = \epsilon$ and $E = \epsilon \pm u_{3}$ respectively. To distinguish the edge state energy from the flat band energy in the topological non-trivial insulating phase ($v<w$), we have to look at the degeneracy of that energy level for two insulating phase (both trivial and non-trivial phases). It is seen that this highly degenerate energy state is $n$-fold degenerate (say) for the topological trivial insulating phase ($v>w$), whereas it shows $n+2$-fold degeneracy in the topological non-trivial insulating phase. This extra two-fold degeneracy of the energy level for the topologically non-trivial insulating phase is actually responsible for the doubly degenerate edge states. So using the degeneracy of the energy levels, it is possible to pick up the edge states from the highly degenerate flat-band energy when gap opening-closing occurs around the flat band (see Fig.~\ref{ek1}(a,c), Fig.~\ref{ek2}(a,c)). 

\subsubsection{Distribution of the edge state wavefunction around the atomic sites}
Fig.~\ref{wave1} shows the distribution of the amplitudes of the wavefunction corresponding to the edge state energies for SSH-Diamond lattice both in the absence and in the presence of a magnetic flux. The wavefunction lives only on both sides of the finite-sized chain, in the bulk portion it has no amplitude. Now if we move to the topologically trivial phase ($v>w$), there are no energy states at these particular eigenvalues (except $E = \epsilon$). As $E=\epsilon$ is the flat band energy, it is located in both insulating phases (trivial and non-trivial). In topological non-trivial insulating phase $v<w$ this energy state only shows the {\it edge localization } for extra two-fold degenerate energy level and in all other states at this energy (both trivial and non-trivial phase) never shows the edge localization which ensures that they are not protected by topology. \par
The distribution of the amplitudes of the wavefunction corresponding to the edge states for the SSH-Hexagon lattice is depicted in Fig.~\ref{wave2} for $u_{1} = u_{2} = u_{3}$. The behavior of the edge state-distribution is similar to the SSH-Diamond lattice. In Fig.~\ref{wave3} edge state distributions are shown when $u_{1} = u_{2} \neq u_{3}$. Now flat band energy and edge state energy are chosen to have different values such that overlapping of edge state with flat band energy is completely removed. \par
So, in both the lattices, a clear existence of edge states is apparent in the topological non-trivial insulating phase. It is also seen that these edge-state energies are exactly equal to the gap-closing energies and there are no such states in the topological trivial insulating phase. So these two seemingly equivalent insulating phases (for $v>w$ and $v<w$) are not equal. They can be distinguished by the existence of the edge states. \par
\subsubsection{Robustness of the edge state against disorder}
To investigate the robustness of the edge states in such decorated lattice models, we introduced disorder in the inter-cell hopping $w$ in all unit cells. The disorder is assigned by the random component $\delta w$ which can take any value between the interval $0$ to $1$. So now the inter-cell coupling becomes $w +\delta w$. Under such applied disorder conditions all inter-cell coupling in such finite-size systems can take various $w$ within the interval $w+\delta w$. The edge state energy and the edge localized behaviour of the wavefunction are found to remain unaltered.\par
\section{Topological Issues related to similar types of decorated lattices}
\label{model}
\subsection{SSH-Square-Hexagon and SSH-Square-Octagon lattices}
\begin{figure}[ht]
\centering
(a)\includegraphics[width=.9\columnwidth]{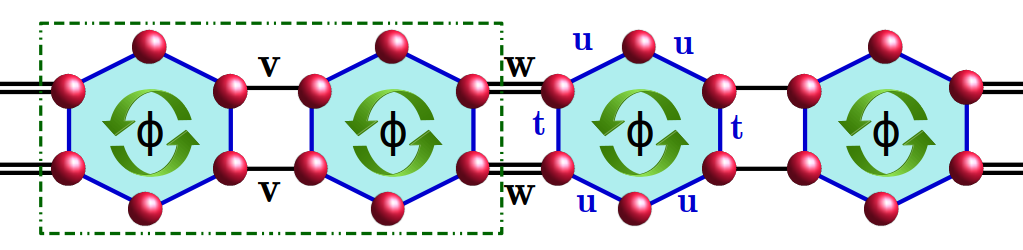}
(b)\includegraphics[width=.9\columnwidth]{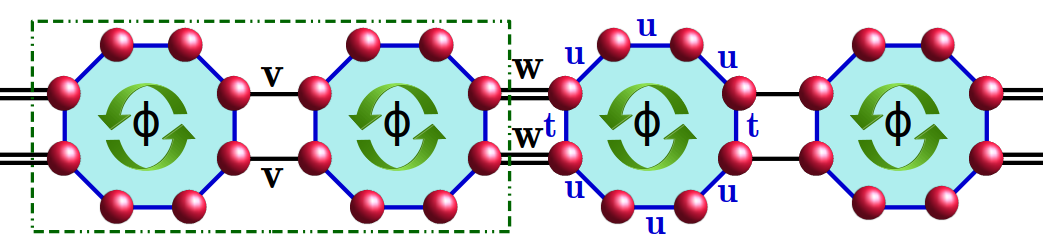}
\caption{(Color online) Schematic diagram of SSH (a) square-hexagon (b) square-octagon array. Each hexagonal or octagonal cavity is trapped by a magnetic flux $\phi$. The unit cells are marked by a green colored box.}  
\label{figG}
\end{figure}
The occurrence of topological phase transition in SSH-Diamond and Hexagon lattices is already discussed in detail for both the presence and absence of magnetic flux. Here we will see that for a similar kind of decorated lattice,(SSH-Square-Hexagon (SSHSH), Fig~\ref{figG}(a) and SSH-Square-Octagon (SSHSO), Fig~\ref{figG} (b) lattices) the TPT is possible only with a special correlation between the numerical values of the external hopping parameter. Along the hexagon or octagon cavity of SSHSH and SSHSO latices, there are two types of hopping parameters $u$ and $t$ as shown in Fig.~\ref{figG}. These hopping integrals pick up a peierls' phase factor in the presence of a non-zero magnetic flux. \par

\begin{figure}[ht]
\centering
(a)\includegraphics[width=.44\columnwidth]{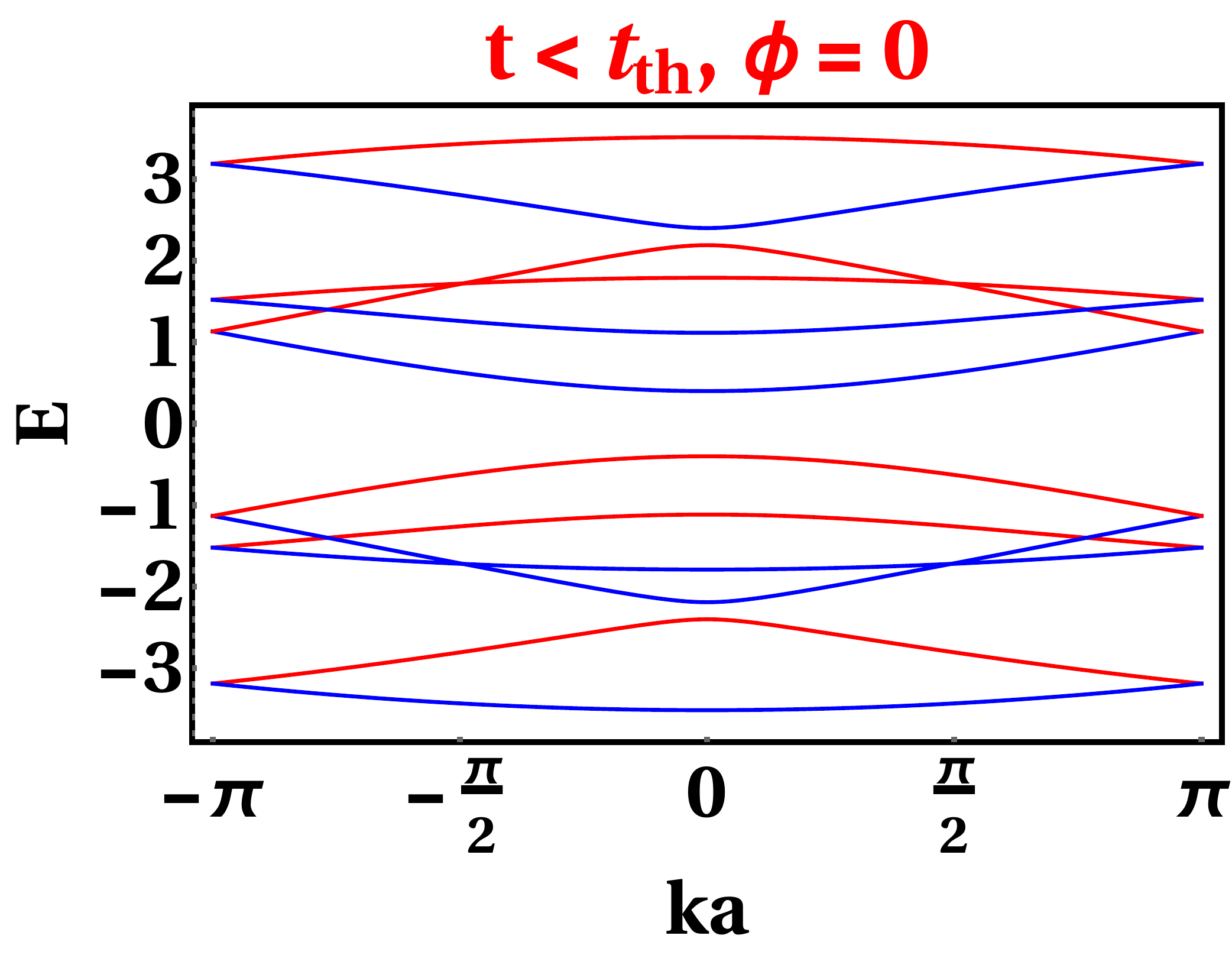}
(b)\includegraphics[width=.44\columnwidth]{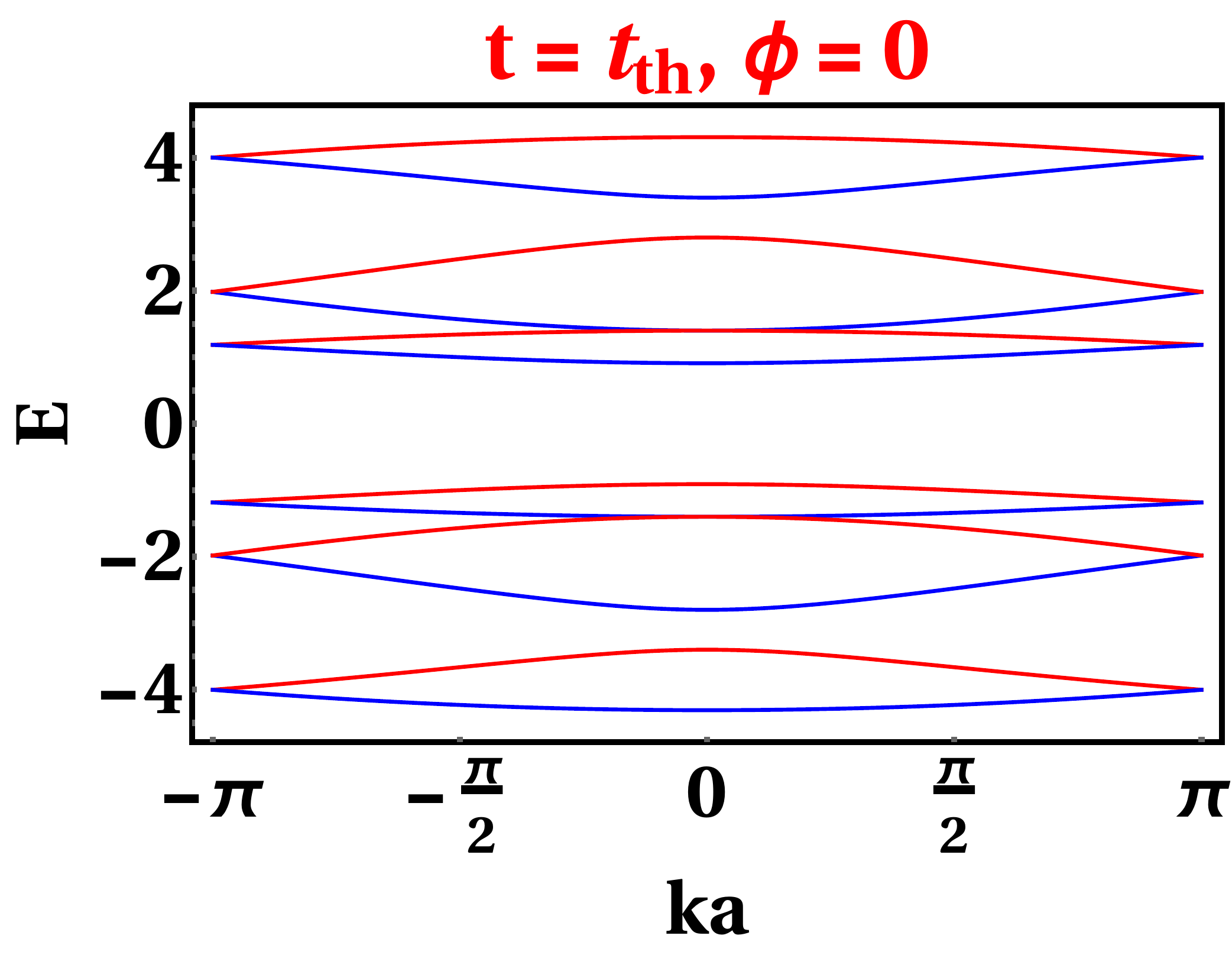}
(c)\includegraphics[width=.44\columnwidth]{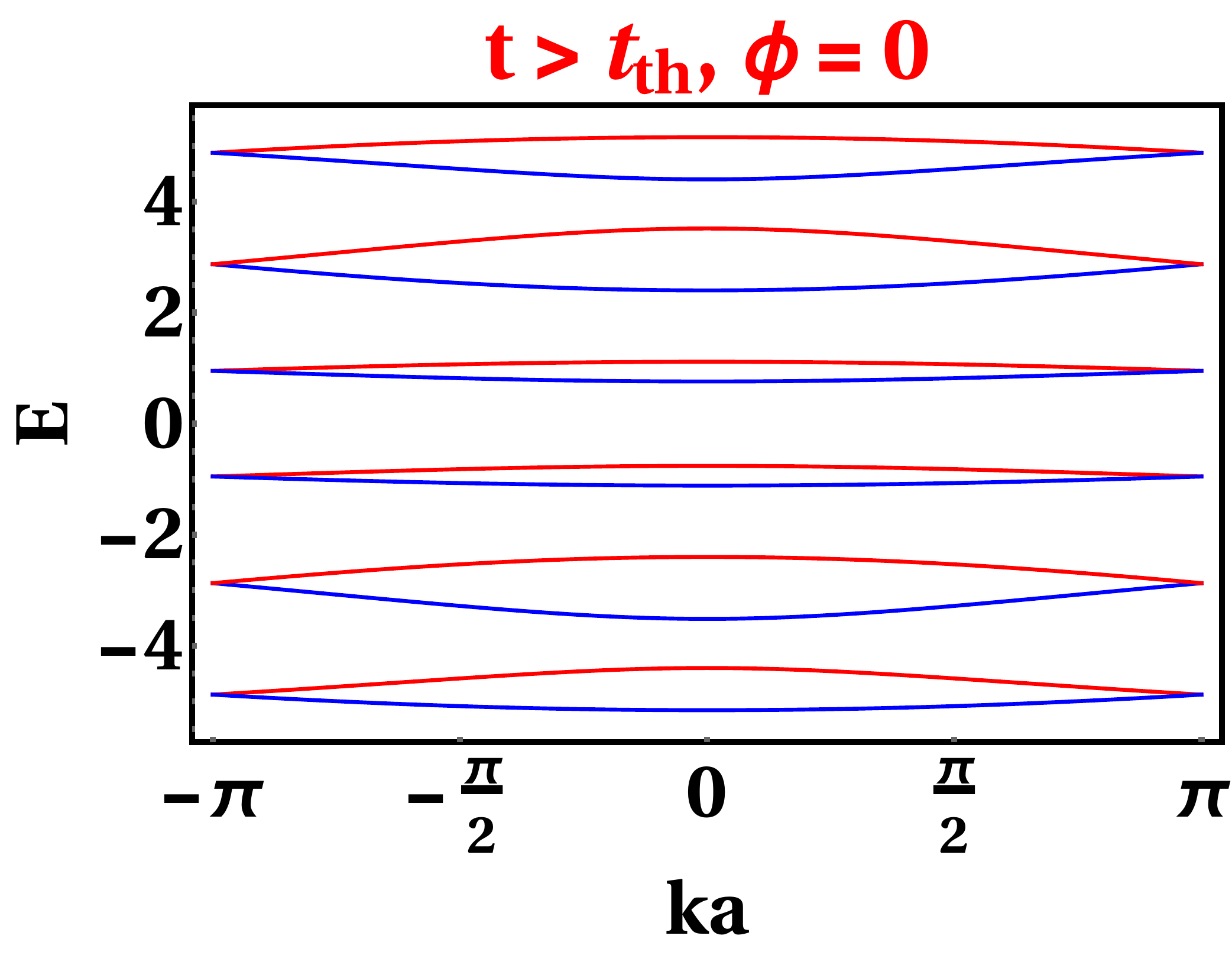}
(d)\includegraphics[width=.44\columnwidth]{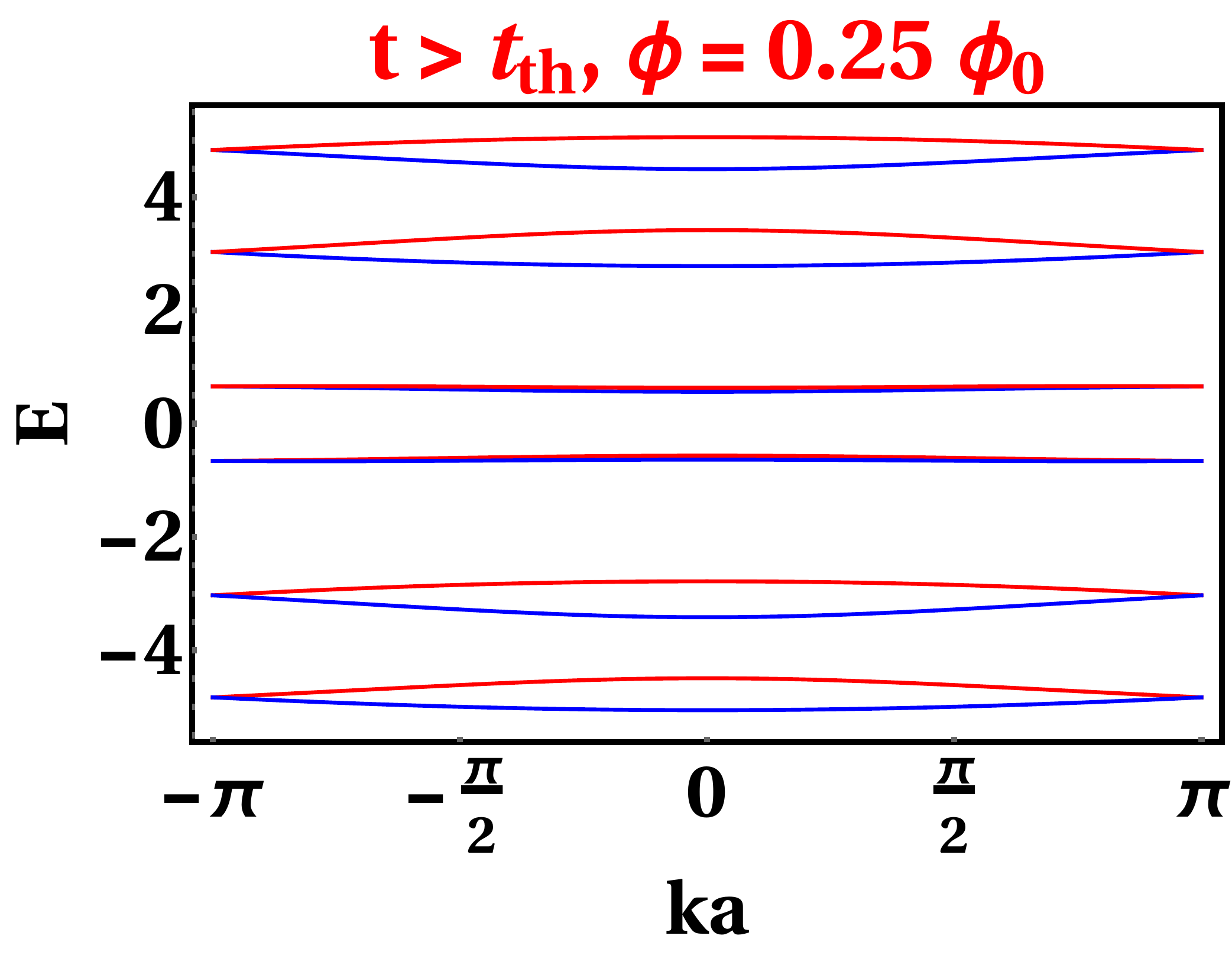}
(e)\includegraphics[width=.44\columnwidth]{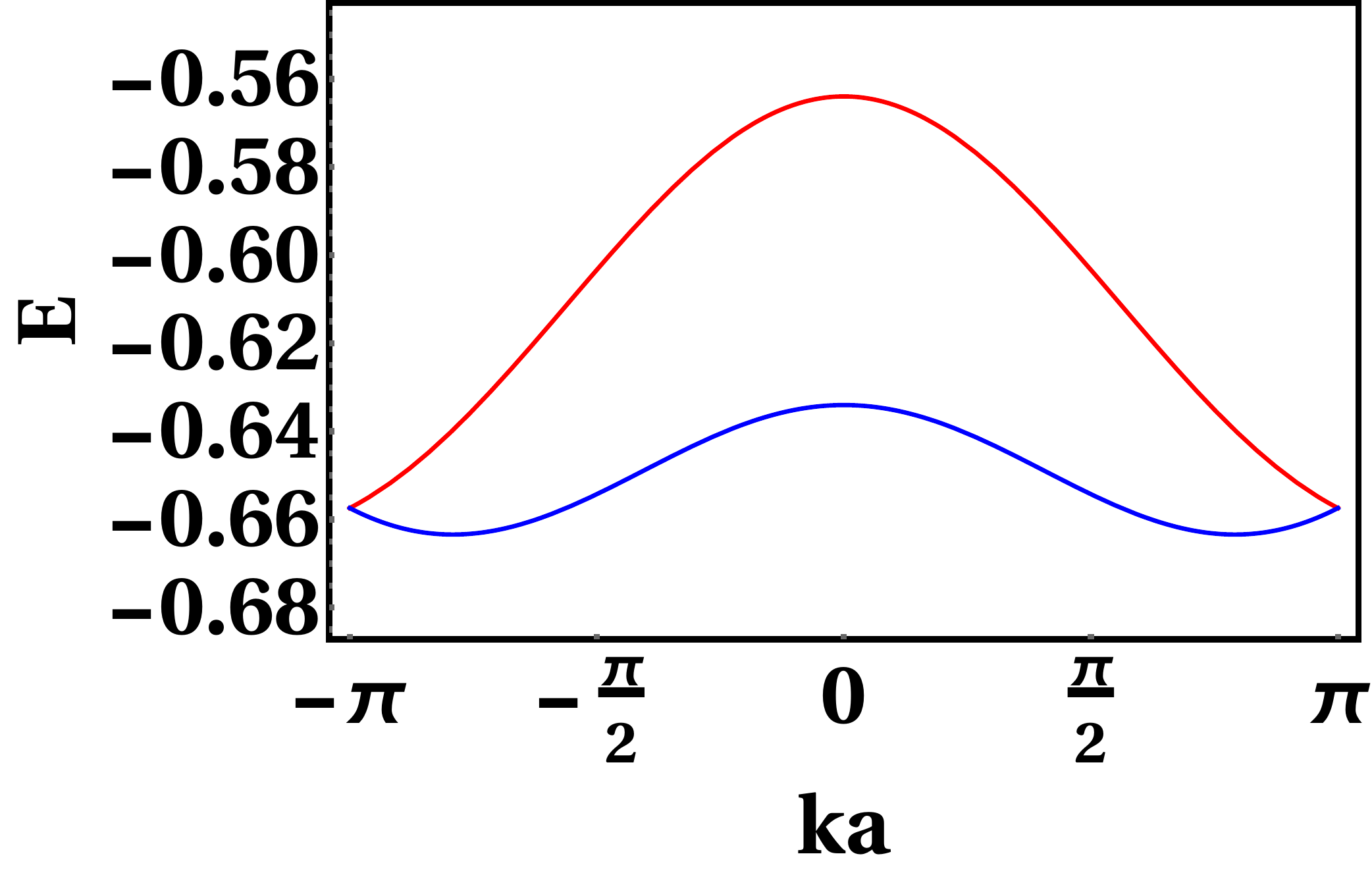}
(f)\includegraphics[width=.44\columnwidth]{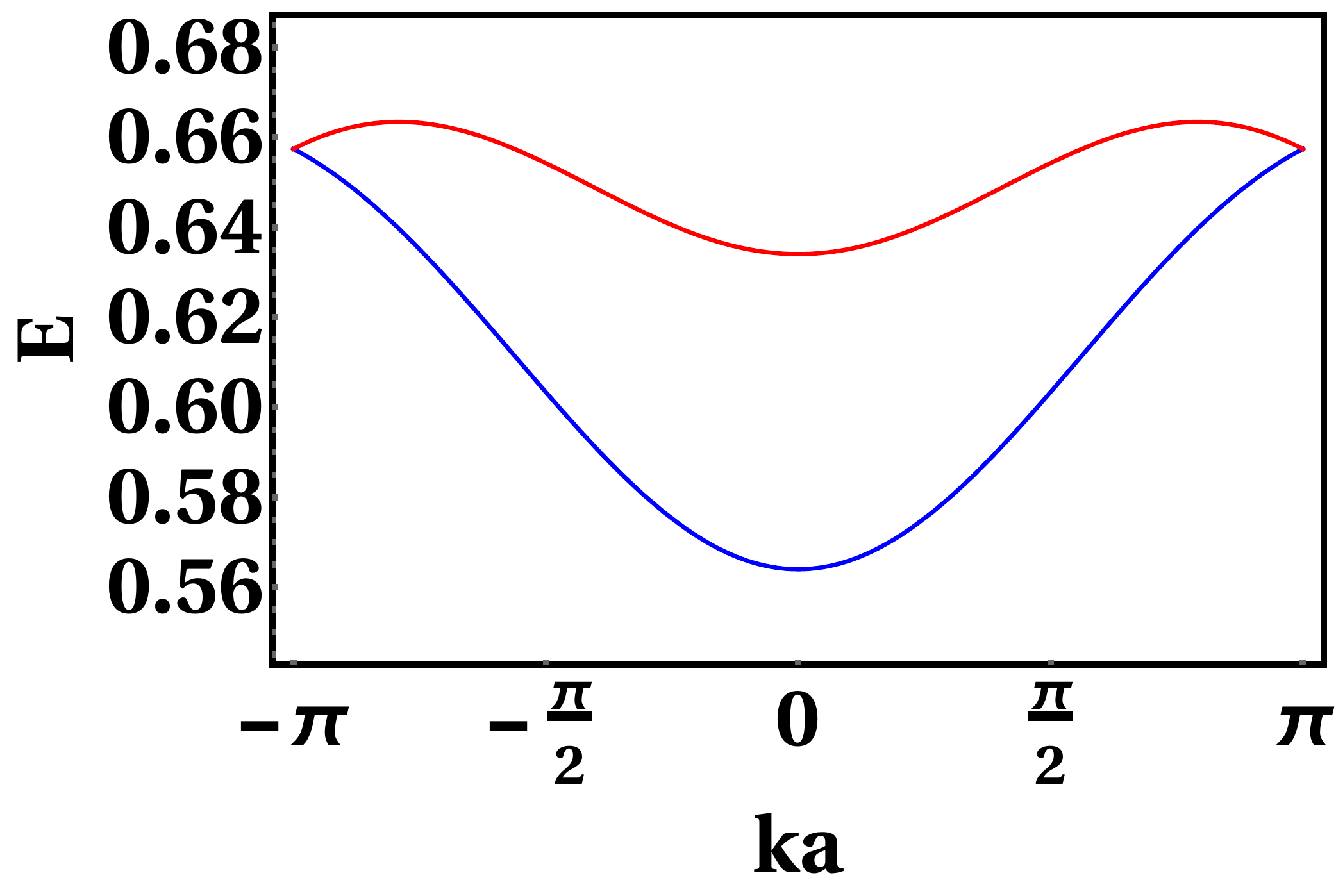}
\caption{(Color online) Energy vs wave vector (E.vs.ka) dispersion relation for the SSH-square-hexagon array. The parameters are chosen as (a) $u = t = 1.4, v = w = 1, \epsilon = 0, \phi = 0$, (b)  $u = 1.4, t = 2.4, v = w = 1, \epsilon = 0, \phi = 0$, (c)  $u = 1.4, t = 3.4, v = w = 1, \epsilon = 0, \phi = 0$, (d)  $u = 1.4, t = 3.4, v = w = 1, \epsilon = 0, \phi = \dfrac{1}{4}\phi_{0}$. (e) and (f) is the magnified version of (d).}  
\label{ek4}
\end{figure}

Under the absence of magnetic flux, the dispersion relation of the SSH-square-hexagon lattice is plotted in Fig.~\ref{ek4}(a). Now a band-overlapped situation is observed. But to see a topological phase transition this overlapping situation must be avoided. Because in that case, the calculation of topological invariant (here the Zak phase) is not possible.\par
The external hopping parameter $t$ plays an important role in the separation of the energy bands. There is a particular value of $t$ above which all bands are get separated. We termed this particular value of $t$ as its threshold value $t_{th}$. This $t_{th}$ can be written as,
\begin{equation}
    t_{th} =\frac{1}{2} \left( v+w+\sqrt{4 u^2+v^2-2 v w+w^2}\right)
    \label{th}
\end{equation}
At $t = t_{th}$ the bands just touch each other as shown in Fig.~\ref{ek4}(b). To observe a topological phase transition (TPT) the value of the external hopping parameter must be set above its threshold value. The separation of the bands is shown in Fig.~\ref{ek4}(c) just imposing $t>t_{th}$ situation. Fig.~\ref{ek4}(d) demonstrates the band diagram under a non-zero magnetic flux $\phi = 0.25 \phi_{0}$.\par
In both cases,(presence and absence of magnetic flux) the gaps are opened at the BZ boundary when $v>w$, closed at $v=w$, and again opened at $v<w$. Here also transition from one insulating phase ($v>w$) to another insulating phase ($v<w$) occur by crossing the metallic phase ($v=w$). This is the primary indication of a topological phase transition.\par
To confirm the TPT a topological invariant associated with each dispersive band must be calculated. It is observed that all Zak phase associated with each dispersive energy band turns out as a quantized value when $t>t_{th}$ situation is achieved. Under non-zero magnetic flux although the time-reversal symmetry is broken still all Zak phases stick with their quantized values. So in such a decorated SSH-square-hexagon lattice, the topological phase transition is possible when the external hopping parameter chooses the appropriate value.\par
For SSH-square-octagon lattice energy band diagram exhibits similar characteristics and all bands contain quantized zak phase when the external parameter is set above its threshold value. So, here also TPT is possible in both the presence and absence of magnetic flux. 
\subsection{SSH-Triangular Lattices}


\begin{figure}[ht]
\centering
\includegraphics[width=.9\columnwidth]{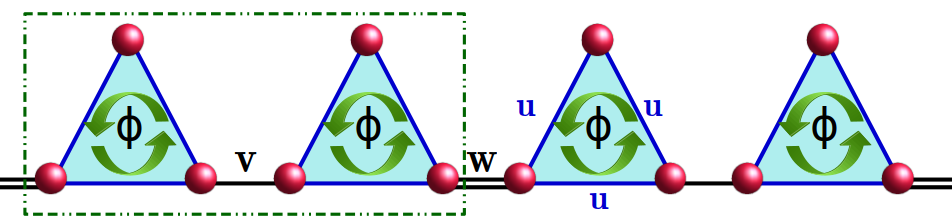}
\caption{(Color online) Schematic diagram of SSH-triangular lattice where each triangular cavity contains a magnetic flux $\phi$. The unit cell is marked by green colored box.}  
\label{figT}
\end{figure}


\begin{figure}[ht]
\centering
(a)\includegraphics[width=.44\columnwidth]{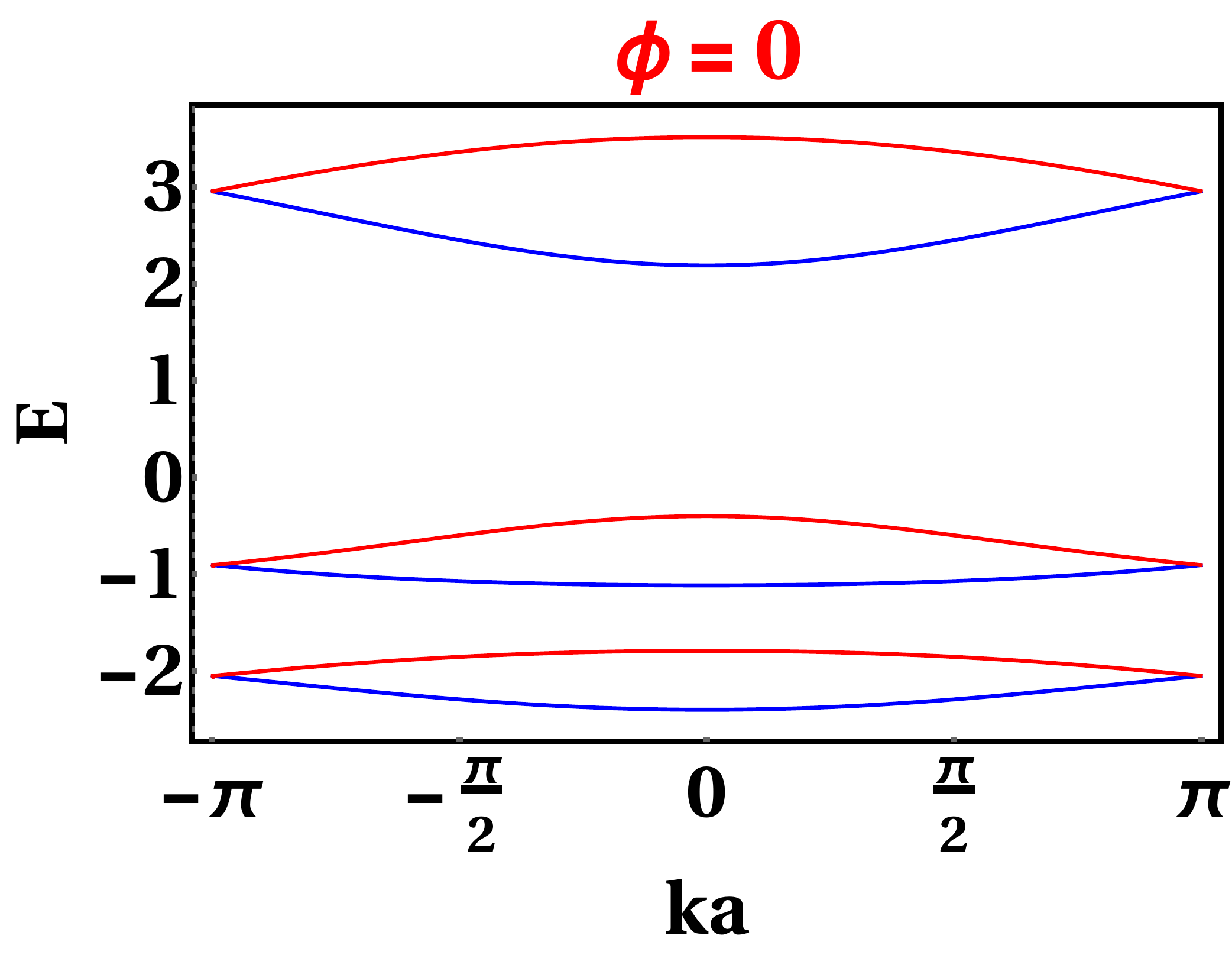}
(b)\includegraphics[width=.44\columnwidth]{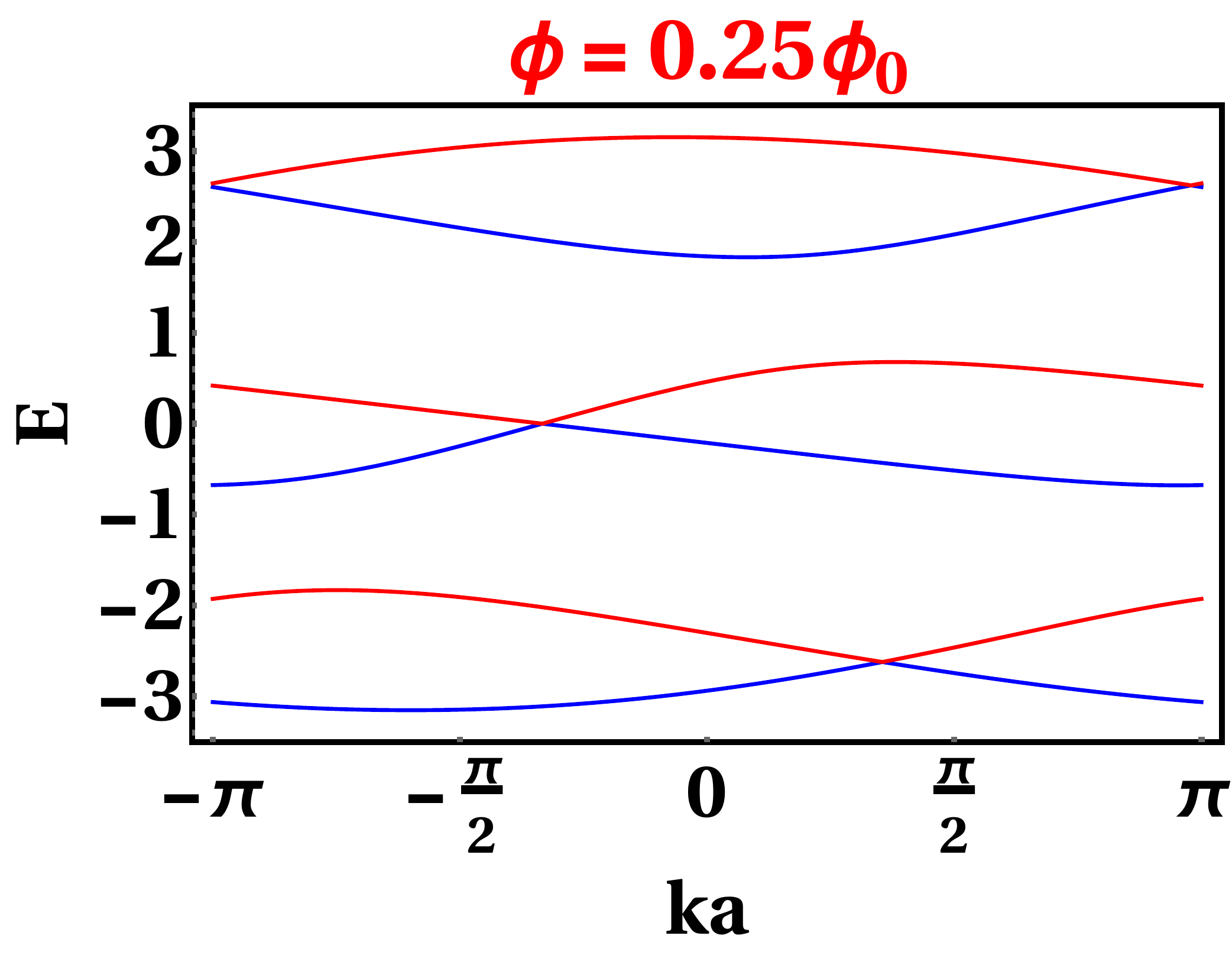}
\caption{(Color online) Energy vs wave vector (E vs ka) dispersion relations for the SSH-Triangular lattice (a) absence of magnetic flux and (b) presence of magnetic flux. The parameters are chosen as (a) $ \epsilon = 0, u = 1.4, v = 1, w = 1,  \phi = 0$, (b)  $ \epsilon = 0, u = 1.4, v = 1, w = 1,  \phi = \frac{1}{4} \phi_{0}$.  }  
\label{ek5}
\end{figure}

In this part, we will investigate the topological phase transition-related issues associated with SSH-triangular (SSHT) lattice geometry (as depicted in Fig.~\ref{figT}). Fig.~\ref{ek5}(a) shows the energy band diagram when no flux is trapped in the triangular cavity. Here the gap opening-closing-opening situation is easily achieved by tuning the inter-cell hopping parameter $w$ with a fixed value of $v$. The topological invariant associated with each dispersive band always gives a quantized value. The edge localized states appear in the topologically non-trivial insulating phase with eigenvalues exactly the same as the gap-closing energy. So under the absence of magnetic flux topological phase transition is possible in SSH-triangular lattice. \par
In Fig.~\ref{ek5}(b) the energy bands are plotted for SSH triangular lattice when a non-zero magnetic flux $\phi = 0.25 \phi_{0}$ is trapped in each triangular cavity. Here gap-opening or closing situation at the BZ boundary which is the primary signal for a TPT, is completely destroyed. So in the presence of magnetic flux the topological phase transition is never possible in such SSH-triangular lattice.\par
\section{Conclusion}
\label{conclusion}
A detailed analysis of topological phases is done in a class of decorated lattices (such as SSH-Diamond lattice and SSH-hexagon lattice) in both the absence and presence of magnetic flux. We describe the systems within a tight binding formalism and using a real space decimation scheme gap-closing, flat band energies are located in an analytical way which supports our graphical results.  In flux-free cases, the occurrence of gap opening-closing-opening at the BZ boundary gives the primary signal for the topological phase transition. The TPT is confirmed by the quantization of Zak phase (topological invariant) associated with each non-dispersive energy band and the appearance of edge localized mode in the topologically non-trivial insulating phase. Such that the {\it Bulk-Boundary-correspondence} is obeyed. The distinguishing of edge states from the flat band energy is done by a proper investigation of the degeneracy of the energy level. For SSH-Hexgon lattices this situation can be overcome just by setting the overlap integral to its appropriate values. In flux-free cases, topological phase transition occurs in the presence of the flat bands which have no significant role in controlling topology due to their $k$-independent eigenvectors. \par
The TPT-related issues in the presence of magnetic flux are discussed. Now TPT is observed without any non-dispersive flat band and the gap opening-closing-opening situation is never affected by the magnetic flux. Although the time-reversal symmetry is broken, still all dispersive bands contain a quantized value of the topological invariant. The topologically non-trivial insulating phase shows a clear appearance of edge states at the gap-closing energies. A common chiral operator is found in both flux-free and non-zero flux cases which can protect the edge states. The robustness of the edge state is examined by assigning a random disorder in the inter-cell coupling.\par
We extended our study to similar types of SSH-square-hexagon and SSH-square-octagon lattices. There we discussed the role of the external hopping parameter in controlling the TPT. The threshold value of this hopping parameter above which all bands get separated is calculated analytically in terms of other parameters.  Finally using the SSH-triangular lattice geometry the impossibility of TPT in the presence of magnetic flux is discussed. So, in this research work, we have examined in detail the (non-)occurrence of TPT in both flux-free and applied flux cases.

\section{Acknowledgement}
I want to express my sincere gratitude to my supervisor Prof. Arunava Chakrabarti for his guidance, invaluable advice, and continuous support while carrying on the present research work. I am thankful to the Government of West Bengal for the SVMCM Scholarship (WBP221657867058). 

\appendix
\section{The kernels of the Hamiltonian}
\begin{widetext}
 The kernels of the Hamiltonian for the unit cell of the  SSHDL and SSHHL are given by,
\begin{equation}
\hat{\mathcal{H}}_{SSHDL}(k) = \left[ \begin{array}{cccccccccccccccc}
 \epsilon & u_{1}e^{-i\theta} & u_{1}e^{i\theta} & 0 & 0 & 0 & 0 & w e^{-ika} \\
 u_{1}e^{i\theta} & \epsilon & 0 & u_{2}e^{-i\theta} & 0 & 0 & 0 & 0 \\
u_{1}e^{-i\theta} & 0 & \epsilon & u_{2}e^{i\theta} & 0 & 0 & 0 & 0 \\
 0 & u_{2}e^{i\theta} & u_{2}e^{-i\theta} & \epsilon & v & 0 & 0 & 0 \\
 0 & 0 & 0 & v & \epsilon & u_{1}e^{-i\theta} & u_{1}e^{i\theta} & 0 \\
 0 & 0 & 0 & 0 & u_{1}e^{i\theta} & \epsilon & 0 & u_{2}e^{-i\theta} \\
 0 & 0 & 0 & 0 & u_{1}e^{-i\theta} & 0 & \epsilon & u_{2}e^{i\theta} \\
 w e^{ika} & 0 & 0 & 0 & 0 & u_{2}e^{i\theta} & u_{2}e^{-i\theta} & \epsilon \\
\end{array}
\right ] 
\label{ham1}
\end{equation}

\begin{equation}
\hat{\mathcal{H}}_{SSHHL}(k) = \left[ \begin{array}{cccccccccccccccc}
 \epsilon & u_{1}e^{i\theta} & u_{1}e^{-i\theta} & 0 & 0 & 0 & 0 & 0 & 0 & 0 & 0 & w e^{-ika} \\
 u_{1}e^{-i\theta} & \epsilon & 0 & u_{3}e^{i\theta} & 0 & 0 & 0 & 0 & 0 & 0 & 0 & 0 \\
 u_{1}e^{i\theta} & 0 & \epsilon & 0 & u_{3}e^{-i\theta} & 0 & 0 & 0 & 0 & 0 & 0 & 0 \\
 0 & u_{3}e^{-i\theta} & 0 & \epsilon & 0 & u_{2}e^{i\theta} & 0 & 0 & 0 & 0 & 0 & 0 \\
 0 & 0 & u_{3}e^{i\theta} & 0 & \epsilon & u_{2}e^{-i\theta} & 0 & 0 & 0 & 0 & 0 & 0 \\
 0 & 0 & 0 & u_{2}e^{-i\theta} & u_{2}e^{i\theta} & \epsilon & v & 0 & 0 & 0 & 0 & 0 \\
 0 & 0 & 0 & 0 & 0 & v & \epsilon & u_{1}e^{i\theta} & u_{1}e^{-i\theta} & 0 & 0 & 0 \\
 0 & 0 & 0 & 0 & 0 & 0 & u_{1}e^{-i\theta} & \epsilon & 0 & u_{3}e^{i\theta} & 0 & 0 \\
 0 & 0 & 0 & 0 & 0 & 0 & u_{1}e^{i\theta} & 0 & \epsilon & 0 & u_{3}e^{-i\theta} & 0 \\
 0 & 0 & 0 & 0 & 0 & 0 & 0 & u_{3}e^{-i\theta} & 0 & \epsilon & 0 & u_{2}e^{i\theta} \\
 0 & 0 & 0 & 0 & 0 & 0 & 0 & 0 & u_{3}e^{i\theta} & 0 & \epsilon & u_{2}e^{-i\theta} \\
 w e^{ika} & 0 & 0 & 0 & 0 & 0 & 0 & 0 & 0 & u_{2}e^{-i\theta} & u_{2}e^{i\theta} & \epsilon \\
\end{array}
\right ] 
\label{ham2}
\end{equation}
where $\theta = \frac{2 \pi \phi}{4 \phi_{0}}$ (for SSHDL) and $\theta = \frac{2 \pi \phi}{6 \phi_{0}}$(For SSHHL).\par
\end{widetext}

\section{Behaviour of the edge state wavefunction under different disorder strength}
\begin{figure}[ht]
\centering
\includegraphics[width=0.95\columnwidth]{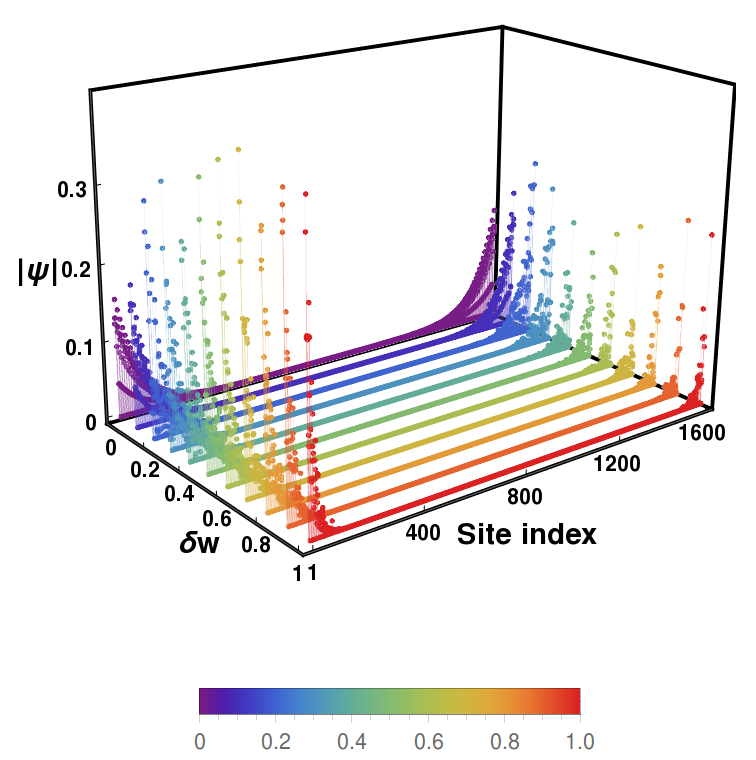}
\caption{(Color online) Distribution of edge sate wavefunction corresponding to edge state energy $E = - 2.97321$ with different disorder strength $\delta w$ for SSH-Diamond lattice. We have used open boundary conditions for $N_x=200$, where $N_x$ denotes the number of unit cells taken along the $x$-direction. The parameters are chosen as $ \epsilon = 0, u_{1} = u_{2} = 1.4, v = 1, w= 1.1, \phi = 0$. This state is doubly degenerate, only one distribution is depicted here. } 
\label{disorder}
\end{figure}
The robustness of the edge states wavefunction has already been discussed previously in details. Here we will investigate it in detail under different disorder strengths. To explore this, first, it is ensured that the system in its topological non-trivial insulating phase ($v < w$) for any strength of disorder. We have put the disorder in the inter-cluster hopping $w$. As already discussed, now this coupling becomes $w+\delta w$, which can take any value within this interval. This disorder is introduced by the random number generation, it can take different disorder distribution in different trials. But interestingly, it is seen that the states always exhibit their edge localized character. In Fig.~\ref{disorder} the amplitude of edge state wavefunction with energy $E = - 2.97321$ for flux-free SSHDL is plotted under different disorder strengths.  Here the disorder window $\delta w$ is varied from $0$ to $1$ within an interval $0.1$ and the corresponding wavefunction distribution is plotted in different colors as shown in Fig~\ref{disorder} for a particular trial. So it is ensured that although a sufficient disorder is applied, still the wavefunction retains its edge localized character. As the disorder is assigned through a random number generation, if we go for another trial, the disorder distribution is changes but the edge localization is not affected. Under an applied magnetic field and in all other cases, it exhibits exactly the similar character.

\end{document}